\documentclass[12pt]{iopart}
\usepackage{graphicx}
\usepackage{pst-all}
\usepackage{pstcol}
\begin{document}

\title{DETERMINATION OF PHASE DIAGRAMS VIA
COMPUTER SIMULATION: METHODOLOGY AND APPLICATIONS TO WATER, ELECTROLYTES AND PROTEINS.}

\author{C. Vega\footnotemark,  E. Sanz, J. L. F. Abascal and E. G. Noya }

\address{Departamento de Qu\'{\i}mica F\'{\i}sica, Facultad de Ciencias
Qu\'{\i}micas,  Universidad Complutense, 28040 Madrid, Spain}

\begin{abstract}
In this review we focus on the determination of phase diagrams by computer simulation
with particular attention to the fluid-solid and solid-solid equilibria. 
The methodology to compute the free energy of solid phases will be discussed. 
In particular, the Einstein crystal and Einstein molecule methodologies are
described in a comprehensive way. It is shown that both methodologies
yield the same free energies and that free energies of solid phases 
present noticeable finite size effects. In fact this is the case for hard spheres 
in the solid phase.  Finite size corrections can be introduced, 
although in an approximate way, to correct for the dependence of the free energy 
on  the size of the system. The computation of free energies of solid phases 
can be extended to molecular fluids. The  procedure to compute free
energies of solid phases of water (ices) will be described in detail. 
The free energies of ices Ih, II, III, IV, V, VI, VII, VIII, IX, XI and XII 
will be presented for the SPC/E and TIP4P models of water. 
Initial coexistence points leading to the determination of the
phase diagram of water for these two models will be provided. 
Other methods to estimate the melting point of a solid, as the direct fluid-solid coexistence
or simulations of the free surface of the solid will be discussed. It will be shown that the 
melting points of ice Ih for several water models, obtained from free energy calculations, direct 
coexistence simulations and free surface simulations, agree within their statistical 
uncertainty. Phase diagram calculations can indeed help to improve
potential models of molecular fluids. For instance, for water, the potential model 
TIP4P/2005 can be regarded as an improved version of TIP4P. 
Here we will review some recent work on the 
phase diagram of the simplest ionic model, the restricted primitive model. Although 
originally devised to describe ionic liquids, the model is becoming quite popular to describe
the behaviour of charged colloids. Besides the possibility of obtaining fluid-solid equilibria
for simple protein models will be discussed. In these primitive models, the protein is 
described by a spherical potential with certain anisotropic bonding sites (patchy sites).

\end{abstract}

\footnotetext{published in J.Phys.Condens.Matter, volume 20, 153101, (2008)}
\newpage 

\tableofcontents
\newpage

\section{Introduction}
One of the first findings of computer simulation was the discovery of a fluid-solid
transition for a system of hard spheres \cite{wood,alder}. At that time the idea of a solid phase
without the presence of attractive forces was not easily accepted. It took some
time to accept it, and it was definitively proved after the work of Hoover and Ree \cite{hoover-ree},
in which the location of the transition was determined beyond any doubt and, even 
more recently, when it was experimentally found for colloidal systems \cite{pusey}.
Certainly the study of phase transitions has always been a hot topic within the 
area of computer simulation. However, fluid-fluid 
phase transitions (liquid immiscibility, vapor-liquid) have received by far more
attention than fluid-solid equilibria \cite{ACP_2000_115_0113_nolotengo}.
The appearance of the Gibbs ensemble \cite{panagiotopoulos87,panagiotopoulos88} in the
late eighties provoked an explosion of papers dealing with vapor-liquid 
equilibria.
The method has been applied to determine vapor-solid equilibria \cite{vapor_solid}, but 
not for studying fluid-solid equilibria.

An interesting approach to the problem of the fluid-solid 
equilibrium was provided  by Wilding and coworkers, who, in 1997, 
proposed the phase switch Monte Carlo method \cite{bruce97,wilding_rev}. This method
was first applied to the study of 
the free energy difference between fcc and hcp close packed structures 
of hard spheres \cite{bruce97,wilding_pre00}. Three years later, Wilding and Bruce showed that the method
could be applied to obtain fluid-solid equilibrium, and the fluid-solid equilibria
of hard spheres was determined for different system sizes \cite{wilding00,wilding02}.
Quite recently Wilding and coworkers and  Errington independently illustrated how the 
method could  also be applied 
to Lennard-Jones (LJ) particles \cite{errington04,wilding06}. In our view, 
the phase switch Monte Carlo is closely related to
the ``Gibbs ensemble method'' for fluid-solid equilibria, because, as in
the Gibbs ensemble method,  
phase equilibria is computed without free energy calculations.
In the phase switch methodology, trial moves are introduced within 
the Monte Carlo program, in which configurations 
obtained from simulations of the liquid are tested for the solid phase and vice-versa (phase
switch). For a certain thermodynamic condition (p and T) the relative probability of 
the system being in the liquid or solid phase is evaluated and that allows one
to 
estimate free energy differences. In the phase switch methodology the system jumps 
suddenly from the liquid to the solid in just one step. 
This method has  been reviewed recently by Bruce and Wilding \cite{wilding_rev}, 
and has proved to be quite 
successful for hard spheres and LJ systems. It is likely that the methodology
can also been applied to molecular systems although results have not been 
presented so far. It is not obvious  whether  the methodology can be used to 
determine  solid-solid equilibria in complex systems. 

Another alternative route has emerged in recent years.
Grochola \cite{grochola1} proposed to establish a 
thermodynamic path connecting the liquid with 
the solid phase.  Of course phase transitions should not occur along the path. If this 
is the case then it is possible to compute  the free energy difference between the 
two phases. 
It is fair to say that Lovett \cite{nato_lovett} was the first to suggest such  
a path 
although it was Grochola who developed into a practical way. 
The system goes from the liquid to the solid not in one step (as in the phase
switching methodology) but in a gradual way. 
Several variations have been proposed so far and the method has been applied successfully 
to  LJ, electrolytes, aluminium \cite{grochola1,grochola2,brennecke} and also
to molecular systems, such as benzene \cite{benzene_aplication}.
At this stage it is not obvious whether it could also be applied to study
solid-solid equilibria. 

Another approach to get free energies of solids is to use lattice dynamic methods \cite{born_book}.
By diagonalizing the quadratic form of the Hamiltonian the system may be transformed
into a collection of independent harmonic oscillators for which the free energy is 
easily obtained. This procedure allows one to estimate free energies at 
low temperatures and  fails for discontinuous potentials and when anharmonic
contributions become important (close to the melting point). The method has been used
by Tanaka {\em et al.} \cite{gao00,tanaka04} to get the melting point of several water models.

In this review we focus on the determination of phase equilibria between two
phases, where at least one of them is a solid phase. Therefore, the goal 
is not just fluid-solid equilibria but also solid-solid equilibria. 
Free energy calculations allow in fact the determination of the global 
phase diagram of a system (fluid-solid and solid-solid). In this methodology 
the free energy is determined for the two coexisting phases and the coexistence
point is obtained by imposing the conditions 
of equal pressure, temperature and chemical potential.
Usually the chemical potential of the liquid is obtained via thermodynamic integration.
Different methods are used to determine the chemical potential of the solid.
In their pioneering work, Hoover and Ree used the so called cell occupancy 
method \cite{JCP_1967_47_04873_nolotengo,hoover-ree}.
In this method each molecule is restricted to its Wigner-Seitz cell, and the 
solid is expanded up to low densities \cite{hoover-ree}.
One of the problems of this method is the appearance of a phase transition in 
the integration path (from the solid to the gas). The method was also applied
to the LJ system \cite{PR_1969_184_000151,hansenmcd}.

In the year 1984, Frenkel and Ladd proposed 
an alternative method, the Einstein crystal method \cite{frenkel84}.
In this method, that has become the standard method for determining 
free energies of solids, the change in free energy from the real
crystal to an ideal Einstein crystal (in which there are not intermolecular interactions
and where each molecule vibrates around its lattice point via an harmonic potential) is computed.
Since the free energy of the reference ideal Einstein crystal is known 
analytically, it is possible to compute the  free energy of the solid.
If the equation of state (EOS) and free energies
of both phases are known  it is then possible to determine the conditions for the 
equilibrium between the two phases. Repeating the calculation at different thermodynamic
conditions then it is possible to determine the phase diagram for a certain 
potential model. This route has often been used in the past for a number 
of simple models including hard ellipsoids \cite{mulder}, the Gay Berne model \cite{demiguel02}, 
the hard Gaussian \cite{saija}, 
hard dumbbells \cite{singer90,vega92b,vega92c,rainwater}, 
hard spherocylinders \cite{JCP_1996_104_06755,JCP_1997_106_00666,JCP_1997_107_02696}, 
diatomic LJ models \cite{2clj_phase_diagram,2clj_hall,lisal_2clj}, quadrupolar 
hard dumbells \cite{vega95qq}, hard flexible chains \cite{malanoski_chains,malanoski_alkanes},
linear rigid chains \cite{lths_fluid_solid,ltlj_phase_diagram}, 
chiral systems \cite{chiral_monson}, quantum hard spheres \cite{sese}, primitive models of 
water \cite{vega98}, electrolytes \cite{bresme00,vega03}, 
benzene \cite{schroer2000,schroer2001}, propane \cite{shen_monson_propane}  and idealised 
models of colloidal particles 
\cite{yukawa_frenkel,yukawa_noe,hynninen,vega03,eva_patchy,sandler2}. Some of the main findings of this research
(up to 2000) have been reviewed by Monson and Kofke \cite{ACP_2000_115_0113_nolotengo}. 
Forty years after the first determination of  
fluid-solid equilibria (for hard spheres \cite{alder}) the number of models for which it 
has been determined is still small. 
The situation is even worse if one considers studies of phase diagram calculations
(including both fluid-solid and solid-solid calculations) for models describing 
real molecules. Then the number of considered systems is quite small, comprising
nitrogen \cite{meijer}, alkanes \cite{polson_alkanes}, fullerenes \cite{c60}, 
ionic salts \cite{JCP_2003_118_00728,kclportugal}
and just in the last years 
carbon \cite{frenkel_carbon}, silicon \cite{hernandez}, silica \cite{silica04,monson_silica},
and hydrates \cite{monson_jpcb07}.

As can be seen, water was missing and this is surprising taking 
into account its importance as solvent and as the medium where life occurs \cite{finneyreview}.
Although water  has been studied in thousands 
of simulation studies since the
pioneering works of Barker and Watts \cite{CPL_1969_3_0144_nolotengo}
and Rahman and Stillinger \cite{JCP_1971_55_03336}, the study of its phase
diagram by computer simulation has not received much attention. Interest has 
focused mainly on the possible existence of a liquid-liquid 
equilibria \cite{poole92,stanley_2,mishima98,debenedetti03,paschek_arxive,brovchenko05,marques07,agua_amorfa_review}. 
The interest in the solid phases of water has been rather limited although
one observes a clear revival in the last 
decade \cite{morse82,JCP_1984_81_06406_nolotengo,svishchev94,matsumoto02,%
gay02,ayala03,tanaka98,gezelter,rick05,baranyai05,baranyai06,vrbka05,carignano05,tanaka_jcp05,%
buch_jcp06,picaud_jcp06,jedlovszky06,tribello06,slater_jacs06,robertocar_ice,piotrovskaya07,brodskaya07,taylor_hydrate,kalinichev06,biomaterial_tip4p_ice}.
The only attempt previous to our work to determine the phase diagram of water was performed
by Baez and Clancy in 1995 \cite{baez95jcp,baez95mp}. Estimates of the melting point
of TIP4P were provided by Tanaka {\em et al.} \cite{gao00,tanaka04},
Vlot {\em et al.} \cite{JCP_1999_110_00055}, and Haymet {\em et al.} \cite{karim88,karim90}.
Motivated by this our group has undertaken the task of determining 
the phase diagram for a number of water models \cite{sanz1,sanz2}. 
The study of water revealed that phase diagram calculations are indeed feasible
for molecular systems  and that they constitute a severe test for potential models.
It is clear that the phase diagram contains information about the intermolecular 
interactions \cite{abascal07a,abascal07b,abascal07c}.

The determination of a phase diagram is not, in principle, a difficult task.
However, it is cumbersome, and somewhat tricky. In this work we will illustrate
the details leading to the determination of the phase diagram of water. They can
indeed be useful for those interested in water and its  phase diagram. But the 
described methodology can be applied to other substances/models as well.
We believe that by describing the calculations for water we are also describing 
how to do it for any other type of molecule. Problems where the determination 
of the fluid-solid equilibria by molecular simulation can indeed bring new 
light are among others, the design of model potentials for water and other
molecules \cite{finneyreview,guillot02,jorgensenreview,benzene_aplication}, the 
study of nucleation \cite{auer_frenkel,JCP_1996_104_09932,trout02,trout03,auer_review} (where the equilibrium conditions should be known in advance), 
the study of the fluid-solid equilibria in colloidal systems, 
and also the very interesting problem of 
protein crystallisation. Our goal here is to describe  all the details
to encourage the reader to implement 
phase diagram calculations (including at least one solid phase) either to 
gain new insight on appealing problems or to improve currently available 
potential models.

\section{Basic definitions }

For a pure substance, two phases (labelled as I and II) are in equilibrium when their pressures, 
temperatures and chemical potentials are equal. 
A phase diagram is just a plot (for instance in the  $p,T$ plane) of the coexistence points between the
different phases of the system (gas, liquid or solid). 
In this paper we shall focus on determining the phase equilibria for  rigid molecules.
Two ensembles are particularly useful to study phase equilibria: the canonical ensemble (NVT) and 
the isobaric isothermal ensemble (NpT). 
In the canonical ensemble  the Helmholtz free energy A is given by 
the following expression \cite{macquarrie,hansenmcd}:
\begin{center}
\begin{equation}
\fl
A = -k_B T \ln (Q(N,V,T))= -k_B T \ln \left( \frac{q^N}{N!} \int 
\exp[-\beta U({\bf r}_1,\omega_1,...,{\bf r}_N,\omega_N)] d1...dN   \right)
\end {equation}
\end{center}
where  $\beta=1/(k_BT)$,  $U$ is the intermolecular energy of the whole system, $q$ is the molecular partition 
function  and  $di$ stands for  $d{\bf r}_i d\omega_i$, where $d{\bf r}_i=dx_i dy_i dz_i$. 
The location of molecule $i$ is given by the Cartesian coordinates $x_i, y_i, z_i$ of the reference
point and a normalised set of angles defining 
the orientation of the molecule ($\omega_i$).  By normalised we mean that $\int d\omega_i = 1$.
For instance a reasonable choice for $\omega_i$ is $\omega_i={\bf \Omega_i}/V_{\Omega}$  where 
${\bf \Omega_i}$ are the 
Euler angles \cite{GrayGubbins} defining the orientation of the molecule and 
$V_{\Omega}=\int d{\bf \Omega_i}=8\pi^2$. 
For a non-linear molecule the partition function can be written as \cite{macquarrie}:
\begin{eqnarray}
\label{fpmol}
\fl
q&=&q_{t'}q_{r}q_{v}q_{e}\\
\fl
q&=&\left[\left(\frac{2\pi mk_BT}{h^2}\right)^{3/2}\right] \left[\frac{(2\pi k_BT)^{3/2}V_{\Omega} (I_1 I_2 I_3)^{1/2}}{s' h^3}\right]  
\left[\prod_j \frac{\exp(-\beta h \nu_j/2)}{1-\exp(\beta h \nu_j)}\right] \left[g e^{-\beta D_e}\right]\nonumber,
\end{eqnarray}

In the previous equation translational and rotational degrees of freedom are treated classically 
(except for the symmetry number $s'$ and for the factor $h$), and vibrational 
and electronic degrees of freedom are described by the quantum partition function.  
$q_{t'}=q_t/V$ is the translational partition function (divided by the volume), 
and $q_{r}$, $q_v$ and $q_{e}$ are the rotational, vibrational and electronic partition functions, 
respectively. The rotational, vibrational and electronic partition functions are dimensionless. 
We shall assume that the rotational, vibrational and electronic partition functions are identical in two coexistence
phases. For this reason their precise value does not affect phase equilibria and we shall simply 
assume that their value is one (we do not pretend to determine absolute free energies but 
rather phase equilibria).  The first
factor $q_{t'}$ has units of inverse volume or inverse  cubic length. 
It is usually denoted as 
the inverse of the cubic de Broglie wave length \cite{macquarrie} (i.e. $1/\Lambda^{3}$). Therefore in 
this work $q_{t'}$ is given by:
\begin{equation}
\label{choiceq}
q_{t'} = \frac{1}{\Lambda^{3}} = \frac{1}
{ ( h^2/(2 \pi m k_B T) )^{3/2} }
\end{equation}
In the $NpT$ ensemble the Gibbs free energy G can be obtained as $G = -k_B T \ln (Q(N,p,T))$ where
$Q(N,p,T)$ is given by :
\begin{equation}
\fl
Q(N,p,T)=\frac{q^N\beta p}{N!}\int \exp(-\beta pV)  dV \int
\exp[-\beta U({\omega_1,\bf s}_1,..,{\bf s}_N,\omega_N;{\bf H})] V^{N} d{\bf s}_1d\omega_1
.d{\bf s}_Nd\omega_N 
\end{equation}
where  ${\bf s}_i$ stands for the coordinates of the reference point of 
molecule $i$ in simulation box units. The conversion from 
simulation box units ${\bf s}_i$ to Cartesian coordinates ${\bf r}_i$
can be performed via the {\bf H} matrix $
{\bf r}_i  =  {\bf H}   {\bf s}_i$ (the volume of the system is just the determinant of the {\bf H} matrix).
When performing Monte Carlo (MC) simulations of solid phases it is important that changes in 
the shape of the simulation box are allowed (i.e., changes in {\bf H}). This is usually denoted as 
anisotropic NpT simulations. They were first introduced within MD simulations by 
Parrinello and Rahman \cite{parrinello80,parrinello81}, and extended to MC simulations by 
Yashonath and Rao \cite{yashonath85}.
In anisotropic NpT Monte Carlo      the elements of the 
{\bf H} matrix undergo random displacements, and that provokes a change both in the volume of 
the system and in the shape of the simulation box. Further details about the methodology
can be found elsewhere \cite{yashonath85,SM_1983_17_1199_nolotengo}. 
The use of the anisotropic version of the NpT ensemble is absolutely required to 
simulate solid phases. It guarantees  that the shape of the simulation box (and 
therefore that of the unit cell of the solid) is the equilibrium one. It also 
guarantees that the solid is 
under hydrostatic pressure and free of stress (the pressure tensor will then be diagonal with 
the three components being identical to the thermodynamic pressure).

\section{ Fluid-solid equilibrium from NpT simulations?}
A possible way to determine the fluid-solid equilibria is by performing simulations at 
constant pressure and cooling the liquid until it freezes. 
However, it is very difficult to  observe in computer
simulations the formation of a crystal and this is 
especially true for molecular fluids \cite{svishchev94,matsumoto02,debenedetti}. 
The nucleation of the solid is an activated process
and it may be difficult to observe within the time scale of the simulation.
In fact even in real experiments super-cooled liquids are often found \cite{debenedetti}. 
The other possibility is to 
heat the  solid until it melts. 
Experimentally, when
a solid is heated at constant pressure it always melts at the
melting temperature (with only a few exceptions to this rule). In fact 
Bridgman \cite{PAAAS_1912_XLVII_13_441_photocopy} stated in 1912:
``It is impossible to superheat a crystalline phase with respect to 
the liquid''. 
Unfortunately in computer simulations 
(in contrast to real experiments)
one may superheat the solid before it melts.
This is well known for hard spheres
\cite{allen_book}
(with pressure being the thermodynamic variable in question)
and  for
Lennard-Jones particles
\cite{stegailov_1}.
The same is true for other systems such as water. 
For water models it has been found that in NpT runs
ices melt at a temperature about 90~K above  
the equilibrium melting point \cite{mcbride05,mcbride04}. Similar results have been obtained for 
nitromethane \cite{superheating_nitromethane} or NaCl \cite{brennecke}.
In that respect NpT simulations provide an  upper
limit of the melting point. Introducing defects within the solid reduces considerably the
amount of superheating \cite{thompson_nitromethane_defects_and_twophases,thompson_review}. 

The difference between the results of NpT simulations and 
those found in experiments (summarised in the Bridgman's 
statement) is striking. The explanation to this puzzle is that in experiments 
melting occurs typically via heterogenous nucleation starting at interfaces (real solids do always have
interfaces) whereas in 
NpT simulations it must occur via homogeneous nucleation 
(due to the absence of the interface), 
requiring a rather long time \cite{stegailov_1,stegailov_2,stegailov_3}. 
Therefore new strategies must be proposed to obtain fluid-solid (or solid-solid) equilibria 
from simulations. 
The first possibility  is to compute separately the free energy of the liquid and of 
the solid and determine the condition of chemical equilibrium. The second is to introduce
a liquid nucleus in contact with the solid (i.e., a seed) since that will eliminate the superheating. 
These two possibilities will be discussed in this paper.

\section{ Thermodynamic integration: a general scheme to obtain free energies. }
In thermodynamic integration the free energy difference between
two states/systems is obtained by integrating a certain thermodynamic function along the 
path connecting both states/systems \cite{frenkelbook}.  
The path connecting the two systems or states must be reversible. No first 
order phase transition should be found along the path. 
We shall distinguish
(somewhat arbitrarily) two types of thermodynamic integration. In the first one
the two states connected by the path possess the same Hamiltonian
(i.e. interaction potential) and they
just differ in the thermodynamic conditions (i.e. $T$, $p$...). 
We shall denote this type of integration as thermodynamic integration.
In the second one, the thermodynamic conditions are the same for the final and 
initial conditions (i.e. the same $p$ and $T$ or same density and $T$), but the 
Hamiltonian (i.e, the intermolecular potential) will be different for the 
initial and final systems. 
This type of integration will be denoted 
as Hamiltonian thermodynamic integration.

\subsection{Thermodynamic integration} 
Assuming that the free energy at a certain thermodynamic state is known, the free energy
at another thermodynamic state is determined by establishing  a reversible path connecting 
both states. For a closed system, with a fixed number of particles, two thermodynamic 
variables are needed to determine the state of the system (for instance p and T or V and T).
In practice, it is convenient to keep one of the thermodynamic variables constant while
performing the integration.

\subsubsection{Keeping T constant (integration along isotherms)} 
Once the  Helmholtz free energy at a certain reference density $\rho_1=N/V_1$ is known, the
free energy at another density $\rho_2=N/V_2$ (T being the same in both cases),
can be obtained as :
\begin{equation}
\label{inttermisot}
\frac{A(\rho_2,T)}{Nk_BT}=\frac{A(\rho_1,T)}{Nk_BT}+\int^{\rho_2}_{\rho_1}\frac{p(\rho)}{k_BT\rho^{2}}d\rho
\end{equation}
The integrand can be obtained in a simple way from NpT runs, isotropic for the fluid
and anisotropic for solid phases \cite{parrinello80,parrinello81,yashonath85} (so that
the equilibrium density is obtained for different pressures). 

\subsubsection {Keeping p constant (integration along isobars)}
In this integration the temperature of the system is modified while keeping 
constant the value of the pressure.  In this way the Gibbs free energy G is obtained for any 
temperature along the isobar starting from an initial known value. 
The working expression is :
\begin{equation}
\label{inttermisob}
\frac{G(T_2,p)}{Nk_BT_2}=\frac{G(T_1,p)}{Nk_BT_1}-\int^{T_2}_{T_1}\frac{H(T)}{Nk_BT^2}dT
\end{equation}
where H is the enthalpy. 
In practice several NpT simulations (anisotropic NpT for solids)
are performed at different temperatures and the 
integrand is determined from the simulations. 

\subsubsection{Keeping the density constant (integration along isochores)}
In this case the density is constant and the temperature is modified. 
The working expression is :
\begin{equation}
\label{inttermisoc}
\frac{A(T_2,V)}{Nk_BT_2}=\frac{A(T_1,V)}{Nk_BT_1}-\int^{T_2}_{T_1}\frac{U(T)}{Nk_BT^2}dT                 
\end{equation}
For fluids the integrand is easily obtained from NVT simulations. 
Although equation (\ref{inttermisoc}) is quite useful for fluid phases, it is not so useful 
for solid phases. The reason is that for solids  the density should be  constant
along the integration but the shape of the simulation box should be not (except for cubic solids).
In fact the equilibrium shape of the unit cell (simulation box shape) changes when 
the temperature is modified at constant density.  \\

\subsection{Hamiltonian integration}
\label{intterhaml}
In this type of integration the Hamiltonian of the system changes between the initial ($\lambda=0$) 
and the final state ($\lambda=1$). This can be accomplished by introducing a coupling parameter 
($\lambda$) into the interaction energy of the system. The interaction energy becomes then a function
of this coupling parameter ($U(\lambda)$). The free energy of the system will be 
a function not only of the thermodynamic variables but also of $\lambda$: 
\begin{equation}
\label{eq1}
A(N,V,T,\lambda)=-k_BT \ln \left[ 
\frac{q^{N}} {N!} \int \exp[-\beta U(\lambda)] d1...dN \right].
\end{equation}
By performing the derivative with respect to $\lambda$ in equation (\ref{eq1}) one obtains:
\begin{equation}
\frac{\partial A(N,V,T,\lambda)}{\partial \lambda}=\left<\frac{\partial U(\lambda)}{\partial \lambda}\right>_{N,V,T,\lambda}.
\end{equation}
By integrating this differential equation one obtains:
\begin{equation}
\label{inttermfree}
A(N,V,T,\lambda=1)=A(N,V,T,\lambda=0)+\int^{\lambda=1}_{\lambda=0}\left<\frac{\partial U(\lambda)}{\partial \lambda}\right>_{N,V,T,\lambda}d \lambda.
\end{equation}
This equation gives the difference in free energy between two states with the same temperature and 
density but with different Hamiltonian (intermolecular potential). 
A similar equation can be obtained
within the isobaric-isothermal  ensemble. In this case the difference in Gibbs free energy between two systems
with the same temperature and pressure and different Hamiltonian is given by:
\begin{equation}
\label{inttermfreeGibbs}
G(N,p,T,\lambda=1)=G(N,p,T,\lambda=0)+\int^{\lambda=1}_{\lambda=0}\left<\frac{\partial U(\lambda)}{\partial
\lambda}\right>_{N,p,T,\lambda}d\lambda. 
\end{equation}

\section{The machinery in action. I. Obtaining the free energy of the liquid phase. }
\label{aliquidsg}
Here we shall briefly describe three possibilities to obtain the free energy of the liquid
phase (there are other possibilities). The three routes considered are :
thermodynamic integration, hamiltonian thermodynamic integration and 
the Widom test particle method. 

\subsection{Thermodynamic integration}
\label{feliinh}
When the density of a fluid tends to zero, the particles are far apart so that 
intermolecular interactions are irrelevant, and the  system tends 
to an ideal gas.  Therefore the free energy of the real fluid at a certain density $\rho$ and 
temperature is given by :
\begin{equation}
\label{itnhgi}
\frac{A(\rho,T)}{Nk_BT}=\ln(\rho \Lambda^{3}) -1  + \frac{\ln(2\pi N)}{2N} +\int^{\rho}_{0}\left[\frac{p}{k_BT\rho^{'2}}-\frac{1}{\rho^{'}}
\right]d\rho^{'}.
\end{equation}
where the first three terms on the right hand side represent the ideal gas contribution to 
the free energy  (a logarithmic correction to the Stirling's approximation was included)
and the last term is the residual part (a residual property is defined as the difference between that of
the system and that of an ideal gas at the same temperature and density). 
To derive equation (\ref{itnhgi}) the rotational, vibrational and electronic 
contributions to the partition function, equation (2), were set to one. 
The integrand in equation (\ref{itnhgi}) tends at low densities to the second virial coefficient.
The first term on the right hand side of equation (\ref{itnhgi}) is just 
a reduced density and of course is dimensionless (although 
its numerical value depends on the value of $\Lambda$).
To avoid phase transitions the integration along the isotherm should be  performed at supercritical 
temperatures. Once the density of the liquid is achieved, one can integrate along an isochore
to low temperatures. 

\subsection{ Free energy of liquids by Hamiltonian thermodynamic integration }
\label{freeliq}
Let us label as A the system for which the free energy is known in the fluid phase, and B the system 
for which the free energy is unknown. By introducing a coupling parameter one can change from the Hamiltonian of A to the Hamiltonian of B :
\begin{equation}
\label{lincop}
U(\lambda)=(1-\lambda) U_A + \lambda U_B.
\end{equation}
where  $\lambda$ is a parameter ranging from 0 (system A) to 1 (system B). 
According to equation (\ref{inttermfree}), the free energy difference between A and B is given by:
\begin{equation}
\label{enlibliq}
A_B(N,V,T) = A_A(N,V,T) + \int^{1}_{0}\left<U_B-U_A\right>_{N,V,T,\lambda}d \lambda.
\end{equation}
where  $<U_B-U_A>_{N,V,T,\lambda}$ can be obtained by performing NVT simulations for a certain 
value of $\lambda$. The value of the integral is then obtained numerically. 

\subsection{Widom test particle method}
\label{widomtest}
The chemical potential can be obtained by the procedure proposed 
by Widom in 1963 \cite{JCP_1963_39_02808_nolotengo} which yields :
\begin{equation}
\mu^{res}=-k_B T \ln \left<\exp(-\beta U_{test})\right>_{N,V,T}. 
\end{equation}
This formula states that the residual value of the chemical potential $\mu^{res}$ is just 
the average of the Boltzmann factor of the interaction 
energy ($U_{test}$ ) of a test particle. Although this formula is quite useful, its 
practical implementation may be problematic when the density of the system is high (so that
inserting a particle becomes difficult). This is especially true for molecular systems and
even more dramatic for systems with important orientational dependence in the pair potential.
This is the case of water, for which it is quite difficult to obtain reliable chemical 
potentials by using the test particle method \cite{widom_water_kofke}.

\section{ The machinery in action. II. Free energy of solids. }


The Einstein crystal method was proposed by Frenkel and
Ladd in the year 1984 \cite{frenkel84} and, since then,
it has become the standard method to compute the free energy
of solids \cite{vega92b,vega95qq,malanoski_chains,polson00,demiguel02,ferrariokf,JCP_2003_118_00728,lths_fluid_solid,barroso,kclportugal}. 
In this method an ideal Einstein crystal is used as the reference
system to compute the free energy of a solid.
An ideal Einstein crystal
is a solid (the word ideal pointing out the absence of intermolecular interactions)
in which the particles (atoms or molecules) are
bounded to their lattice positions and orientations by an
harmonic potential and in which there are not interparticle
interactions. The free energy of an ideal Einstein crystal can
be computed analytically for atomic solids and numerically
for molecular solids. 

For practical reasons, that will be clarified later,
it is convenient to use an Einstein crystal where a certain reference point of the whole
crystal is fixed. Two choices are possible:
\begin{itemize}
\item{ Fixing the position of the center of mass.
That was the original choice of Frenkel and Ladd \cite{frenkel84}. We shall 
denote the reference system as an ideal Einstein crystal with fixed center of mass and the 
technique will be referred to as the Einstein crystal approach. }
\item{ The second choice is to fix the position of just one of the molecules of the system, for
instance molecule 1. In this second case we shall denote the reference system as an 
ideal Einstein molecule with fixed molecule 1. This methodology has been proposed quite 
recently \cite{vega_noya} and will be denoted as the Einstein molecule approach.}
\end{itemize}
Since the determination of free energy for solids is rather involved and not many
examples can be found in the literature describing the details, we shall describe 
both methodologies in detail. Obviously both approaches are quite similar and indeed 
provide identical values of the free energy of the solid phase.

\subsection{ The Einstein crystal method}
\begin{figure}\centering
\includegraphics[clip,height=0.35\textheight,width=0.60\textwidth,angle=-0]{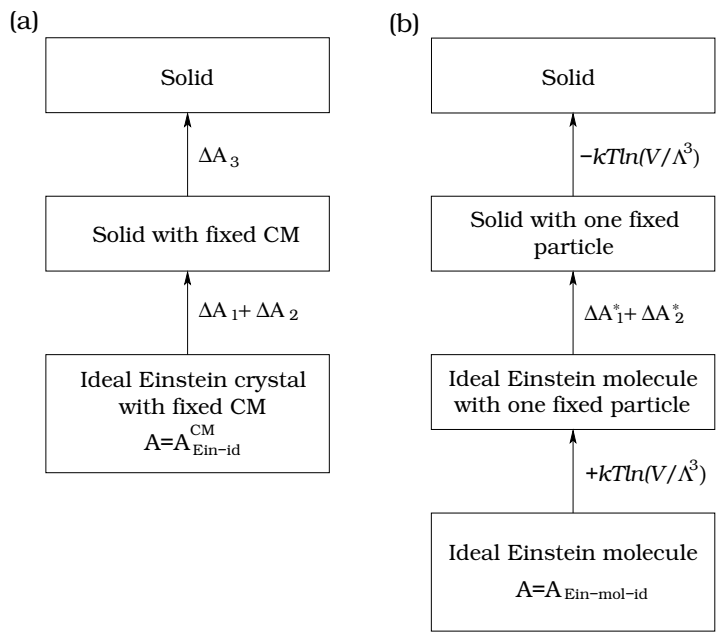}
\caption{\small\sf Thermodynamic path used in (a) the Einstein crystal 
method (Frenkel and Ladd \cite{frenkel84} and Polson {\em et al.} \cite{polson00}), 
and in (b) the Einstein molecule approach \cite{vega_noya}. }
\label{esquema_eva}
\end{figure}

In the Einstein crystal approach the reference system is an ideal 
Einstein crystal with fixed center of mass. 
Let us describe briefly the notation that will be used in this
section.  The superscript $CM$ indicates that the center
of mass is fixed. The  subscript specifies the interactions
present in the system. In particular, the subscript $Ein-id$ stands for ideal Einstein
crystal (without intermolecular interactions), the subscript $Ein-sol$ means that
both the harmonic springs and the intermolecular interactions are
present and the subscript $sol$ indicates that only the intermolecular interactions
are present (without the harmonic springs).

The whole path from the reference ideal Einstein crystal with fixed 
center of mass to the crystal of interest can be described as :
\begin{equation}
\label{ahivaeso}
A_{sol}=A^{CM}_{Ein-id}+ [(A^{CM}_{Ein-sol}-A^{CM}_{Ein-id}) + (A^{CM}_{sol}-A^{CM}_{Ein-sol})]+
(A_{sol}-A^{CM}_{sol}).
\end{equation}
Here $A^{CM}_{Ein-id}$ is the free energy of the reference system (i.e. 
the ideal Einstein crystal with fixed center of mass).
The first step  is the computation of the free energy difference 
between the ideal Einstein crystal and the interacting Einstein
crystal both with 
center of mass fixed ($A^{CM}_{Ein-sol}-A^{CM}_{Ein-id}$). 
In the second step  ($A^{CM}_{sol}-A^{CM}_{Ein-sol}$), the springs of the interacting Einstein crystal are gradually turned off to obtain the crystal of interest (both systems with fixed center of mass). 
Finally the solid with fixed center of mass is transformed into a solid with no fixed 
center of mass ($A_{sol}-A^{CM}_{sol}$). Equation (\ref{ahivaeso}) can be 
written in a more simple way as :
\begin{equation}
\label{ahivaeso_simple}
A_{sol}=A^{CM}_{Ein-id}+[\Delta A_1+\Delta A_2] + \Delta A_{3}.
\end{equation}
By comparing equations (\ref{ahivaeso_simple}) and (\ref{ahivaeso}) 
the meaning of the terms $\Delta A_1$, $\Delta A_2$,  $\Delta A_3$ is clarified.
Basically obtaining $A_{sol}$ is a four step process, since you need to obtain 
$A^{CM}_{Ein-id}$ (step 0), $\Delta A_1$ (step 1), $\Delta A_2$ (step 2), and $\Delta A_3$ (step 3). 
This integration path is
schematically shown in figure \ref{esquema_eva} (a).

\subsubsection{Step 0. Obtaining the free energy of the ideal Einstein crystal with fixed center of mass: 
$A^{CM}_{Ein-id}$.} 

\begin{figure}\centering
\includegraphics[clip,height=0.25\textheight,width=0.36\textwidth,angle=-0]{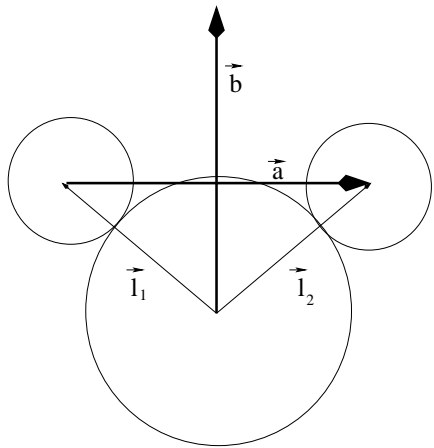}
\caption{\small\sf Definition of the vectors $\vec{a}$ and $\vec{b}$ in a triatomic
rigid molecule with a twofold symmetry axis. The vectors $\vec{a}$ and $\vec{b}$ should be 
normalised to have modulus one.}
\label{vecori}
\end{figure}

As mentioned above, an ideal Einstein crystal is a solid
in which the molecules are bounded to their lattice positions and
orientations by harmonic springs. 
We will focus on rigid non-linear
molecular solids. Although the translational field is always applied in the same form, the expression 
of the orientational field depends on the geometry of the considered molecule.
We shall describe here the procedure for a molecule with point group $C_{2v}$, as for instance
water. The appropriate expression of the orientational field for other geometries will be given 
later on. 

The energy of the ideal Einstein crystal is given by:
\begin{equation}
\label{ete_total}
U_{Ein-id} = U_{Ein-id,t}+U_{Ein,or}    
\end{equation}
\begin{equation}
\label{ete_t}
U_{Ein-id,t}=\sum ^{N}_{i=1} \left[\Lambda_E({\bf r}_{i}-{\bf r}_{io})^{2} \right] 
\end{equation}
\begin{equation}
\label{ete_r}
U_{Ein,or}(C_{2v}) = \sum^{N}_{i=1} u_{Ein,or,i} =  \sum^{N}_{i=1} \left[ \Lambda_{E,a} \sin^2 \left( \psi_{a,i}  \right) + 
\Lambda_{E,b} \left( \frac{\psi_{b,i}}{\pi} \right)^2 \right].
\end{equation}

In the preceding equation ${\bf r}_{i}$ represents the instantaneous 
location of the reference point of molecule $i$, 
and ${\bf r}_{io}$ is the equilibrium position of this reference point of molecule i 
in the crystal (i.e. ${\bf r}_{i}$ will fluctuate along the simulation run but ${\bf r}_{io}$ not).
A possible choice for the reference point (which defines the location of the molecule) 
is the molecular center of mass.
 In fact the rotational partition function of the molecule $q_r$  
is computed by using the principal moments of inertia ($I_1$, $I_2$ and $I_3$) with respects to 
a body frame with origin at the center of mass of the molecule. One could also use the center of 
mass of the molecule as the reference point to compute configurational properties. However, it should
be pointed out that configurational properties do not depend on the choice of the reference point. 
For this reason, to compute configurational properties,
 there is a certain degree of freedom in choosing the reference point of the molecule. 
For free energy calculations it is very convenient (the reasons will be clarified later)
if the reference point is chosen so that all elements of symmetry pass through it.
This requirement is satisfied by the center of mass, but it may also be satisfied by other points.
  For instance, in the case of water, all elements of symmetry 
pass through the oxygen atom so that its choice as reference point is 
also quite convenient. 
Alternatively, one could argue that the oxygen would become the center of mass of 
the molecule if the hydrogen atoms would have zero mass. Since the phase diagram of a water 
model does not depend on the masses of the atoms forming the molecule, setting the masses of
the hydrogen atoms to zero would not affect the phase equilibria. 
In this work we shall use the oxygen as the reference point of the 
water molecule and this choice will not affect the phase equilibria (take the reason you prefer either
because configurational properties do not depend on the choice of the reference point, or because
there is a certain combination of masses of the atoms of the molecule that render the center of mass
on the oxygen atom ). 
The term $U_{Ein-id,t}$ in equation (\ref{ete_t})
is a harmonic field that tends to keep the particles 
at their lattice positions (${\bf r}_{io}$), while 
$U_{Ein,or}$ forces the particles to have the right orientation.
$\Lambda_E$, $\Lambda_{E,a}$ and $\Lambda_{E,b}$ are the coupling parameters
of the springs (not to be confused with the thermal de Broglie wave length $\Lambda$). Notice that $\Lambda_{E,a}$ and $\Lambda_{E,b}$ have energy units whereas 
$\Lambda_{E}$ has units of energy over a squared length. 
The angles $\psi_{a,i}$ and $\psi_{b,i}$ are
defined in terms of two unit vectors, $\vec{a}$ and $\vec{b}$, that
specify  the orientation of the molecule.
$\psi_{a,i}$ is the angle formed by the unit vector $\vec{a}$ of molecule $i$ 
in a given configuration ($\vec{a}_{i}$) and the unit vector  ($\vec{a}_{io}$) of that 
molecule in the reference lattice. 
The angle $\psi_{b,i}$ is defined analogously but with vector $\vec{b}$.
The definition of vectors $\vec{a}$ and $\vec{b}$ for a rigid
triatomic molecule is shown in Figure \ref{vecori}. This form of the orientational field (equation (\ref{ete_r})) was 
used by Vega and Monson \cite{vega98} to get the free energy of a primitive model of water \cite{nezbeda_pm}.
The vector $\vec{a}$ is calculated as the subtraction of
the bond vectors $\vec{a}=(\vec{l}_2 - \vec{l}_1)/\mid \vec{l}_{2}-  \vec{l}_{1} \mid $, while
$\vec{b}= (\vec{l}_2 + \vec{l}_1)/\mid \vec{l}_{2} + \vec{l}_{1}\mid $.
The angles $\psi_{a,i}$ and $\psi_{b,i}$ can be obtained simply from  the scalar 
product of vectors $\vec{a}_{i}$ and $\vec{a}_{io}$ (both of them being unitary vectors), and $\vec{b}_{i}$ 
and $\vec{b}_{io}$ (both of them being unitary vectors) respectively as:
\begin{eqnarray}
\psi_{a,i}=\arccos \left( \vec{a}_{i} \cdot \vec{a}_{io} \right) \nonumber\\
\psi_{b,i}=\arccos \left( \vec{b}_{i} \cdot \vec{b}_{io} \right)
\end{eqnarray}
so that $\psi_{a,i}$ and $\psi_{b,i}$ will adopt values between 0 and $\pi$.
Notice that in the orientational field along the $\vec{b}$ direction (see equation (\ref{ete_r})),
the angle $\psi_{b,i}$ is divided by a factor of $\pi$. 
In this way, 
this term $(\psi_{b,i}/\pi)^{2}$ also takes values between 0 (when $\vec{b}$ is parallel to
$\vec{b}_{io}$) and 1 (when $\vec{b}_{i}$ and $\vec{b}_{io}$ form
an angle of $\pi$ radians), and both orientational 
fields have the same strength (the $sin^2(\psi_a)$ field changes from 0 when $\psi_a=0$
or $\psi_a=\pi$ to 1 when $\psi_a=\pi/2$).

The partition function of the ideal Einstein 
crystal in the canonical ensemble (after integrating over the rotational momenta) is given by:
\begin{equation}
\label{eins_id_1}
\fl
Q_{Ein-id}= \frac{1}{N!} \frac{1}{h^{3N}} (q_r q_v q_e)^{N} \int \exp\left[-\beta \sum^{N}_{i=1} \frac{{\bf p}^{2}_{i}}{2m_{i}}\right]d{\bf p}_{1}...d{\bf p}_{N}
\int \exp\left[-\beta U_{Ein-id}\right]d1...dN,
\end{equation}
where ${\bf p}_i = (p_{xi}, p_{yi}, p_{zi})$ represents the momentum of molecule $i$  
and $di = d{\bf r}_i d{\bf \omega}_i$, being ${\bf r}_i$ the position vector
of the reference point of molecule $i$ 
and ${\bf \omega}_i$ its normalised angular 
coordinates.  Consistent with our choice for the fluid phase, $q_r, q_v, q_e$ will 
be set arbitrarily to one (they will be omitted in what follows).
Now a subtle issue appears. In the Einstein crystal approach each molecule (via the reference 
point) is attached to a lattice point. One can compute the free energy for a solid where 
each molecule is attached to one and only one lattice point. 
However one should not forget that there are 
$N!$ possible permutations. Therefore, the true free energy of the system is that obtained 
for a certain field where each molecule is attached to one lattice site multiplied by the 
number of possible permutations (i.e., $N!$). For this reason the partition function 
is :
\begin{equation}
\label{eins_id_2}
\fl
Q_{Ein-id}=  \frac{1}{h^{3N}} \int \exp\left[-\beta \sum^{N}_{i=1} \frac{{\bf p}^{2}_{i}}{2m_{i}}\right]d{\bf p}_{1}...d{\bf p}_{N}
\int_{one\;\;\;permutation} \exp\left[-\beta U_{Ein-id}\right]d1...dN,
\end{equation}
where the integral over coordinates is now computed for just one permutation 
(and hence the label one permutation in the integral over coordinates). 
The expression {\em one permutation} in equation (\ref{eins_id_2}) reminds that 
each molecule is attached (via $U_{Ein-id}$) to one and only one lattice point.
Let us now impose mathematically the condition of fixed center of mass of the 
reference points. For water we are fixing the center of mass of 
the oxygen atoms (the O will act as the reference point of the molecule) 
rather than fixing the center of mass of the whole system (including the hydrogens). 
It is simpler and more convenient for molecular fluids to fix the center of 
mass of the reference points, rather than fixing the
center of mass of all the atoms of the system. 
In the configurational space, the restriction implies that:
\begin{eqnarray}
\label{cmfijo}
& &{\bf R}_{CM} ( {\bf r}_{1}, {\bf r}_{2}... {\bf r}_{N} ) -{\bf R}_{CM}^{0}=0\nonumber\\
& &\sum^{N}_{i=1} \mu_{i} ({\bf r}_{i}-{\bf r}_{io})=0
\end{eqnarray}
where, if the mass assigned to all reference points (one per molecule) is the same,
$\mu_{i}=1/N$, $i=1,...N$.
In the previous equation ${\bf R}_{CM}$ is the center of mass of the reference points 
of the system
(there is one reference point per molecule) in an instantaneous configuration and ${\bf R}_{CM}^{0}$ is 
the center of mass of the
reference points of the system when the molecules stand on the lattice 
positions of the Einstein crystal field. ${\bf R}_{CM}$ is a function of the coordinates of the 
particles of the system whereas ${\bf R}_{CM}^{0}$ is one parameter. Due to thermal vibration, in general
${\bf R}_{CM}$ will be different from ${\bf R}_{CM}^{0}$. 
The constraint given by  equation (\ref{cmfijo}) means that from all possible 
configurations of the particles of the system only those
satisfying ${\bf R}_{CM}={\bf R}_{CM}^{0}$  will be allowed.  

A comment is in order here. The value of the molecular mass does not affect the phase
equilibria (i.e., the molecular mass is irrelevant to determine phase transitions). 
For instance for a LJ system, the triple point does not depend on the particular value of 
the mass of the system (for Ar and Kr the melting point is different but not because of their
mass but for the different value of the parameters of the LJ potential). 
For this reason, for phase diagram calculations it is quite convenient to assign the same
mass to all molecules of the system (regardless of whether this is true or not in 
real experiments). For instance for NaCl, it is possible to assign the same mass to 
Na and Cl, without affecting the phase equilibria of the model. In fact we have used 
this strategy to determine its melting point \cite{sanz_nacl}. Therefore the 
simple choice $\mu_{i}=1/N$, $i=1,...N$ 
(i.e., assigning the same mass to all particles of 
the system or similarly to all reference points ) can be used to determine phase equilibria without affecting the results. We strongly recommend this choice.
Of course dynamic properties depend on the mass, but not phase equilibria which is the
main focus of this paper. 

As a consequence of the centre of mass constraint, the space of momenta is
constrained to
\begin{equation}
\sum^{N}_{i=1}{\bf p}_{i}=0
\end{equation}
The partition function of an ideal Einstein crystal with fixed center of
mass $Q_{Ein-id}^{CM}$ can be written as:
\begin{equation}
Q_{Ein-id}^{CM} =  Q^{CM}_{Ein,t}  Q_{Ein,or}
\end{equation}
Then the free energy is simply
obtained as:
\begin{eqnarray}
\label{vdpsductyoo}
A^{CM}_{Ein-id}&=&A^{CM}_{Ein,t}+A_{Ein,or}\nonumber\\
&=&-k_BT\ln Q^{CM}_{Ein,t}-k_BT\ln Q_{Ein,or}
\end{eqnarray}

The orientational term $Q_{Ein,or}$ will be computed by evaluating numerically the 
following integral :
\begin{equation}
\label{orientationalintegral}
Q_{Ein,or}=\left[ \frac{1}{8\pi^2}
\int \exp \left( -\beta u_{Ein,or} \right)
\sin\theta d\phi d\theta d\gamma \right ]^N
\end{equation}
where $\phi$, $\theta$ and $\gamma$ stand for the Euler angles defining 
the orientation of the molecule and $u_{Ein,or}$ is the orientational Einstein field for just
one molecule (see equation (\ref{ete_r})).  We have chosen to
use the definition of Gray and Gubbins of the Euler angles \cite{GrayGubbins}.
In the particular case of a molecule with $C_{2v}$ symmetry (for instance
water) it reads:
\begin{equation}
\label{orientationalintegral_c2v}
\fl \qquad
Q_{Ein,or}=\left[\frac{1}{8\pi^2}\int
\exp\left(-\beta\Lambda_{E,a}\sin^2
\left(\psi_{a}\right)-\beta\Lambda_{E,b}
\left(\frac{\psi_{b}}{\pi}\right)^2\right)\sin\theta 
d\phi d\theta d\gamma\right]^N,
\end{equation}
Notice that $\psi_{a}$ and $\psi_{b}$ are functions of the Euler angles.
The integral given by equation (\ref{orientationalintegral}) or equation 
(\ref{orientationalintegral_c2v}) can be evaluated numerically
(for instance using a Monte Carlo numerical integration methodology). 
An approximate analytical expression \cite{vega98} has been provided for $C_{2v}$ which is valid
in the limit of large coupling constants ($\Lambda_{E,a},\Lambda_{E,b}$).
The translational term $Q^{CM}_{Ein,t}$ is given by the following expression:
\begin{eqnarray}
\label{q_transl}
Q_{Ein,t}^{CM}&=&\frac{1}{h^{3(N-1)}} \int \exp\left[-\beta \sum ^{N}_{i=1} 
\frac {{\bf p}^{2}_{i}}{2m_{i}}\right]\delta(\sum^{N}_{i=1}{\bf p}_{i})
d{\bf p}_{1}...d{\bf p}_{N}\nonumber\\
& &\int\exp\left[-\beta \Lambda_E \sum ^{N}_{i=1} 
({\bf r}_{i}-{\bf r}_{io})^{2}\right]\delta(\sum^{N}_{i=1} \frac{1}{N}
({\bf r}_{i}-{\bf r}_{io}))d{\bf r}_1...d{\bf r}_N.
\end{eqnarray}
Notice that to simplify the notation we have dropped the subindex "one permutation" in the integration over
coordinates since it is sufficiently clear that this is indeed the case when each molecule is attached by
harmonic springs to just one lattice points. 
This integral (equation (\ref{q_transl})) can be solved analytically \cite{polson00} (see Appendix A 
for the details) and the result is:
\begin{equation}
Q_{Ein,t}^{CM}= P^{CM}\left(\frac{\pi}
{\beta \Lambda_E}\right)^{3(N-1)/2}
\left(N\right)^{3/2}
\end{equation}
The factor $P^{CM}$ accounts for the contribution of the integral over 
the space of momenta. Its 
value is not given explicitly, because we will see later that it
cancels out with another similar term.

\subsubsection{Step 1. Free energy change between an interacting Einstein crystal and
a non-interacting Einstein crystal (both with fixed center of mass): evaluating $\Delta A_1$.}

The free energy difference between two arbitrary systems 1 and 2
is given by:
\begin{equation}
\label{deltaA}
A_{2}-A_{1}=-k_BT \ln \frac{\int \exp(-\beta U_2)d1...dN}
{\int \exp(-\beta U_1)d1...dN}.
\end{equation}
Multiplying and dividing the numerator of the integrand by the
factor $\exp(-\beta U_1)$, it is obtained that:
\begin{equation}
\label{enigma}
A_{2}-A_{1}=-k_BT \ln \left<\exp\left[-\beta(U_2-U_1)\right]\right>_1
\end{equation}
where $\left<\exp\left[-\beta(U_2-U_1)\right]\right>_1$ is an average over
the configurations visited by the system 1.
Taking $U_2=U_{Ein-id}+U_{sol}$ and $U_1=U_{Ein-id}$ (being $U_{sol}$ the intermolecular 
potential of the solid),
the previous expression
can be written:
\begin{equation}
\label{psfmn}
A^{CM}_{Ein-sol}-A^{CM}_{Ein-id}=
-k_BT \ln \left<\exp\left[-\beta(U_{sol})\right]\right>_{Ein-id}
\end{equation}
Therefore, the free energy change can be computed simply 
as the ensemble average of the factor $\exp\left[-\beta(U_{sol})\right]$
along a simulation of the ideal Einstein crystal with fixed center of mass. This
average is evaluated in a NVT MC simulation \cite{vega95qq,vega98,eva_patchy,vega03}. 
Note that
this calculation must be done with the center
of mass fixed.  Often it is not possible to evaluate the free energy change 
as expressed in equation (\ref{psfmn}), because the exponential
$exp(-\beta U_{sol})$ takes values larger than those that
can be handled by a computer. This problem can be avoided if the
expression is rewritten in such a way that the exponent does not
take large values, for example, adding and subtracting from the energy
of the solid $U_{sol}$ the constant lattice energy
$U_{lattice}$:
\begin{equation}
\label{deltaa1}
\fl \qquad
\Delta A_{1}= A^{CM}_{Ein-sol}-A^{CM}_{Ein-id}= U_{lattice}-k_BT \ln 
\left<\exp\left[-\beta(U_{sol}-U_{lattice})\right]\right>_{Ein-id}.
\end{equation}
One of the parameters that needs to be fixed when implementing
the Einstein crystal method is the value of the spring constant 
(we will choose  $\Lambda_E=\Lambda_{E,a}=\Lambda_{E,b}$).
A convenient choice for  $\Lambda_E$ is one that guarantees a small
value (of about $0.02 Nk_BT$) for the second term on the right hand side
of equation (\ref{deltaa1}). When this is the case $\Delta A_{1}$ is quite close
to the lattice energy  $U_{lattice}$, defined as the intermolecular
energy of the system when the molecules stand on the positions and
orientations  of the external Einstein field. 

\subsubsection{Step 2. Free energy change between the solid and the 
interacting Einstein crystal (both with fixed center of mass) : evaluating $\Delta A_2$.}
The free energy change between the solid 
and the interacting Einstein crystal 
(both with fixed center of mass) will be 
computed by Hamiltonian thermodynamic integration. The harmonic springs
are turned off gradually, and the total potential
energy can be given by:
\begin{equation}
\label{desenchufo_muelles_crystal}
U(\lambda)=\lambda U_{sol}+( 1 - \lambda) (U_{Ein-id}+U_{sol}).
\end{equation}
The parameter $\lambda $ is defined between 0 and 1, so that
when $\lambda =0$ one has the Einstein solid and when $\lambda=1$ one obtains the 
solid of interest.  The free energy change along this path will be 
given by:
\begin{eqnarray}
\fl
\Delta A_2 = A(N,V,T,\lambda=1)-A(N,V,T,\lambda=0)&=&\int^{\lambda=1}_{\lambda=0}\left<\frac{\partial
U(\lambda)}{\partial\lambda}\right>_{N,V,T,\lambda}d\lambda \nonumber\\
\fl
&=& - \int^{\lambda=1}_{\lambda=0}\left<U_{Ein-id}\right>_{N,V,T,\lambda}d\lambda.
\label{delta_a2}
\end{eqnarray}
It is a good idea to use the same values for 
$\Lambda_E,\Lambda_{E,a}, \Lambda_{E,b}$.
Then the spring constant along the integration are given by 
$\lambda \Lambda_E$, $\lambda \Lambda_{E,a}$ and
$\lambda \Lambda_{E,b}$ (being all of them equal). 
It is convenient to perform a change in the independent variable from $\lambda$ to 
$\lambda \Lambda_E$ so that the integral of equation (\ref{delta_a2}) can be rewritten as:
\begin{equation}
\label{delta_a2_integral}
\Delta A_2 =A^{CM}_{sol}-A^{CM}_{Ein-sol}=-\int^{\Lambda_E}_{0}\frac{\left<U_{Ein-id}\right>_{N,V,T,\lambda}}{\Lambda_E}d(\lambda \Lambda_E).
\end{equation}
Since the integrand changes several orders of magnitude it is convenient to perform a new
change of variable \cite{frenkel84,frenkelbook} $\lambda \Lambda_E$ to $\ln(\lambda \Lambda_E+c)$ where
$c$ is a constant:
\begin{equation}
\label{lqvi}
\fl
\Delta A_2 = A^{CM}_{sol}-A^{CM}_{Ein-sol}=\Delta A_{2}= -\int^{\ln(\Lambda_E+c)}_{\ln(c)}
\frac{\left<U_{Ein-id}\right>_{N,V,T,\lambda}
(\lambda \Lambda_E+c)}{\Lambda_E}d(\ln(\lambda \Lambda_E+c)).
\end{equation}
The integrand is now a smooth function of the variable $\ln(\lambda \Lambda_E+c)$.
A value \cite{frenkel84} of $c=\exp(3.5)$ provides a 
good estimate of the integral (although the 
optimum value of $c$ may depend on the particular considered problem). 
The integral of this smoother function
can be accurately computed using, for example, the Gauss-Legendre 
quadrature formula \cite{numericalrecipes}. It is usual to
use between ten  to twenty points to evaluate the integral.
Fixing the position of the center of mass avoids the quasi-divergence 
of the integrand of equation (\ref{delta_a2_integral}) when the coupling parameter $\lambda$ tends 
to zero.  Without this constraint, the integrand would increase sharply in this limit (although it
would remain finite), making the evaluation  of the integral (equation (\ref{delta_a2_integral})) 
numerically involved, and  making the accurate evaluation of the integrand at low
values of the coupling parameter somewhat difficult. For this reason, it 
is numerically convenient to avoid the
translation of the crystal as a whole for low values of $\lambda$ and this is achieved, 
either by fixing the center of mass, as in the Einstein crystal technique or by fixing
the position of one molecule of the system as in the Einstein molecule approach to be 
described below. 
In Appendix B, the procedure to implement the somewhat unpleasant condition 
of fixed center of mass within a Monte Carlo simulation is described. This
is important since the calculations leading to $\Delta A_1$ and $\Delta A_2$ should
be done with the center of mass fixed. 


\subsubsection{Step 3. Free energy change between an unconstrained solid and the solid
with fixed center of mass: evaluating $\Delta A_3$.}

As we have seen before (see equation (\ref{deltaA})) the free energy change 
between two systems can be obtained as:
\begin{equation}
\label{asol_asolcm}
\Delta A_3 = A_{sol}-A^{CM}_{sol}=-k_BT \ln \frac{Q_{sol}}{Q^{CM}_{sol}}
=k_BT\ln \frac{Q^{CM}_{sol}}{Q_{sol}}
\end{equation}
where $Q_{sol}^{CM}$ is given (after integration over rotational momenta) by :
\begin{eqnarray}
\fl
Q_{sol}^{CM} = & \frac{ (q_{r}q_{v}q_{e})^{N} }{N! h^{3(N-1)}} \int \exp\left[-\beta \sum ^{N}_{i=1} 
\frac {{\bf p}^{2}_{i}}{2m_{i}}\right]\delta(\sum^{N}_{i=1}{\bf p}_{i})
d{\bf p}_{1}...d{\bf p}_{N} \nonumber \\
\fl
&   \int\exp\left[-\beta U_{sol}({\bf r}_{1},\omega_{1}...{\bf r}_{N},\omega_{N})\right]
\delta(\sum^{N}_{i=1} \mu_{i}
({\bf r}_{i}-{\bf r}_{io})) d{\bf r}_1 d\omega_{1} ...d{\bf r}_N d\omega_{N}.
\end{eqnarray}
and $Q_{sol}$ is given by an expression similar to that of $Q_{sol}^{CM}$ but 
without the delta functions (and with $h^{3N}$ in the denominator instead of $h^{3(N-1)}$ ).  Notice that the factor $N!$ cancels out
when computing the free energy change (it appears both in 
$Q_{sol}$ and $Q^{CM}_{sol}$ ).
The integration over the space of momenta of the unconstrained solid 
is simply the integral of a product of Gaussian functions, whose
solution is (when all molecules have the same mass):
\begin{equation}
P= \left(  \frac{ 2\pi m k_B T }{ h^{2} } \right) ^{(3N)/2} = 
\left(\frac{1}{\Lambda}\right)^{3N} 
\end{equation}
The integral over the space of momenta of the solid
with fixed center of mass is equal to the integral of momenta
of the ideal Einstein crystal with fixed center
of mass which was denoted as $P^{CM}$.
Substituting the partition functions in equation (\ref{asol_asolcm}),
we arrive to the following expression:
\begin{eqnarray}
\fl \qquad
\Delta A_3  = A_{sol}-A^{CM}_{sol} = 
k_BT \ln \left( \frac{P^{CM}}{P} \right)  \nonumber \\
\fl
\qquad + k_BT \ln  \frac{ \int \exp\left[-\beta U_{sol}({\bf r}_{1},\omega_1...{\bf r}_{N},\omega_N)\right]\delta(\sum^{N}_{i=1}
(1/N)({\bf r}_{i}-{\bf r}_{io}))d{\bf r}_1\omega_1...d{\bf r}_N\omega_N }
{ \int\exp\left[-\beta U_{sol}({\bf r}_{1},\omega_1,..{\bf r}_{N},\omega_N)\right]
d{\bf r}_1 d\omega_1...d{\bf r}_N d\omega_N } 
\label{deltaa3larga}
\end{eqnarray}
The energy of a system is not modified if the system is translated (while keeping the
relative orientation of the molecules). The mathematical consequence
of that is that $U_{sol}( {\bf r}_{1},\omega_1,..{\bf r}_{N},\omega_N )$ can be 
rewritten as  
$U_{sol}(\omega_1,{\bf r}_{2}^{'},\omega_2,...{\bf r}_{N}^{'}\omega_N$) where ${\bf r}_{i}^{'}= {\bf r}_{i}- {\bf r}_{1}$.
Let us locate the center of mass of the lattice point at the origin of the coordinates 
system so that ( $\sum (1/N){\bf r}_{io} = {\bf R}^0_{CM}=0$). 
Let us perform a
change of variables from ${\bf r}_{1},{\bf r}_{2}, ... {\bf r}_{N}$ to 
${\bf r}_{2}^{'}, ... {\bf R}_{CM}$ where ${\bf R}_{CM}$ is the position of the 
center of mass of the
reference points. The Jacobian
of this transformation is N. With these changes one obtains for 
the second term on the right 
hand side of equation (\ref{deltaa3larga}):
\begin{equation}
\fl
k_BT\ln\frac{\int\exp(-\beta U_{sol}(\omega_1,{\bf r}_{2}^{'},\omega_2,{\bf r}_{3}^{'},...{\bf r}_{N}^{'},\omega_N )) 
\delta({\bf R}_{CM}) N d\omega_1 d{\bf r}_{2}^{'} d\omega_2 d{\bf r}_{3}^{'}..  d{\bf r}_{N}^{'}d\omega_N d{\bf R}_{CM}}
{\int\exp(-\beta U_{sol}(\omega_1,{\bf r}_{2}^{'},\omega_2,{\bf r}_{3}^{'},...{\bf r}_{N}^{'},\omega_N ))  
N d\omega_1 d{\bf r}_{2}^{'}d\omega_2 d{\bf r}_{3}^{'}..d{\bf r}_{N}^{'} d\omega_N d{\bf R}_{CM}}
\end{equation}
After integrating with respect to $\omega_1 {\bf r}_{2}^{'}\omega_2... {\bf r}_{N}^{'}\omega_N$ one obtains:
\begin{equation}
\label{cm_espacio}
k_BT \ln \frac{ \int  \delta({\bf R}_{CM})  d{\bf R}_{CM} }
{ \int d{\bf R}_{CM} } =  k_BT \ln \frac{1}{  \int d{\bf R}_{CM} } 
\end{equation}
since the Dirac Delta is normalised to one. Now there is a quite subtle issue. 
The integral in the denominator of equation (\ref{cm_espacio}) is just the volume 
available to the center of mass. What is the value of this volume? 
An interesting comment pointed out explicitly by Wilding \cite{wilding_pre00,wilding02,wilding06,wilding_rev} is that the translation of a 
crystal as a whole under 
periodical boundary conditions generates $N$ permutations between the particles.
This is illustrated in fig.\ref{einsmol} for a two dimensional model. 
When counting the number of possible configurations we used the value $N!$ when 
going from equation (\ref{eins_id_1} to equation (\ref{eins_id_2}). 
Therefore, we counted all possible permutations. 
Therefore the integral in the denominator of equation (\ref{cm_espacio}) is the
volume available to the center of mass, within one given permutation. 
This value is simply $V/N$. Using V instead of (V/N) in the denominator 
of equation (\ref{cm_espacio}) is incorrect if the value $N!$ was used to count the number of 
permutations. In this case certain permutations would be counted twice, the first time in the
factor $N!$ and the second via the translation of the whole crystal (in the volume V).
Therefore :
\begin{equation}
\label{deltaa3}
\Delta A_{3} =   
A_{sol}-A^{CM}_{sol}=k_BT \left[ \ln ( P^{CM}/P )  -   \ln (V/N) \right]
\end{equation}
As can be seen the expression for $\Delta A_{3}$ is general and does not depend on the
particular form of the intermolecular potential $U_{sol}$.
Notice that correct results would also be  obtained if one uses $V$ in the 
denominator of equation (\ref{cm_espacio}) (so that the center of mass moves in the 
whole simulation box) but uses $(N-1)!$ when counting the number of permutations
(i.e. count all permutations between particles except those obtained  via the 
translation of the whole crystal through the periodical boundary conditions). 
In equation (\ref{eins_id_2}) one then would obtain a term  $(N-1)!/N!$ which 
provides an $1/N$ factor that could be joined with the $\ln(1/V)$ term of 
equation (\ref{cm_espacio}) to give a contribution $-kT \ln(V/N)$ which is 
identical to that given in equation (\ref{deltaa3}).
Thus $\Delta A_{3}$ will have a term of the form $-kT\ln (V/N)$ if $N!$ 
permutations were included in 
$A_{Eins-id}^{CM}$ (as done by Polson {\em et al.} \cite{polson00}, and described here)
or will have a term of the form $-kT\ln (V)$ if 
$(N-1)!$ permutations were included in $A_{Eins-id}^{CM}$. Both choices are possible and provide
the same total free energy. However when presenting results it is important to state clearly 
the choice not to confuse the reader.
A sentence like that could be useful:
\begin{itemize}
\item{ All permutations were included in the reference ideal Einstein crystal. That would
indicate that a term $N!$ was used, and therefore  $\Delta A_3$ contains a term of the form 
$-kT \ln(V/N)$}
\item { All permutations except those obtained by translation of the crystal under periodical
boundary conditions were included in the reference ideal Einstein crystal. 
That would indicate that a term $(N-1)!$ was used, and therefore  $\Delta A_3$ contains a term of the form 
$-kT \ln(V)$ }
\end{itemize}
However when presenting results we recommend to join $A_{Eins-id}^{CM}$ and $\Delta A_3$ into a 
unique term since the sum of both terms is unique and does not depend on the choice of the number of
permutations included in $A_{Eins-id}^{CM}$. 
It is fair to say that Wilding \cite{wilding00} was the first pointing out explictly that N permutations were generated 
by translation under periodical boundary conditions. This has been taken into account implicitly by 
Polson {\em et al.} \cite{polson00}, since they used a term of the form $-kT \ln(V/N)$ for $\Delta A_3$. 

\subsubsection{Final expression}
The final expression of the free energy of the solid is :
\begin{equation}
A_{sol}=(A_{Ein-id}^{CM}+\Delta A_3)+\Delta A_1+\Delta A_2=A_{0}+\Delta A_1+\Delta A_2 
\end{equation}
where we have defined  $A_{0}$ as  $A_0= A_{Ein-id}^{CM} + \Delta A_3$. Taking into account all the contributions to the free energy,
the free energy of a molecular solid can be computed
using the following expression:
\begin{eqnarray}
\fl \qquad
\frac{A_{sol}}{Nk_BT}=
-\frac{1}{N} \ln \left[ \left(  \frac{1}{\Lambda} \right)^{3N}
\left( \frac{\pi}{\beta \Lambda_E} \right) ^{3(N-1)/2}
(N)^{3/2} \frac{V}{N} \right]
+\frac{A_{Ein,or}}{Nk_BT}\nonumber \\
\fl \qquad 
+\left[ \frac{U_{lattice}}{Nk_BT}-\frac{1}{N}\ln \left<\exp\left[-\beta(U_{sol}-U_{lattice})\right]\right>_{Ein-id} \right]
-\int^{\lambda=1}_{\lambda=0}\left<\frac{U_{Ein-id}}{Nk_BT}\right>_{N,V,T,\lambda} d\lambda
\label{atotal}
\end{eqnarray}
Notice that $P^{CM}$ does not appear in the final expression (so that its value
is irrelevant for free energy calculations). 
The first two terms in equation (\ref{atotal}) correspond to $A_0$.
The last two terms on equation (\ref{atotal}) are $\Delta A_1$ and $\Delta A_2$
respectively.  
The argument of the logarithm in the first term on the right hand side (embraced by 
brackets) is adimensional. In fact it has three factors, the first factor having 
dimensions of L$^{-3N}$, the second factor having dimensions of L$^{3(N-1)}$ and the last 
factor having dimensions of  L$^3$. 
In any computer program a unit of length ${\it l}$ is selected.
It is quite convenient to set the thermal de Broglie wave length
to $\Lambda={\it l}$, and this choice should be used for the solid (in equation (\ref{atotal}))
and for the liquid (in equation (\ref{itnhgi})). 
Then the volume of the simulation box V (in equation (\ref{atotal})) should be 
given in ${\it l^{3}}$ units and the value of the translational 
spring $\Lambda_{E}$ should be given in Energy$/({\it l^{2}} )$ units. 
Notice that assigning an arbitrary value to $\Lambda$ affects the
absolute value of the free energies but it does not affect the coexistence
properties.

An important final comment is in order. Free energy calculations are usually performed
in the NVT ensemble (with temperature and density fixed). It is quite important
that the shape of the  simulation box used in  free energy calculations
corresponds to that adopted by the system at equilibrium. It is
not valid  to impose (for instance from experiment) the shape of the simulation 
box since that will give free energies
higher than the correct ones (the equilibrium shape minimises the free energy of the system 
for a certain density). 
Rather one should first perform
NpT anisotropic Monte Carlo simulations \cite{parrinello80,parrinello81,yashonath85}
, and determine the shape at equilibrium 
of the simulation box at a certain p,T  and Hamiltonian (the density will be obtained as an average 
of the run) and then to perform free energy calculations in the NVT ensemble using the density and
equilibrium shape of the simulation box obtained from the NpT runs. This remark is important for
solids belonging to any crystalline class but cubic. 
A convenient choice for the vectors  ${\bf r}_{i0}$,  $\vec{a}_{i,0}$ and 
$\vec{b}_{i,0}$ that define the Einstein crystal field  (equations 
(\ref{ete_total}), (\ref{ete_t}) and (\ref{ete_r})), is to use the equilibrium 
positions (to determine ${\bf r}_{i0}$)
and orientations (to determine $\vec{a}_{i,0}$ and $\vec{b}_{i,0}$) 
of the molecules of the system. Other choices are also possible (for
instance fields driving the molecules into configurations slightly distorted from 
the equilibrium one). However the choice of the equilibrium configuration has the  
advantage that a lower value of the external field $\Lambda_E$ is needed 
to obtain reliable results. Obviously the free energy of the solid should not 
depend on the particular choice of the vectors  ${\bf r}_{i0}$,  $\vec{a}_{i,0}$ and $\vec{b}_{i,0}$
that define the Einstein crystal field.

\subsection{The Einstein molecule approach}

Quite recently Vega and Noya 
have proposed \cite{vega_noya} a slightly different version of the Frenkel Ladd method. The method
has been denoted as the Einstein molecule approach.
The idea behind the Einstein molecule approach is to fix 
the position of one molecule of the system (say molecule 1) instead of fixing the 
center of mass. More precisely, by fixing the position of molecule 1, we mean
that we fix the position of its reference point.  
The molecule can still rotate as far as its reference point remains fixed. 
Therefore we fix the position of molecule 1 (as given by the reference point) but 
we do not fix the orientation of molecule 1. Of course for a simple fluid (HS, LJ) there
are no orientational degrees of freedom so that in the Einstein molecule approach, atom 1 is fixed. 
Fixing the position of one molecule avoids the quasi-divergence of the integrand
of equation (\ref{delta_a2_integral}) when the coupling parameter $\lambda$ 
tends to zero. 
The computational implementation of the method  as well as the
derivation of the main equations is rather simple.  

\subsubsection{The ideal Einstein molecule: definition and free energy}

The partition function in the canonical
ensemble (after integrating over the space of momenta) is given by
the following expression:
\begin{equation}
Q= \frac{ (q_r q_v q_e)^N  }{N! \Lambda^{3N}}
\int \exp\left[-\beta U({\bf r}_1,\omega_1,...,{\bf r}_N,\omega_N) \right]d{\bf r}_1 d\omega_1...d{\bf r}_N d\omega_N
\end{equation}
We shall assign $q_r,q_v,q_e$ the arbitrary value of one. 
This expression can be written in a more convenient way by exploiting the
fact that the potential energy of the system $U$ depends only
on the relative positions of the particles, but not on their absolute
positions, i.e., it is invariant under translations of the whole system (while keeping
the orientations of all the molecules in the translation).
We will perform a change of variables from $({\bf r}_1,{\bf r}_2, ...,{\bf r}_N)$
to $({\bf r}_1, {{\bf r}_2}^{'}={\bf r}_2-{\bf r}_1, ...,  {{\bf r}_N}^{'}={\bf r}_N-{\bf r}_1)$.
Under periodic boundary conditions and the minimum image
convention, this change of variables leaves the limits of the integrals
unchanged, because the maximum distance between two particles in any of
the three directions of the space is always less than the length of
the simulation box. Therefore:
\begin{eqnarray}
\fl \qquad
Q & = & \frac{1}{N!\Lambda^{3N}} \int d{\bf r}_1
\int \exp\left[-\beta  U(\omega_1,{\bf r}'_2,\omega_2,...,{\bf r}'_N,\omega_N) \right] d\omega_1 d{\bf r}'_2 d\omega_2...d{\bf r}'_N d\omega_N  \nonumber \\
\fl \qquad
& = & \frac{1}{N! \Lambda^{3N}}
\int d{\bf r}_1 \, \kappa
\end{eqnarray}
The value of the integral $\kappa$ is independent of the value of ${\bf r}_1$ and,
therefore, we can integrate over ${\bf r}_1$:
\begin{equation}
Q= 
\frac{1}{N! \Lambda^{3N}} V \, \kappa
\end{equation}
The whole partition function ($\kappa$) can be computed by multiplying the 
integral corresponding to one permutation ($\kappa'$) by the number of possible
permutations, which, for a given fixed position of particle 1, is equal to 
$(N-1)!$. Therefore, the partition function can be written as:
\begin{equation}
Q= \frac{1}{N!\Lambda^{3N}}V(N-1)!\, \kappa'
= \frac{1}{N\Lambda^{3N}} V \, \kappa' 
\label{eq_general_partition}
\end{equation}

\begin{figure}[!h]
\begin{center}
\begin{pspicture}(0,0)(6,6)
\pscoil[coilheight=0.3,coilwidth=0.6]{*-*}(1.5,4.5)(0.5,3.5)
\pscoil[coilheight=0.3,coilwidth=0.6]{*-*}(4.5,4.5)(5.5,5.5)
\pscoil[coilheight=0.3,coilwidth=0.5]{*-*}(4.5,1.5)(5.5,0.5)
\cnode[fillstyle=solid,fillcolor=black](0.5,3.5){0.30cm}{P}
\cnode[fillstyle=solid,fillcolor=black](5.5,5.5){0.30cm}{P}
\cnode[fillstyle=solid,fillcolor=black](5.5,0.5){0.30cm}{P}
\cnode[fillstyle=solid,fillcolor=black](1.5,1.5){0.30cm}{P}
\psline[linewidth=1pt](1.5,1.5)(1.5,4.5)
\psline[linewidth=1pt](1.5,1.5)(4.5,1.5)
\psline[linewidth=1pt](4.5,1.5)(4.5,4.5)
\psline[linewidth=1pt](4.5,4.5)(1.5,4.5)
\uput[r](5.5,5.0){{\large\sf 3}}
\uput[r](0.0,4.3){{\large\sf 2}}
\uput[r](5.8,0.5){{\large\sf 4}}
\uput[r](1.7,1.2){{\large\sf 1}}
\uput[r](1.0,0.7){{\large\sf Carrier}}
\end{pspicture}
\qquad \qquad
\includegraphics[clip,width=0.4\textwidth,angle=-0]{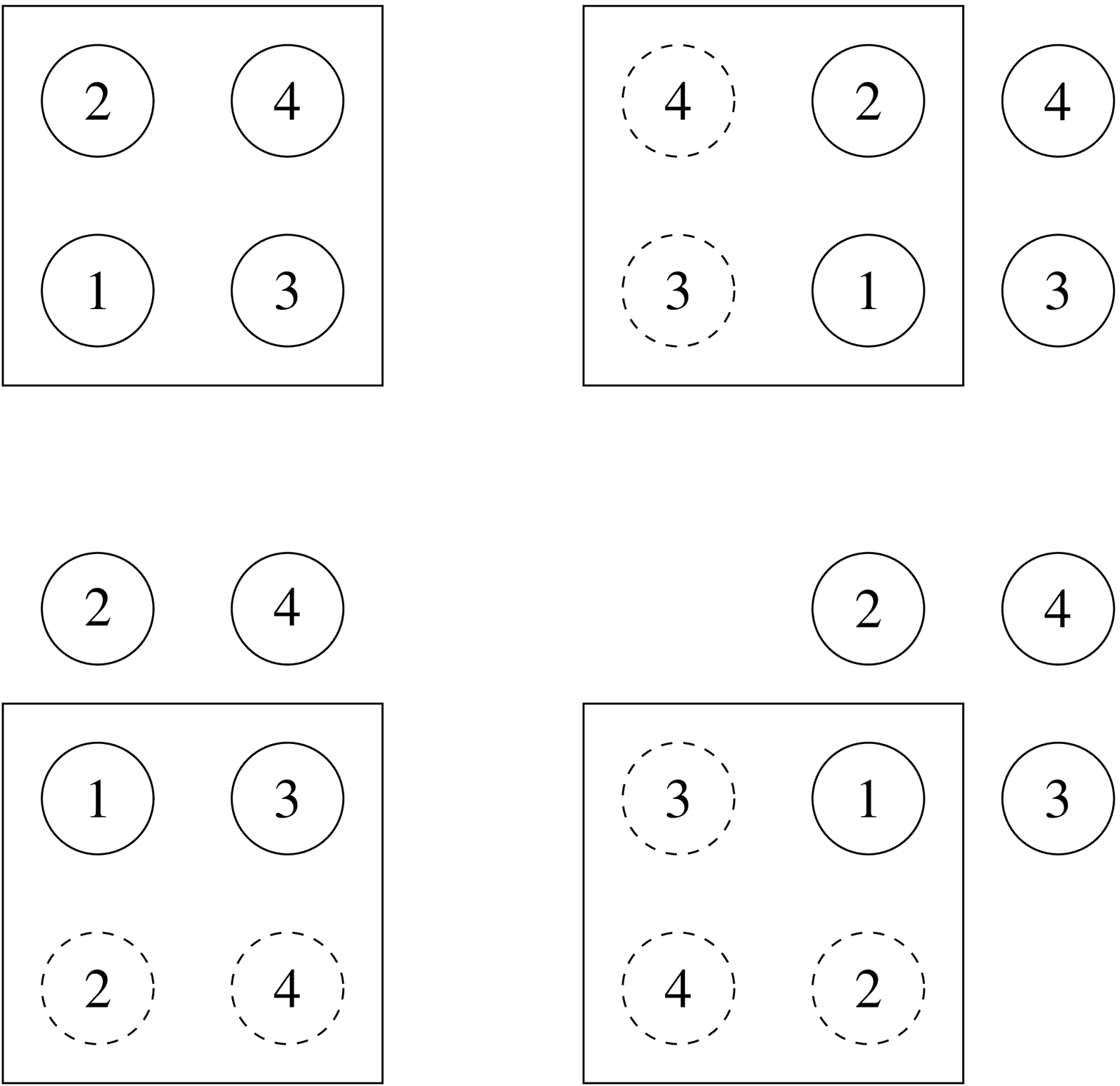}
\end{center}

\caption{\label{einsmol} Left : schematic representation of the Einstein
molecule, in which particle 1 is fixed and acts as the carrier
of the lattice. The movement of all the remaining particles
is given relative to the position of particle 1. Right: Permutations generated
through periodical boundary conditions by the motion of particle 1.}
\end{figure}

Let us now define the ideal Einstein molecule. The ideal Einstein molecule
is an ideal system (without intermolecular interactions) 
where the reference point of one of the 
molecules (e.g., molecule 1) does not vibrate and acts
as reference, while the rest of the molecules of the system (i.e., molecules 2,3,..N) 
vibrate around their equilibrium configurations (see figure \ref{einsmol} for a schematic
representation). 
The reference point of molecule 1 is called the carrier, because this point 
transports the lattice. Notice that in the Einstein molecule, molecule 1 can 
undergo orientational vibrations, as far as its reference 
point remains in a fixed position (obviously for
a simple fluid there is no such a rotation, and the carrier is just the position of
atom 1). 
The lattice(crystal) is uniquely determined by the position of the carrier.
The Einstein molecule can move as a whole, and this motion is represented by the motion of 
the carrier, which  is able to move and occupy any 
position in the simulation box. The expression of the energy of the ideal
Einstein molecule is:
\begin{eqnarray}
\label{ete_einstein_molecule}
U_{Ein-mol-id} = U_{Eins-mol-id,t}+U_{Ein,or}       \nonumber \\
U_{Ein-mol-id,t}= \sum ^{N}_{i=2} \left[\Lambda_E({\bf r}_{i}-{\bf r}_{io})^{2} \right]
\end{eqnarray}
Notice that the main difference with equation (\ref{ete_t}) is the absence 
of an harmonic term for the reference point of molecule 1. 
The orientational part of the potential is identical 
in the Einstein crystal and in the Einstein molecule approach. 
The partition function of the ideal Einstein molecule can be obtained by 
performing the integral $\kappa'$ for this particular case and substituting the
value in equation (\ref{eq_general_partition}).
The translational integral is particularly simple since is just a set of 
$3(N-1)$ oscillators. 
The orientational contribution is obtained as in the Einstein crystal 
approach.  Therefore, the Helmholtz free energy $A_{Ein-mol-id}$ of 
the ideal Einstein molecule is given by:
\begin{equation}
\fl
\frac{\beta A_{Ein-mol-id}}{N} = - \frac{1}{N} \ln (Q) = \nonumber \\
\frac{1}{N} \, \ln \left( \frac{N\Lambda^{3}}{V} \right) + \frac{3}{2}
\left(1-\frac{1}{N}\right) \hspace{0.2cm} \ln \left( \frac{ \Lambda^{2} \beta \Lambda_{E} }{\pi} \right)
- \frac{1}{N} \ln ( Q_{Ein,or} )
\end{equation}
In the case that the carrier molecule (molecule 1) is
fixed, the free energy will be equal to the free energy
of the ideal Einstein molecule plus a term $kT\ln(V/\Lambda^3)$ (where the term
$V$ comes from the constraint on the position of molecule 1, and the
term $\Lambda^3$ comes from the constraint on the momentum).

\subsubsection{Integration path and computation of the free energy in each step.}

In the ideal Einstein molecule approach, the free energy of a given solid
will be computed from integration to the ideal Einstein molecule. 
This integral is performed in several steps, that are summarised in
the scheme shown in figure \ref{esquema_eva}. First the ideal Einstein molecule 
is transformed into an ideal Einstein molecule with one molecule fixed (what we 
mean by particle fixed is that its reference point remains fixed). 
Then the ideal Einstein molecule with one particle  
fixed is transformed
into the real solid with one particle fixed. In the last step this fixed
particle is allowed to move to obtain the real solid.
As it is shown in the scheme, the factor $kT\ln(V/\Lambda^3)$ that
appears as a result of fixing one molecule in the ideal Einstein molecule
cancels out with the free energy contribution of allowing molecule 1
to move to recover the real solid. As a result, the free energy 
of a solid can be computed simply by adding to the free energy of
the ideal Einstein molecule, the free energy change between an ideal
Einstein molecule with one fixed atom and the solid with one
fixed atom (given by $\Delta A_1^*+\Delta A_2^*$): 
\begin{equation}
A_{sol}=A_{Ein-mol-id} + \Delta A_1^* + \Delta A_2^* = A_0^{*} +  \Delta A_1^* + \Delta A_2^*
\end{equation}
where the asterisk in $\Delta A_1^*$ and $\Delta A_2^*$ serves to remind us that
the integral should be performed while keeping the position of the reference 
point of molecule 1 fixed (and $A_0^{*}$ is just $A_{Ein-mol-id}$). 
The computation of the free energy change between the solid and the
ideal Einstein molecule keeping one particle fixed is completely analogous
to the computation of the free energy change between the solid 
and the ideal Einstein crystal keeping the center of mass of the system fixed.
As in the Einstein crystal method, this free energy change will be calculated
in two steps, represented by the terms $\Delta A_1^*$ and $\Delta A_2^*$. 
In particular, 
in the first step ($\Delta A_1^*$) we will compute the free energy change 
between the interacting Einstein molecule with one fixed particle
and the ideal Einstein molecule with one fixed particle.
This free energy change is evaluated using the same procedure
as in the Einstein crystal method
with the difference that, instead of fixing the center of mass, the position
of molecule 1 is kept fixed:
\begin{equation}
\label{deltaa1p}
\Delta A_{1}^{*}=  U_{lattice}-k_BT \ln 
\left<\exp\left[-\beta(U_{sol}-U_{lattice})\right]\right>_{Ein-mol-id}.
\end{equation}
which is formally identical to equation (\ref{deltaa1}) except for the fact that
averages should be computed over the ideal Einstein molecule system, rather than
over the ideal Einstein crystal, and that molecule 1 will be fixed instead of 
the center of mass.

In the second step, the free energy change between the
interacting Einstein molecule with one fixed particle and the solid with 
one fixed particle is computed ($\Delta A_2^*$). This will be done by slowly
switching off the springs of the interacting Einstein molecule :
\begin{equation}
U(\lambda)=\lambda U_{sol}+( 1 - \lambda) ( U_{Ein-mol-id} + U_{sol} )
\end{equation}
The parameter $\lambda $ is defined between 0 and 1, so that
when $\lambda =0$ one has the interacting Einstein molecule, and when 
$\lambda=1$ one obtains the solid of interest (both with the position of molecule 
1 fixed). $U_{sol}$ is the potential
of the system under consideration. This equation is equivalent to 
equation (\ref{desenchufo_muelles_crystal}) for the Einstein crystal. 
The free energy change in this first step will be calculated from
the following expression:
\begin{equation}
\label{delta_a2_integral_einstein_molecule}
\Delta A_2^* =-\int^{\Lambda_E}_{0}\frac{\left<U_{Ein-mol-id}\right>_{N,V,T,\lambda}}{\Lambda_E}d(\lambda \Lambda_E).
\end{equation}
which is identical to equation (\ref{delta_a2_integral}) except for the replacement 
of $U_{Ein-id}$ by  $U_{Ein-mol-id}$.
The asterisk indicates that the reference point of molecule 1 is fixed in the integration.  Notice  
that this integral does not diverge at low values of
$\lambda$, because the translations of the system
as a whole are prevented by fixing the reference point of molecule 1. 


\subsection{ Calculations for the hard sphere solid}
\begin{figure}[!hbt]
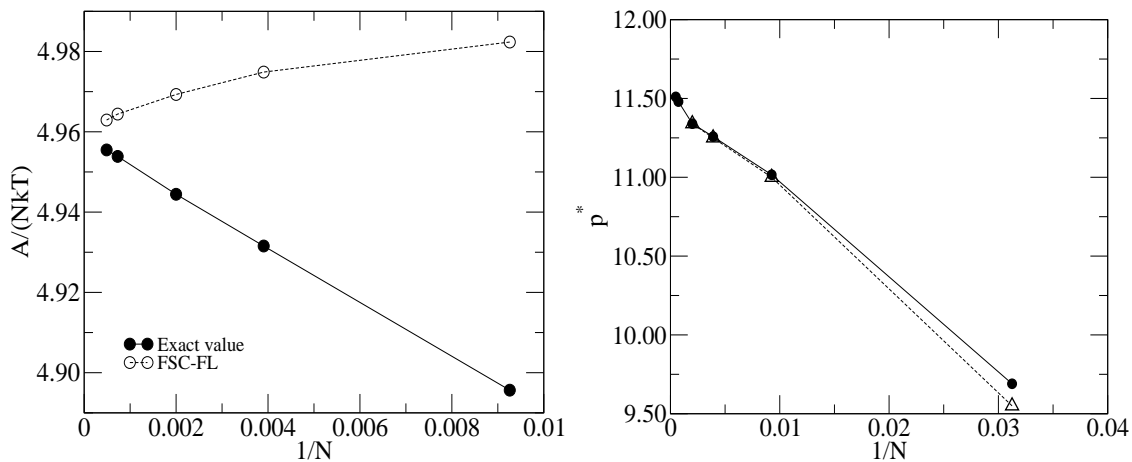
\centering
\includegraphics[clip,height=6cm,width=0.47\textwidth,angle=-0]{a_hs_review.eps}
\includegraphics[clip,height=6cm,width=0.47\textwidth,angle=-0]{pressure_review.eps}
\caption{\small\sf 
Left. Free energies of HS in the fcc solid for $\rho^{*}=1.04086$ as
a function of system size (filled circles). 
The open circles represent the free energies after including the Frenkel Ladd 
finite size corrections (i.e., adding $(2/N) \ln N$ to the free energies
of the solid). 
For all values of N the free energies obtained here (black circles) are in 
excellent agreement with those reported by Polson {\em et al.} \cite{polson00} and 
de Miguel {\em et al.} \cite{enrique07}. 
Right. Coexistence pressure of the fluid-solid equilibria of hard spheres
as a function of the system size as obtained from free energy calculations 
(filled circles) or from phase switching simulations as reported by Wilding 
\cite{wilding00} and Errington \cite{errington04} (open triangles).}
\label{a_hs_review} 
\end{figure}

Let us now present some results for the free energy 
of a fcc solid of hard spheres at a density $\rho^*=1.04086$. We shall compute
the free energy using both, the Einstein crystal methodology \cite{frenkel84,polson00} described 
extensively in this paper and the Einstein molecule approach. 
Results are presented in Table  \ref{tbl_free_terms}. 
The first  point to be noted is that $\Delta A_1$ and $\Delta A_1^*$ (and $\Delta A_2$ and $\Delta A_2^*$ )
are similar but not identical (reflecting the fact that it is not exactly 
the same fixing the center of mass as fixing molecule 1). However, the sum of all terms 
contributing to $A_{sol}$ gives the same value, so that the estimated free energy is the same (within 
statistical errors) with both  methodologies. 
Obviously the free energy of a well-defined state should not depend on the procedure chosen to compute it. 
Since $\Delta A_1$ and $\Delta A_1^*$ are quite similar,
and the free energy of the system must be the same computed by both routes (fixing the center of mass or
fixing molecule 1), then $\Delta A_2$ and $\Delta A_2^*$ must differ in about $3\ln(N)/(2N)$ which 
is the analytical difference between  $A_{0}^{*}$ and  $A_{0}$.
This is indeed the case as it can be seen in Table \ref{tbl_free_terms}.
The third aspect to be considered from the results of Table \ref{tbl_free_terms}
is that the total free energy presents a strong size dependence. 
Notice that this is not a problem of the methodology chosen to compute the free
energy, but it is an intrinsic property of the HS solid (and likely of 
other solids  as well). In other words, the free energy of solids presents important finite size effects.
This is further illustrated in figure \ref{a_hs_review} where the free energy is 
plotted as a function of $1/N$. 
The estimated value of $A/(Nk_BT)$ in the thermodynamic limit from our results 
is  $4.9590(2)$, which is in good agreement with the estimates of Polson 
{\em et al.} \cite{polson00} (4.9589), Chang and Sandler \cite{sandler1} 
(4.9591), Almarza \cite{noe} (4.9589) and de Miguel {\em et al.} (4.9586) 
\cite{enrique07} (all obtained from free energy calculations although with
different implementations).
Therefore, the value of the free energy
of hard spheres in the thermodynamic limit for the
density  $\rho^{*}=1.04086$ seems to be firmly established.

A consequence of the strong N dependence of the solid free energy is that the coexistence pressure
$p^{*}$ also presents a strong N dependence as illustrated in figure \ref{a_hs_review}.
It is of interest to estimate the properties at coexistence in the thermodynamic limit. 
We found  \cite{vega_noya}, $p^{*}=p/(k_BT/\sigma^3)=11.54(4)$, 
$\rho^*_{s}=\rho_{s}\sigma^3=1.0372$, $\rho^*_{l}=\rho_{l}\sigma^3=0.9387$, and 
$\mu^*=\mu/(k_BT)=16.04$. The coexistence pressure
is in agreement with estimates by Frenkel and Smit \cite{frenkelbook} (11.567),
Wilding \cite{wilding00} (11.50(9)),  Speedy \cite{speedy97} (11.55(11)) 
and Davidchack and Laird \cite{laird98} (11.55).
The  Hoover and Ree  estimate (11.70) seems now to be a little bit high. 
The chemical potential at coexistence obtained here is  consistent with the
value reported by Sweatman \cite{sweatman}  ($\mu^*=15.99-16.08$) obtained 
using the self referential methodology to compute fluid-solid equilibria.

Although finite size effects are present both in fluid and solid phases, they seem to be more 
pronounced in the solid (probably due to the coupling between the periodical boundary conditions
and the geometry of the solid). In principle one is interested in properties of the system
in the thermodynamic limit rather than for a finite size system.  
To estimate free energies in the thermodynamic limit one should repeat 
the free energy calculations for several system sizes and extrapolate to the thermodynamic
limit. This is quite involved and time consuming. 
For this reason it is of practical interest to introduce
finite size corrections (FSC) that allow the estimation (although in an 
approximate way) of large systems free energies, by performing simulations of 
small systems (something similar to the g(r)=1 approximation \cite{allen_book} used to
correct for the introduction of the cut-off). 
Several recipes have been proposed recently \cite{vega_noya}.
Here we shall describe one of them, namely the Frenkel-Ladd FSC. 
\begin{table}[!h]
\centering
\caption{\label{tbl_free_terms} \small\sf Free energy of the fcc hard sphere solid at
a density $\rho^*=1.04086$. The value of the different terms that contribute
to the free energy 
in the Einstein molecule and in the Einstein crystal methods, are also shown.
All free energies are given in $Nk_{B}T$ units. The thermal de Broglie wave length was 
set to $\Lambda=\sigma$, the hard sphere diameter.\vspace{0.2cm}}
\small
\begin{tabular}{ccccccccccc}
\hline\hline
 && \multicolumn{4}{c}{Einstein molecule} &  & \multicolumn{4}{c}{Einstein crystal}  \\
\cline{3-6}  \cline{8-11}
$N$  &  $\Lambda_{E}/(kT/\sigma^{2})$  & $\Delta A_1^*$ & $\Delta A_2^*$ & $A_{0}^{*}$ &  $A_{sol}$ &
&$\Delta A_1$&$\Delta A_2$&$A_{0}$&$A_{sol}$\\
\hline
108  & 632.026 & 0.0172  &  -3.0046 & 7.8830  &  4.896 & & 0.0175  &   -2.9400  & 7.8180 &  4.895  \\
256  & 632.026 & 0.0174  &  -3.0116 & 7.9254  &  4.931 & & 0.0175  &   -2.9797  & 7.8929 &  4.931  \\
1372  & 1000.00 & 0.0018  &  -3.6862 & 8.6383  &  4.955 & & 0.0018  &   -3.6802  & 8.6304 &  4.952  \\
2048  & 1000.00 & 0.0018  &  -3.6866 & 8.6403  &  4.955 & & 0.0015  &   -3.6819  & 8.6347 &  4.954\\
\hline\hline
\end{tabular}
\end{table}

\subsection{Finite size corrections: the Frenkel-Ladd approach.}
In the original paper of 1984, Frenkel and Ladd (FL) provided an  
expression for the free energy of the solid
$(2/N) \ln N$ higher than the correct free energy.
That was first pointed out by Polson {\em et al.} \cite{polson00}.
In Appendix C the reasons for the appearance of the extra term  $(2/N) \ln N$ will be
described. Thus the FL free energy $A_{sol}^{FL}/(Nk_BT)$ is given by:
\begin{equation}
A_{sol}^{FL}/(Nk_BT) = A_{sol}/(Nk_BT) + (2/N) \ln N 
\end{equation}
Notice that the term $(2/N) \ln (N)$ tends to zero in the thermodynamic limit, and therefore
the FL expression is valid in this limit. 
However, for finite systems the FL expression gives a free energy higher than the true
free energy of the system. 
For a typical system size $N=$350
the difference between both values is on the order of $0.03Nk_BT$. 
The interesting issue is that the FL free energies although incorrect (for a certain value
of $N$) are relatively close to the value of the free energy in the thermodynamic limit. 
This is illustrated in figure \ref{a_hs_review} for the HS system.
For this reason, one may simply view the FL expression as containing 
an approximate prescription for the finite size corrections,
providing  free energies
closer to the thermodynamic limit than the correct free energies of 
the system of finite size. Other approximate expressions for the finite size corrections (FSC) have been
proposed recently \cite{vega_noya}.

\subsection{The symmetry of the orientational field in Einstein crystal calculations}
For molecular fluids, the choice of the orientational external field used within
the Einstein crystal (or Einstein molecule) simulations should be done with care
since this can be a source of  methodological errors.
The position of the molecule is given by the position of the reference point.
Things become simpler if the reference point is chosen in such a way that all
elements of symmetry of the molecule contain this point. For instance, a convenient 
choice for $H_{2}O$, $NH_{3}$, benzene and $N_2$  are the O, the N, the center of the 
hexagon, and the geometrical center of the $N_2$, respectively.  
Now two  orthogonal unit vectors $\vec{a}$ and $\vec{b}$ are  attached 
to the reference point, and these 
two vectors are sufficient to define the orientation of the molecule (in fact two degrees of 
freedom are needed to locate the orientation of a unit vector $\vec{a}$, and just one degree of freedom 
to locate the vector $\vec{b}$, which is perpendicular to $\vec{a}$). Therefore 
the unit vectors $\vec{a}$ and $\vec{b}$ are
a useful way of defining the orientation of the molecule (the Euler angles could be used 
as well but it is more convenient to use  $\vec{a}$ and  $\vec{b}$). For convenience 
the vector  $\vec{b}$ is  chosen 
along the principal symmetry axis of the molecule (the $C_{n}$ with the highest value of n).
When the molecule i stands on its equilibrium 
position and orientation in the crystal then the vectors  $\vec{a}_i$ and  $\vec{b}_i$ 
adopt the values  $\vec{a}_{io}$ and $\vec{b}_{io}$ respectively. Thus the 
subindex 0 will refer to the orientation of the molecule in the equilibrium 
lattice position. Let us denote as $\psi_{a,i}$ the angle between the vectors  $\vec{a}_i$ 
and  $\vec{a}_{io}$ and 
$\psi_{b,i}$ the angle between the vectors $\vec{b}_i$ and $\vec{b}_{io}$ in an 
instantaneous configuration. Which expression should be used for the orientational field?
The translational part will always be given as in equation (\ref{ete_t}) for the Einstein crystal 
approach or as in equation (\ref{ete_einstein_molecule}) for the Einstein molecule approach. 
The orientational part will be the same for the Einstein crystal or Einstein molecule
approaches. 
We have already given a convenient expression for the orientational 
field of a molecule with point group $C_{2v}$ (as, for example, water). 
Let us now give a convenient expression for other symmetries. 
For a molecule with a point group of type $C_{nv}$ a convenient expression for $U_{Ein-id,or}$ is :
\begin{equation}
U_{Ein,or} =   \sum^{N}_{i=1} \left[ \Lambda_{E,a} \sin^2 \left( \frac{n \psi_{a,i}}{2} \right) + 
\Lambda_{E,b} \left( \frac{\psi_{b,i}}{\pi} \right)^2 \right].
\end{equation}
For a molecule with a point group of type $D_{nh}$ a convenient expression for $U_{Ein-id,or}$ is :
\begin{equation}
U_{Ein,or}= \sum ^{N}_{i=1} \left[
\Lambda_{E,a}\sin^2\left(\frac{n\psi_{a,i}}{2}\right) 
+ \Lambda_{E,b}\sin^2\left( \psi_{b,i} \right) 
\right].
\end{equation}
For a molecule with point group $O_{h}$, a convenient expression \cite{eva_patchy} 
for  $U_{Ein-id,or}$ is:
\begin{equation}
U_{Ein,or}=\sum ^{N}_{i=1} \left[
\Lambda_{E,a}\sin^2\left(\psi_{a,i,min}\right) 
+ \Lambda_{E,b}\sin^2\left( \psi_{b,i,min} \right) 
\right].
\end{equation}
where $\psi_{b,i,min}$ stands for the minimum angle 
between  $\vec{b}_{io}$  and the six vectors connecting the reference point of the molecule 
with the six octahedral atoms/sites and an analogous definition for  $\psi_{a,i,min}$. 
For a linear molecule only one vector (i.e. vector  $\vec{b}_{i}$)  is needed
and the applied field should 
be of the form, for a $D_{\infty,h}$ :
\begin{equation}
U_{Ein,or}=\sum ^{N}_{i=1} \left[ \Lambda_{E,b}\sin^2\left( \psi_{b,i} \right) 
\right].
\end{equation}
For a $C_{\infty,v}$ molecule a convenient choice is :
\begin{equation}
U_{Ein,or}=\sum ^{N}_{i=1} \left[
\Lambda_{E,b}\left(\frac{\psi_{b,i}}{\pi}\right)^2\right].
\end{equation}
When performing a MC run, it is convenient to introduce two different types of moves, translations
and rotations. In a translation move the molecule moves as a whole and there is no change in 
the orientation of the molecule. Only the change in the translational energy with respect to the
reference Einstein crystal (or molecule) needs to be computed. 
In a second type of move the molecule is rotated in a random direction and angle with respect to an axis 
passing through the reference point  of the molecule. Since the reference point does not change the
position under such a rotation, only the orientational energy with respect to the reference Einstein 
crystal (or molecule needs to be computed). 

The choice of an orientational field adapted to the symmetry of the molecule 
as the ones proposed here is 
highly recommended. When this is done the energy with the external field is invariant to any
of the symmetry operations of the molecule. Thus, a standard MC or MD 
program will provide correct values of the orientational contribution to the free energy. 
One interesting questions is: is it possible to use an external 
orientational field that does not reflect the symmetry of the molecule?
The answer is : in principle, yes, but you should write a special MC or MD code for that purpose. 
Special moves should be added where the symmetry operations of the molecule are implemented.
For instance for water, one should incorporate the $C_2$ operation that exchanges the positions 
of the two H atoms.  
Of course the energy of the molecule with the rest of the system is not affected by this operation.
However, the energy of the molecule with the external orientational field may change when 
the external orientational field does not reflect the symmetry of the molecule
(see the interesting paper by Schroer and Monson \cite{schroer2000} illustrating this problem 
for benzene).

For this reason, it is by far more convenient and simpler to use an orientational
field that respects the symmetry of the molecule (examples for  $C_{nv}$, $D_{nh}$, $O_{h}$ and 
linear molecules have been given here). 
This subtle issue of the symmetry of the orientational field may have been an important source of 
errors in free energy calculations for molecular fluids. Let us just finish by saying that 
although we found convenient to have the vectors $\vec{a}$ and  $\vec{b}$ orthogonal, other
choices (as far as they are not colinear) are also valid and correct.

\subsection{ Einstein crystal calculations for disordered systems}
Let us now discuss briefly the case of disordered solids. When implementing the Einstein 
crystal harmonic springs are incorporated to fix the position (as given 
by the reference point) and the orientation of the 
molecules of the system to the equilibrium configuration. In a disordered solid, 
there may be many possible ``equilibrium configurations'' differing in
a significant way (not just differing 
in the labelling of the molecules). Let us just give 
three examples:

{\em Plastic crystals}.
Molecules with an almost spherical shape tend to form plastic crystals when freezing. In these
plastic crystals the reference points of the molecules form a true lattice but the other atoms
of the molecule are able to rotate (either with a free or with a hindered rotation) around the
reference points.  In principle the Einstein crystal 
methodology described in the previous section can also 
be applied to determine free energies 
for plastic crystals \cite{mulder,singer90,vega92b,vega92c,JCP_1997_106_00666,JCP_1997_107_02696,eva_patchy}. In addition to the translational field, an 
orientational field is included forcing
the molecules to adopt an orientationally ordered solid 
for large values of the orientational field. 
Some issues should be taken into account when evaluating the free energy for 
a plastic crystal phase. At low values of the orientational field very long runs 
should be performed to guarantee that the molecules 
are able to rotate. Many values (20-30) of the coupling parameter 
$\Lambda_E$  should be used to compute 
the integral of equations (\ref{delta_a2_integral}) or (\ref{delta_a2_integral_einstein_molecule})
(the orientational contribution to the integrand increases quite rapidly for low values 
of the orientational field). Finally  the absence of phase transitions along the 
integration path should be checked ( the external field should lead the system from an 
orientationally disordered solid at low values of coupling parameter to an 
orientationally ordered solid for large values of the coupling parameter without
undergoing any phase transition).

{\em Water}.  In the case of solid water (say ice Ih) 
while the oxygens are ordered (i.e., they form a lattice) the hydrogens 
are disordered.  However, Bernal and Fowler \cite{JCP_1933_01_00515_nolotengo} and 
Pauling \cite{pauling35} suggested that configurations 
satisfying the so called Bernal and Fowler rules have the same statistical weight and 
that configurations violating the Bernal Fowler rules can be neglected. The correct estimate 
of the experimental residual entropy of ice at 0~K by using these two assumptions was 
a major achievement.  Therefore the free energy of ice is approximated by :
\begin{equation}
\label{paulingidea}
A = -k_B T \ln (  \Omega  A_{configuration}  ) = -k_B T \ln(\Omega) - k_B T \ln( A_{configuration})
\end{equation}
where  $A_{configuration}$ is the free energy (obtained via the Einstein crystal methodology
for a certain configuration satisfying the Bernal Fowler rules) 
and $\Omega$ is the degeneracy. Pauling estimates 
$-k_B T \ln(\Omega)$ as $-k_B T \ln ( (3/2)^N )$. Therefore for ices one computes the free energy for 
a certain configuration and then adds the Pauling contribution to the free energy.
This entropy can also be computed numerically 
(see for instance the work by Berg and Yang \cite{berg_ice_entropy}). 
Notice that when  MC or MD runs are performed for a certain configuration of ice 
satisfying the Bernal Fowler 
rules, the system remains in this configuration along the run. This is because the time required
by the system to jump from a configuration to another (both satisfying the Bernal Fowler rules)
is beyond the typical time of a simulation run. 
equation (\ref{paulingidea}) is useful not only for ice but for other disordered solids 
as well. In fact it can be applied 
successfully \cite{Wojciechowski87,Wojciechowski,vega92b,JML_2004_113_0037,2clj_phase_diagram}
to tangent dimers, formed by two tangent spheres, where 
Nagle \cite{nagle} has estimated $\Omega$,
and for fully flexible hard sphere chains \cite{malanoski_chains}(where Flory and 
Huggins \cite{JCP_1941_09_00660_nolotengo,JCP_1941_09_00440_nolotengo}
have estimated $\Omega$). 

{\em Partially disordered phases}.
In certain cases the system possess disorder but, still certain configurations are more 
likely than others. Getting the free energy in such a situation is especially difficult.
Firstly it is important to sample the configurational space properly to obtain equilibrium
configurations of the system representative of the partial disorder. Secondly these 
configurations will differ in statistical weight, so that it does not seem a good idea
to perform Einstein crystal calculations for just one configuration, since its statistical weight is 
unknown. In this case thermodynamic integration can be a more adequate route. 
An example of a partially disordered phase is the fcc disordered structure of the RPM model
(see discussion on this later on).


\section{ The machinery in action. III. Obtaining coexistence lines: the Gibbs Duhem integration.}
\label{g-dint}
Once  the free energy of the liquid and the solid has been obtained for a reference state it is 
relatively straightforward to perform thermodynamic integration to obtain it for other
thermodynamic states and locate a coexistence point between two phases (in case where it exists).
The Gibbs-Duhem integration allows the determination of the coexistence 
lines once an initial coexistence point is known.

\subsection{Gibbs Duhem integration}
\label{g-d}
In the year 1993 Kofke realized that the Clapeyron equation can be integrated to determine
coexistence lines  \cite{MP_1993_78_1331_nolotengo,JCP_1993_98_04149,ACP_1998_105_0405__coll_nolotengo}. 
The Clapeyron equation between two coexistence phases (labelled as I and II) can be 
written as:
\begin{equation}
\frac{dp}{dT}=\frac{s_{II}-s_I}{v_{II}-v_I}=\frac{h_{II}-h_I}{T(v_{II}-v_I)}
\end{equation}
where we use lower case for thermodynamic properties per particle.
Since the difference in enthalpy and volume between two phases can be determined easily
(at a certain T and p) the equation can be integrated numerically. 
When implementing the Gibbs Duhem integration one obtains the coexistence pressure for the 
selected temperatures (the temperature acting as the independent variable).
This is quite convenient when the coexistence line does not present a large slope 
in the $p-T$ plane. When the slope of the coexistence line is large  within a $p-T$ 
representation then it may be more convenient to integrate the Clapeyron 
equation in a different way: $\frac{dT}{dp}=\frac{T\Delta v}{\Delta h}$.
In this case the coexistence temperatures are determined for a set of selected 
pressures (the pressure acting as the independent variable).
A fourth order Runge-Kutta algorithm is quite useful to integrate the differential equation. 
It is important to stress that anisotropic NpT simulations should be used for the solid 
phase within Gibbs Duhem calculations. Isotropic NpT simulations could be used for 
fluid phases and for solids of cubic symmetry. 

\subsection{Hamiltonian Gibbs Duhem integration}
\label{g-d-f}
When a coupling parameter $\lambda$ is introduced within the expression of the potential energy
of the system, then a set of Generalised Clapeyron equations 
can be derived \cite{singer90,kofkeljmelting,laird}. 
For two phases at coexistence:
\begin{equation}
g_I(T,p,\lambda)=g_{II}(T,p,\lambda).
\end{equation}
If the system is perturbed slightly while preserving the coexistence it must hold that:
\begin{equation}
\label{gdnorm}
v_Idp-s_IdT+\left(\frac{\partial g_I}{\partial \lambda}\right)d\lambda=v_{II}dp-s_{II}dT+
\left(\frac{\partial g_{II}}{\partial \lambda}\right)d\lambda,
\end{equation}
the last terms appearing in equation (\ref{gdnorm}) are  due to the presence of the  new 
intensive thermodynamic variable $\lambda$. If $\lambda$ is constant when performing 
the perturbation then one recovers the traditional Clapeyron equation. 
If the pressure remains constant when the perturbation is performed (so that T and  $\lambda$
are changed) then one obtains :
\begin{equation}
\frac{dT}{d\lambda}=\frac{T[(\partial g_{II}/\partial \lambda)-(\partial g_{I}/\partial \lambda)]}{h_{II}-h_I}
\end{equation}
It is simple to show (within the NpT ensemble ) that $\partial g/\partial \lambda$ is 
nothing but $
\frac{\partial g}{\partial \lambda}=\left<\frac{\partial u(\lambda)}{\partial \lambda}\right>_{N,p,T,\lambda}$,
which can be determined within an NpT simulation. 
The final working expression of the Generalised Clapeyron equation (for perturbations of 
T and $\lambda$ while keeping p constant) is :
\begin{equation}
\label{gdfconpcon}
\frac{dT}{d\lambda}=\frac{T(<\partial u_{II}(\lambda)/\partial \lambda>_{N,p,T,\lambda}
-<\partial u_{I}(\lambda)/\partial \lambda>_{N,p,T,\lambda})}{h_{II}-h_I}
\end{equation}
This Generalised Clapeyron equation can be integrated numerically yielding the change in 
coexistence temperature (at a certain pressure) due to a perturbation of the Hamiltonian 
of the system (i.e., of the potential energy). 
A similar expression can be obtained for the case in which the system is perturbed at 
constant T (changing the pressure and $\lambda$ ). In this case one obtains :
\begin{equation}
\label{gdfcontcon}
\frac{dp}{d\lambda}=-\frac{<\partial u_{II}(\lambda)/\partial \lambda>_{N,p,T,\lambda}
-<\partial u_{I}(\lambda)/\partial \lambda>_{N,p,T,\lambda}}{v_{II}-v_I}
\end{equation}
The change in the coexistence pressure (at a certain temperature) due to 
a change in the Hamiltonian of the system (potential energy) is then obtained. 
Equations (\ref{gdfcontcon}) and  (\ref{gdfconpcon}) will be denoted as 
Hamiltonian Gibbs Duhem integration. 
Hamiltonian Gibbs Duhem integration is a very powerful technique since it allows one to 
analyse the influence of the parameters of the potential on the coexistence properties. 
It also allows one to change the parameters of the potential to improve phase 
diagram predictions. These two possible applications will be illustrated later on 
for the case of water.

In the particular case in which $\lambda$ is used as a coupling parameter taking the 
system from a certain potential to another (by changing $\lambda$ from zero to one):
\begin{equation}
U(\lambda)=\lambda U_{B} + (1-\lambda) U_{A}.
\end{equation}
Then the generalised Clapeyron equations can be written as:
\begin{equation}
\label{gdftemp}
\frac{dT}{d\lambda}=\frac{T(<u_{B}-u_{A}>_{N,p,T,\lambda}^{II}
-<u_{B}-u_{A}>_{N,p,T,\lambda}^{I})}{h_{II}-h_I}
\end{equation}
\begin{equation}
\label{gdfpres}
\frac{dp}{d\lambda}=-\frac{<u_{B}-u_{A}>_{N,p,T,\lambda}^{II}
-<u_{B}-u_{A}>_{N,p,T,\lambda}^{I}}{v_{II}-v_I}
\end{equation}
where $u_{B}$ is the internal energy per molecule when the interaction between particles
is described by $U_{B}$ (with a similar definition for $u_{A}$ ).
If a coexistence point is known for the system with potential A, then it is possible to 
determine the coexistence conditions for the system with potential B (it is just sufficient
to integrate the previous equations changing $\lambda$ from zero to one). 
In this way the task of determining the phase diagram of system B (unknown) from the 
phase diagram of system A (known) is simplified considerably.

\section{Coexistence by interfaces}

\subsection{Direct fluid-solid coexistence}
In 1978 Ladd and Woodcock
devised a method to obtain fluid-solid equilibria, without free energy calculations, 
the direct coexistence method \cite{woodcock1,woodcock2,woodcock3}.
In this method, the  fluid and the solid phases are introduced 
into the simulation box, and simulations are performed (NVE MD) to achieve 
equilibrium between the two coexistence phases. The coexistence conditions
can then be obtained easily. 
Although the initial results for LJ and inverse twelve power were not very
successful (probably due to the small size of the systems and to the short
length of the runs), the method is becoming more popular in the last few
years. In fact it has been applied to simple fluids \cite{broughton5,morris02,mori,kyrlidis,laird98,laird02},
metals \cite{belonoshko00,morris94,madden00,alfe03}, silicon  \cite{morris04}, ionic systems \cite{madden04,nacl_surface_tension},
hard dumbells \cite{songhd}, nitromethane \cite{thompson_nitromethane_defects_and_twophases}
and water \cite{karim88,karim90,bryk02,bryk04,wang05,kusalik_hydrate,ramon06,abascal06,kroes_hydrate}.  
Two simulation boxes, having an equilibrated solid and liquid respectively, are joined 
along the z axis (the direction perpendicular to the plane of the interface). That could generate overlapping at the
interface, and this overlapping should be relaxed/removed. 
The coexistence conditions (i.e., pressure, temperature) will be independent on 
the plane selected for the interface, but the dynamic behavior (and of course the interface 
properties) will be different for different planes \cite{tension_ice_water_zeng,tension_ice_water_luo,abascal06}.

The direct coexistence method can be implemented either
within Molecular Dynamics or Monte Carlo simulations. 
Both are equally valid, although if dynamical properties are of 
interest (for instance crystal growth rates) then MD is the only choice. 
The direct coexistence method was firstly  used  in the NVE ensemble 
, but other ensembles as 
$NVT$, $NpH$, $NpT$, $Np_zT$ can be applied.  
Each ensemble will have its advantages and
disadvantages, and the election
of one ensemble or another depends on the information that one
wants to obtain. Broadly speaking there are two kinds of ensembles, those 
at which it is possible to reach equilibrium having two coexistence phases
at equilibrium and those for which it is not possible to have two phases 
at equilibrium. Obviously for the study of interface properties only those
ensembles that lead to equilibrium should be used.

{\bf A. NVE ensemble}

This is the simplest approach. 
The idea behind the method is that  
the system will evolve to the equilibrium temperature and pressure
by moving the interface (so that either the amount of solid or the amount of liquid 
increases). 
If the system is above the melting
temperature, the solid will melt provoking (at constant E) a decrease of temperature. 
If the system is below the melting
temperature, the fluid will freeze provoking (at constant E) an increase of temperature. 
In $NVE$ runs the initial configuration should not be too far from equilibrium
to guarantee some portion of liquid and solid in the final configuration.
The equilibrium temperature and pressure are obtained in $NVE$ simulations at the end of the run. 
The knowledge of the coexistence pressure only at the end of the run is a serious problem.
In fact the lattice parameter used in the $xy$ plane (which remains fixed along the run)
may not correspond to the equilibrium value for the solid at the coexistence pressure. 
In other words, stress was introduced, and that may affect the free energy of the solid and,
therefore, the melting point. This can be adjusted by trial an error \cite{broughton5,laird98}.

{\bf B. NVT ensemble}

In the $NVT$ ensemble the system also evolves
to equilibrium by changing the relative amount of liquid and solid phases, 
in this case to adjust the densities and pressure to their coexistence values.
One important difference with the $NVE$ ensemble is that, in the $NVT$, the heat released or
absorbed by the crystallization or the melting is immediately accommodated
by the thermostat and, therefore, it is expected that the
system will attain the equilibrium faster than in the $NVE$ simulations.
Actually, heat transfer is usually the determining rate in
the crystallization or melting process, and
it has been seen that the presence of a thermostat in the
simulations leads to crystallization rates much higher 
than those measured experimentally \cite{nada04}.
However, as in the $NVE$ case, the solid is not able to
relax in the $xy$ plane and, therefore, the system might
be under some stress.

{\bf C. NpH ensemble}

 A less common approach is to perform the simulations in the
isobaric-isoenthalpic $NpH$ ensemble \cite{JCP_1980_72_02384},
with anisotropic scaling, i.e., the three edges of the simulation
box change independently 
(see, for example, Ref. \cite{wang05}). 
In this ensemble, the system will also attain the equilibrium, in this
case by evolving towards the coexistence temperature. One advantage with respect to the
previous ensembles is that now  
the fluctuations of the volume will allow the solid to relax,
removing the presence of stress. Moreover, as the volume of the box is allowed to change,
the system can adapt more easily to changes in the relative ratio 
of the amount of solid and liquid phases,
especially when the densities of the solid and liquid phases are
very different. 
The problem in this case is that
it is not strictly correct to use simulations under constant pressure
in the presence of an interface, because, due to the contribution of the 
interface, the normal and tangential components of the stress tensor 
are not equal. However, if the system 
is chosen to be very large in the direction perpendicular to
the interface, it is expected that the error introduced by
the presence of the interface will be small.
Another disadvantage of this ensemble is that, as in the $NVE$,
the transfer of heat is not very efficient and, therefore, long
simulations are needed to obtain the equilibrium.

{\bf D. NpT ensemble}

It is possible to tackle both the problem of having stress
in the solid and of slow heat transfer  by performing simulations
in the anisotropic $NpT$ ensemble where each of the edges of
the box changes independently. In this case, as the volume is able 
to fluctuate, the solid can relax to equilibrium and, as
the temperature is fixed, the transfer of heat will occur
very rapidly. However, as in the $NpH$ ensemble, the use of the
$NpT$ ensemble in the presence of an interface is not strictly correct,
although, as mentioned before, it is expected that the error is small
for a system sufficiently large along the $z$ axis.
One important difference of this ensemble with the previous ones
is that, as both the pressure and temperature
are set, it is not possible to have the interface at equilibrium.
The procedure to determine the coexistence properties is as follows. 
At a given pressure, different simulations
are performed at a few temperatures.
If the temperature is
above the melting temperature the solid will melt (i.e. the total energy
of the system will increase) and on the contrary, 
if the temperature is below the melting temperature
the fluid will freeze (i.e. the total energy will decrease). 
In this way, it is possible to establish
a lower and upper limit to the 
melting temperature. 

{\bf E. $Np_zT$ ensemble}

We have mentioned before that, due to the contribution of the
interface, the stress tensor has different normal and transversal 
components and, therefore, it is not correct to perform simulations
under constant pressure in the presence of an interface.
The correct way would be to allow the size of
the box to change only along the axis normal to the 
surface, i.e., the $z$ axis. We will call such ensemble $Np_zT$.
The procedure to determine the coexistence properties is completely
analogous to the procedure followed in the $NpT$ ensemble. The only
difference is that now a new starting configuration 
with the corresponding bulk lattice parameter  at pressure $p_z$ and temperature
$T$ must be generated for each 
simulation at different temperature and/or pressure in order to
avoid having stress in the solid. This was
not necessary in the $NpT$ ensemble, as the fluctuations along the
$x$ and $y$ planes allowed the solid to relax to equilibrium.

Two issues that deserve special attention when implementing
the direct coexistence method are the system size and the length of
the simulation. As with regards to the system size, a typical simulation 
box could have 10 molecular diameters in the $x$ and $y$ direction and about 
30 in the $z$ direction. Accordingly, studies by direct coexistence 
used typically 1000-3000 molecules, and these sizes provide results relatively close
to the thermodynamic limit \cite{morris02,morris94,thompson_nitromethane_defects_and_twophases,ramon06,abascal06}. 
Besides large system sizes, extremely long simulations are also needed (10 millions of 
time steps or more may be needed in many systems). Systems without a thermostat ($NVE$, $NpH$) 
may require even longer runs, since heat transfer along the interface may be quite slow. 

\subsection{Estimating melting points by studying the free surface}

It is now commonly accepted that melting starts at the surface and, already at 
temperatures lower than the bulk melting point, solids exhibit a liquid-like layer
at the surface \cite{abraham1981,Dash_89,Bienfait_92,broughton1983,Nenov_84,Fran_Veen_85,frenken_prb,thompson_surface_melting,mecanismo_melting_LJ}.
Thus for most substances a liquid layer is presented in the surface even at 
temperatures below the melting point. When the thickness of this layer diverges at 
the melting point this is denoted as surface melting. When the thickness of the liquid
layer remains finite (or even zero) at the melting point this is denoted as 
incomplete surface melting   \cite{Carnaveli_etal_87,pluis1987,Tartaglino_etal_05}. 
The thickness of the quasi-liquid layer  for a certain T depends on the considered 
material and on the exposed plane (as labelled by the Miller indexes). 
When the size of the liquid layer is sufficiently large either because the 
system has surface melting or incomplete surface melting (with an significant 
thickness of the liquid layer) then it is not possible to superheat a solid. 
This is the reason why experimentally solids usually melt at the melting
point (at least one plane of the crystals of the powder 
presents a large quasi-liquid layer provoking the melting of
the whole sample). For instance for ice it is not
possible to superheat the solid, except for a few nanoseconds \cite{iglev_1}.
Superheating of solids (over macroscopic times) has been found experimentally 
only for monocrystals when the exposed planes have no liquid layer at all.  
Since for most of substances a  quasi-liquid layer will be present at the melting point
it is expected 
that for most of materials, the melting will occur at the melting point (when having
a free surface). That provides a remarkably simple methodology to estimate 
melting points (at zero pressure).
NVT simulations of the solid exhibiting a free surface are performed 
at different temperatures.
A convenient geometry is to locate a
slab of solid in the center of an orthorhombic simulation box. 
The pressure will be essentially zero since no vapor
was introduced in the simulation box, and besides the vapor pressure is typically so small that
the sublimation of a molecule from the solid will be a rare event within the typical simulation 
times.  
The lattice parameter of the solid in the direction perpendicular to the interface 
should correspond to the equilibrium values at zero pressure for the studied temperature.
At temperatures below the melting point a stable thin liquid layer will be formed in the 
surface.  At temperatures above the melting 
point the solid will melt. The simulations should be rather long to allow the system to 
melt completely. In the case of water, the ice took about 10-20ns to melt in the presence
of a free surface. This technique has been applied successfully to estimate the melting point
of the LJ model, water \cite{maria06} and nitro-methane \cite{thompson_surface_melting,mas_nitromethane_surface_melting}. 
Notice that the technique will fail, for instance for NaCl, a substance 
having no liquid layer on the free surface \cite{nacl_surface_tension}. 

\section {Consistency checks}

Evaluation of free energy and coexistence points requires more effort than 
performing simple NpT runs. Besides, the possibility of introducing errors
in the calculations is relatively high. For this reason it is a good idea
to introduce several tests to guarantee the accuracy of the calculations. 

\subsection {Thermodynamic consistency}
In a number of cases it is possible to determine the free energy for two different 
thermodynamic states. For instance for a solid the free energy at two different 
densities along an isotherm can be determined by using Einstein crystal calculations.
The free energy difference between these two states (as obtained from free energy
calculations) should be identical to that obtained
by thermodynamic integration (integrating the EOS along the isotherm). We indeed recommend
the implementation of this test before performing any further calculation. 
Failing in the test indicates an error in the free energy calculations. 
However, passing the test it is not a definite proof of the correctness of the free energy
calculations. They could still be wrong by a constant (being the
constant identical for the two thermodynamic states). 
It is clear that other tests should also be introduced to guarantee 
the correctness of the calculations. 

\subsection{Consistency in the melting point obtained from different routes}
The melting point obtained from free energy calculations should be similar to that
obtained from direct coexistence simulations (where the fluid and the solid phases
coexist) within the simulation box. Notice that 
only fluid-solid equilibria can be studied by direct coexistence (it is not obvious
how to implement solid-solid equilibria by direct coexistence). 
This is indeed a useful test. 
An incorrect prediction of the free energy of the solid phase (the typical source of errors)
will provoke an incorrect prediction of the fluid-solid equilibria as compared to the estimate
obtained from direct coexistence techniques. 
Important differences (above 2-3\%) in the melting point estimated from free 
energy calculations and from direct coexistence are not acceptable. 
The reader may be surprised by the fact that 
we stated that both melting points should be similar instead of stating that they 
should be identical. 
In the thermodynamic limit (for very large system sizes) they should indeed become identical.
However for finite systems some technical aspects may provoke small differences. 
There are at least two reasons:

(a) {\em System size effects}. The fluid-solid equilibria may have a strong N dependence. 
For this reason the coexistence pressure/temperature 
obtained  from free energy calculations for a system of  N  molecules
will not be identical to that
obtained from direct coexistence simulations obtained with $N^{*}$ molecules. 
The case of HS was illustrated in a previous section. In general the stability of the solid
increases as the system becomes smaller. Since typically $N^{*} > N$, then the solid will appear
slightly more stable in the free energy calculations than in the direct coexistence results (assuming
that no FSC corrections were introduced).
This is what one may expect when the only difference between both type of calculations is 
the size of the system. 

(b) {\em Cut-off effects}. This is important when the cut-off used in free energy calculations is 
different from that used in direct coexistence simulations. The difference in the melting 
point may just be due to the fact that we are simulating two sightly different potentials. 
Even if the potentials were truncated at the same distance in both methodologies there may
be differences. For instance, in free energy calculations, the potential may be truncated 
at a distance $r_{c}$, but long range corrections can be incorporated in 
the calculations \cite{allen_book}.
In direct coexistence simulations, the potential may have been truncated at the same $r_{c}$ but 
in this technique it is difficult to incorporate long range corrections. 
Therefore certain differences in the melting point between free energy simulations and direct 
coexistence calculations can be due to a different implementation of the potential.

Notice that effects a) and b) may occur simultaneously. For instance in the case of water,
different number of particles were used in free energy calculations and in direct coexistence 
simulations (effect a). But also the truncation of the potential (LJ contribution) was 
different in the free energy simulations (where long range corrections were included) and in the direct coexistence simulations (where they were not included). In any case for water models 
we found differences in the melting point as estimated from both techniques of about 1-2\% \cite{vegafilosofico,ramon06}.

\subsection{Some useful tests involving Gibbs Duhem integration}
Some useful tests that can be performed when using the Gibbs-Duhem technique are:
\begin{itemize}
\item{ The coexistence lines should be identical (within statistical uncertainty) 
when the integration is performed forward (say by increasing the T) and when it is
performed backward (say by decreasing the T).}
\item{ If free energy calculations were used to locate an initial coexistence point 
between phases  I-II, I-III and II-III, then the three coexistence lines obtained with 
Gibbs Duhem integration should cross at a point (the triple point).}
\item{ If the melting point of two models has been determined by free energy calculations then 
it is also possible to use Hamiltonian Gibbs Duhem calculations to check that the melting point 
of model B is obtained starting from the melting point of model A and viceversa.}
\item { If the melting point of models A and B is known, then Hamiltonian Gibbs Duhem integration 
could be performed to estimate the melting point of model C. Both integrations (one starting from 
A and the other starting from B) should provide the same estimate of the
melting point of C.}
\item{If two coexistence points between phases I and II are known (either by free energy calculations or by direct coexistence) a Gibbs-Duhem integration 
starting from one of them should pass through the other one. }
\end{itemize}

\subsection{Consistency checks at 0 K}

At zero temperature the condition of chemical equilibrium (i.e., the
equilibrium pressure $p_{eq}$) between two phases,
labelled as phase I and II, respectively, is given by:
\begin{equation}
\fl \qquad
U_I(p_{eq},T=0)+p_{eq}V_{I}(p_{eq},T=0)=U_{II}(p_{eq},T=0)+p_{eq}V_{II}(p_{eq},T=0)
\label{exact_p}
\end{equation}
Hence, phase transitions between solid phases at zero temperature occur
with zero enthalpy change.
This is really useful
since it means that phase transitions at 0~K can be estimated without free energy 
calculations (just computing mechanical properties as densities and internal energies). 
By performing several NpT simulations where the temperature is reduced up 
to zero it is possible to obtain the EOS (and internal energy) of 
each solid phase at 0~K. Then by equating the enthalpy, it is possible to locate 
phase transitions (at 0~K).  This can be used as a consistency check. By performing 
free energy calculations it is possible to locate the coexistence pressure between 
two phases (I and II) at a finite non zero temperature. Then, by performing 
Gibbs Duhem simulations it is possible to determine the coexistence line up 
to 0~K. The coexistence pressure at 0~K obtained from this long route (free energy
calculations+Gibbs Duhem integration) should be identical to that obtained from the short route (estimating
the properties of the system at 0~K). This is again a severe consistency check. 
Although runs at 0~K enable a check to be made on the consistency of phase 
diagram calculations, they do not allow by themselves to draw the phase 
diagram of a certain model. Gibbs Duhem simulations can not be initiated from 
a known coexistence point at 0~K since both $\Delta H$ and T are null so that its ratio
$\Delta S$, which within classical statistical thermodynamics is finite even at 0~K,
can not be determined. 

\section{ A worked example. The phase diagram of water for the TIP4P and SPC/E models. }

We shall now illustrate how the previously described methodology can be 
applied to determine the phase diagram for two popular water models: SPC/E \cite{berendsen87} and
TIP4P \cite{jorgensen83}. We believe that they illustrate quite well 
the typical difficulties found when determining by computer simulation free energies of 
solid phases and phase diagram calculations. 
The SPC/E and TIP4P models are presented in Table \ref{tabl_params} (along with two
other recently proposed models TIP4P/Ice \cite{abascal05a} and TIP4P/2005 \cite{abascal05b}). 
In these models a LJ center is located on the O atom, and positive charges are located 
on the H atoms. The negative charge is located at a distance $d_{OM}$ from the O along the
H-O-H bisector. Let us now describe briefly some of the simulation details. 

\begin{table}[h]
\centering
\caption{\small\sf Potential parameters for several water potentials. Notice that the $OH$ bond length and the 
$HOH$ angle are different for the SPC/E and TIP4P models.\vspace{0.2cm}}
\label{tabl_params}
\begin{tabular}{lcccc}
\hline \hline
Model  & $\epsilon/k$(K)&$\sigma$(\AA)&$q_H$(e)&$d_{OM}$(\AA) \\
\hline
SPC/E \cite{berendsen87}        &  78.20    &   3.1656  &  0.4238 &  0     \\
TIP4P \cite{jorgensen83}       &   78.0    &   3.154   &  0.520  & 0.150  \\
TIP4P/2005 \cite{abascal05b}  &   93.2    &   3.1589  &  0.5564 & 0.1546 \\
TIP4P/Ice \cite{abascal05a}   &  106.1    &   3.1668  &  0.5897 & 0.1577 \\
\hline \hline
\end{tabular}
\end{table}

\subsection{Simulation details} 
In our Monte Carlo simulations, the LJ potential was truncated for all phases at $8.5$ \AA. 
Standard long range corrections to the LJ energy were added.
The importance of an adequate treatment of the long range coulombic forces when 
dealing with water simulations has been pointed out in recent 
studies \cite{JCP_1998_108_10220,lisal02,rick04JCP_2004_120_06085,yonetani} and it is likely 
that this is even more crucial when considering solid phases. 
In this work, the Ewald summation technique has been employed for the 
calculation of the long range electrostatic forces.
The real space contribution of the Ewald sum was also truncated at $8.5$ \AA.
The screening parameter and the number of vectors of reciprocal
space considered had to be carefully selected for each 
crystal phase \cite{allen_book,frenkelbook}.
For the fluid phase we used 360 molecules. 
The number of molecules for each solid phase was chosen so as to fit at least twice 
the cutoff distance in each direction. 
Three different types of runs were performed: NpT, NVT Einstein crystal calculations 
and Gibbs Duhem integration. 
The equation of state (EOS) of the fluid was obtained from isotropic NpT runs, whereas 
anisotropic Monte Carlo simulations (Parrinello-Rahman like)
\cite{parrinello80,parrinello81,yashonath85} were used for the solid phases. 
In the NpT runs, about 40000 cycles were used to obtain 
averages, after an equilibration period of about 40000 cycles. However longer
runs were used for the fluid phase at low temperatures. 
A cycle is defined as a trial move per particle plus 
a trial volume change.
To evaluate the free energy of the solid, Einstein crystal (NVT) calculations were
performed (with fixed center of mass). Also the length of the runs was of 
about 40000 cycles to obtain averages after an equilibration of 40000 cycles. 
In the Gibbs Duhem simulations a fourth order Runge-Kutta was used to integrate the 
Clapeyron equation. In total about 60000 cycles were used to pass from a coexistence point 
to the next one. When using Hamiltonian Gibbs Duhem integration 5-10 values of $\lambda$
where used to connect the initial to the final Hamiltonian. In the Gibbs Duhem simulations,
the fluid, and cubic solid were studied with isotropic NpT runs whereas the solid phases
were studied with anisotropic NpT runs.

\subsection{Free energy of liquid water}

The free energy of the liquid is calculated by integrating the free energy 
along a reversible path in which
the water molecules are transformed into Lennard-Jones spheres by switching the charges off. 
The free energy of the reference Lennard-Jones fluid is reported in the work of 
Johnson {\em et al.} \cite{MP_1993_78_0591_photocopy}.
The energy (say, for the TIP4P model of water, being the treatment for SPC/E fully 
equivalent) of the system for a given point of the path, $\lambda$, is given by:
\begin{equation}
U(\lambda)=\lambda U_{TIP4P} + (1-\lambda) U_{LJ} 
\end{equation}
where $\lambda$ varies between 0 (LJ) and 1 (water) along the integration path. 
Given that $\partial A(\lambda)/\partial \lambda = <\partial U (\lambda)/\partial \lambda>_{NVT}$,
the free energy difference between liquid water and the Lennard-Jones fluid is given by:
\begin{equation}
\label{enlibliq_LJ}
A_{TIP4P}(N,V,T)-A_{LJ}(N,V,T)=\int^{\lambda=1}_{\lambda=0}\left<U_{TIP4P}-U_{LJ}\right>_{N,V,T,\lambda}d \lambda.
\end{equation}
where $<U_{TIP4P}-U_{LJ}>_{N,V,T,\lambda}$ is an $NVT$ simulation average computed for a given value of $\lambda$.
The integral is solved numerically (using Gauss-Legendre quadrature) 
by calculating the integrand at different 
values of $\lambda$.  In practice
we perform the MC runs starting from $\lambda=1$ and going to $\lambda=0$.
The final configuration of a run was used as the input configuration of the next run.
The LJ fluid chosen as a reference state 
has the same LJ parameters ($\epsilon/k_B$ and $\sigma$) as
the water model. Therefore, the difference $U_{TIP4P}-U_{LJ}$ is just the electrostatic energy. 
Once the Helmholtz free energy, $A$, is known, the chemical potential is obtained simply as:
$\frac{\mu}{k_BT} = \frac{A}{Nk_BT} + \frac{pV}{Nk_BT}$.
In this way we have computed the free energy of the liquid at 225~K
and 443 K (see table \ref{tablaAliqaq}). 
In addition to the total free energy we report 
the free energy difference with respect to the reference LJ fluid ($\Delta A$), the residual free energy 
for the LJ fluid and the ideal free energy ($Nk_BT(\ln (\rho \Lambda^3) - 1)$).
In this work, for water, we shall assign the thermal de Broglie wave length to $\Lambda=1$
\AA  $\;$ both for the liquid and for the solid phases. 

\begin{table}[!hbt]\centering
\caption{\small\sf 
Free energy of liquid water ($A_{liquid}$) for the SPC/E and TIP4P  models 
(with  $q_r=q_v=q_e=1$ ). 
The residual and ideal contributions to the free energy of the reference LJ 
fluid are given. 
The residual term of the LJ fluid as obtained from the EOS 
of Johnson {\em et al.} \cite{MP_1993_78_0591_photocopy}.  
The ideal term was obtained (in $Nk_BT$ units) as $\ln(\rho \Lambda^{3}) - 1 $ 
where $\Lambda=1$ \AA. 
The difference in free energy between the reference fluid and the water model 
$\Delta A$ is given. 
Simulations were performed in the NVT ensemble for the density $d$.\vspace{0.2cm}}
\label{tablaAliqaq}
\footnotesize
\begin{tabular}{cccccccc}
\hline
Model & $ T(K)$& $p(bar)$&$d(g/cm^{3})$ & $A_{liquid}/(Nk_BT)$& $A^{res}_{LJ}/(Nk_BT)$ & $A^{id}_{LJ}/(Nk_BT)$ & $\Delta A/(Nk_BT)$\\ 
\hline
SPC/E & 225 & 564 & 1.05 & -21.82 &  2.500 & -4.350 & -19.97 \\
TIP4P & 225 & 743 & 1.05 & -19.48 &  2.401 & -4.350 & -17.52 \\
SPC/E & 443 & 4010& 1.05 & -9.53 & 2.856 & -4.350 & -8.04 \\
TIP4P & 443 & 4280& 1.05 & -8.58 & 2.777 & -4.350 & -7.01 \\ 
\hline
\end{tabular}
\end{table}

Let us now present some consistency checks for the free energies.
We shall only discuss it  for the TIP4P model. 
We have computed the free energy at 225~K and 443~K
for the density  d$=1.05$ g/cm$^3$.
The free energy difference between both states is $A_{443~K}/(Nk_BT) -
A_{225~K}/(Nk_BT)= -8.58+19.48=10.90$. 
Then  we calculated the same difference by means of thermodynamic integration 
along an isochore (eq. \ref{inttermisoc}), obtaining again 10.90 $Nk_BT$.
Besides Jorgensen  {\em et al.} have estimated the free energy for the TIP4P 
model at p=1bar and T=298~K to be G = -6.1 kcal/mol \cite{CP_1989_129_193}.
Starting from the free energy at 225~K and d$=1.05$ g/cm$^3$ and performing 
thermodynamic integration we obtained  -6.09 kcal/mol, which is in excellent
agreement with the value of Jorgensen  {\em et al.} \cite{CP_1989_129_193}.

Instead of using the LJ fluid, it is also possible to compute the free energy 
of the liquid taking the ideal gas as a reference system. We obtain 
the free energy of TIP4P water at 240 K and 1.0174 g/cm$^3$ by two different routes.
In the first one the TIP4P is transformed into a LJ model. We obtained for the 
free energy of TIP4P in this state $A$(240~K,1.0174~g/cm$^3$)$/Nk_BT=$ -20.15. 
In the second route, a supercritical isotherm (T=900 K) is used from zero 
density to 1.0174 g/cm$^3$ (obtaining using equation (12) -4.699 for $A/Nk_BT$ 
of this intermediate state).
Then we integrate along the isochore up to 240~K (with a free energy change 
computed by equation (7) of -15.434 $NK_BT$).
By adding these two numbers together we obtain 
$A$(240~K,1.0174~g/cm$^3$)$/Nk_BT_2=$-20.13 from this second route, 
in very good agreement with that obtained by the first one.
The computation of the free energy using the LJ fluid as a reference system 
is considerably shorter than using the ideal gas (besides this last route is especially difficult 
since the parameters of the Ewald sum should be chosen carefully along the supercritical isotherm
integration). Values of the free energy of liquid water for other potential models have
been reported recently \cite{free_energy_water_henchman}.

\subsection{Free energy of ice polymorphs}

The free energy of the different ice polymorphs is calculated 
using the Einstein crystal method (with fixed center of mass).
The O is used as reference point and the field used is that described by 
equations (\ref{ete_total}), (\ref{ete_t}) and (\ref{ete_r}).  
For disordered ices \cite{petrenko99,eisenberg69,lobban98} (Ih, Ic, III, IV, V, 
VI, VII, XII) the oxygens form a well defined lattice, but the water molecules 
can orient in different ways for a given oxygen lattice provided that
the Bernal-Fowler ice rules \cite{JCP_1933_01_00515_nolotengo} are satisfied. 
We have generated the disordered solid structures (with almost zero dipole 
moment) using the algorithm proposed by Buch {\em et al.} \cite{buch98} (see Ref.
\cite{hayward97} for other algorithm). 
For proton disordered ices 
we calculated the free energy for a particular proton disordered configuration.
Due to the fact that there are many configurations 
compatible with a given oxygen lattice, there is a degenerational 
contribution to the free energy. The degenerational entropy of ice was estimated 
by Pauling in 1935 \cite{pauling35} as 
$S_{deg}=k_B \ln \Omega=k_BN \ln (3/2)$. Therefore, the disordered ice phases have an extra contribution to the free energy of $-Nk_BT \ln(3/2)$. Ices III and V present partial proton ordering \cite{lobban00}, and that decreases
a little bit the Pauling estimate\cite{macdowell04,berg07}. For ices III and V we shall use the 
sligthly lower value of the degenerational entropy estimated by MacDowell et al.\cite{macdowell04}. 
For  ices II,IX,VIII and the antiferroelectric analogous \cite{morokuma84}
of ice XI the protons are ordered, and there is no degenerational entropy contribution. Generating 
an initial configuration for proton ordered ices is relatively straightforward. 

The free energy calculations were performed in the NVT ensemble using the equilibrium shape of the 
simulation box obtained in anisotropic  NpT  runs. The location of the springs of the 
ideal Einstein crystal
field were chosen to be close to the  equilibrium positions/orientations of the molecules. 
The computed free energy should not depend on the particular choice of the positions and orientations
of the ideal Einstein crystal field (provided that they are sufficiently close to the equilibrium
position and orientations of the molecules in the absence of the external field).
Several strategies are possible. For instance one could choose the position/orientations of 
the external field as those obtained from an energy minimisation at constant density (using 
the equilibrium box shape of the system). Another possibility is to use the experimental
crystallographic positions (if available) of the atoms of the molecule and modify them slightly 
to satisfy the bond lengths and angles of the model ( the bond lengths may be different in the model 
and in the real molecule, as for instance in  SPC/E water). Also, experimental crystallographic positions (if
available)  could be used for the reference point (oxygen in the case of water) 
, and the orientations could be optimised from 
an energy minimization. We have used this last approach for ices. This approach has also been used 
recently by Wierzchowski and Monson for gas hydrates \cite{monson_jpcb07}. 
To obtain $\Delta A_2$ we used Gaussian integration with 12 points. 
$\Delta A_1$ was evaluated from runs (200000 cycles) of the ideal Einstein crystal with fixed center
of mass.  The value of $\Lambda_E$ (in $k_B T/(\AA)^2$) was identical
to the value selected for $\Lambda_{E,a}$ and $\Lambda_{E,b}$ (in $k_B T$ units).
The value of $\Lambda_E$ was chosen so that the computed value of  $\Delta A_1$ differs by 
about $0.02N_BT$ units from the lattice energy of the solid (defined as the intermolecular energy 
of the system when the molecules occupy the positions and orientations of the external Einstein field).

Coexistence between phases at 150, 225 and 443 K were investigated.
Therefore, the free energy calculations have been performed
at those temperatures. In tables \ref{tablaAspce} and  \ref{tablaAtip4p} we report the 
free energy ($A_{sol}/(Nk_BT)$) calculated for different ice phases for  SPC/E and TIP4P respectively. 
The different contributions to the free energy ($A_0$, $\Delta A_1$, $\Delta A_2$) are given. 
The term $A_0$ is the sum of $A_{Eins-id}^{CM}$ plus $\Delta A_3$ plus a finite size 
correction (Frenkel-Ladd type, $(2/N) \ln N$). 
For proton disordered ices the value of $A_{sol}$ is the sum of $A_0$, 
$\Delta A_1$, $\Delta A_2$ and the Pauling degeneracy entropy $-Nk_BT\ln(3/2)$.
For proton ordered ices (XI,II,IX,VIII) the total value of A is just the sum of
$A_0$, $\Delta A_1$, $\Delta A_2$. 
The lattice energy $U_{lattice}/Nk_BT$ is the energy of the solid when all water
molecules remain fixed on the position/orientations of the Einstein crystal 
field (being the value of $\Delta A_1/(Nk_BT)$  quite close to  $U_{lattice}/Nk_BT$).
The free energies of the SPC/E model are lower than those of 
the TIP4P (this is consistent with their
lower internal energies). 

\begin{table}[!hbt]\centering
\caption{\small\sf 
Free energy of the ice polymorphs for the SPC/E model (with $q_r=q_v=q_e=1$).
The free energy reported in the last column corresponds to the sum of all the 
terms ($A_{0}+\Delta A_1+\Delta A_2$) plus the degenerational free energy 
($-Nk_BT \ln(3/2)$) for the case
of orientationally disordered phases (the typical uncertainty of the resulting 
solid free energies is about $0.05Nk_BT$). 
For ices III and V we did not use 
Pauling estimate for the degenerational free energy ($-Nk_BT \ln(3/2)$) since these
ices present partial proton disorder, but the slightly lower value reported by 
MacDowell et al.\cite{macdowell04}. 
The  number of molecules used for each solid phase is indicated in parenthesis
just after the Roman numeral of the phase. A finite size correction (Frenkel Ladd type) has been included 
in $A_0$. The thermal de Broglie wave length  $\Lambda$ was set to $\Lambda=1$ \AA. The residual
internal energy of the ice $U$ is reported, so that the entropy of the solid can be obtained
easily from the relation $S=(U-A_{sol})/T$.
The orientational contribution to $A_0$ (equation (29)) was computed from the 
approximate expresion given in Reference [48].\vspace{0.2cm}}
\label{tablaAspce}
\begin{tabular}{cccccccccc}
\hline 
Ice & 
$p(bar)$  & 
$T(K)$ & 
$d(g/cm^{3}$)  & 
$ \frac{U}{Nk_BT} $  & 
$\frac{\Lambda_E}{k_BT}$({\AA}$^{-2}$) & 
$\frac{A_{0}}{Nk_BT}$ & 
$\frac{\Delta A_{2}}{Nk_BT}$ & 
$\frac{\Delta A_{1}}{Nk_BT}$ & 
$\frac{A_{sol}}{Nk_BT}$\\
\hline
Ih(288) & 500  & 150 & 0.965 &-46.08  &25000 & 29.46 & -16.92 & -48.94 & -36.84 \\
Ic(216) & 2620 & 150 & 0.983 &-46.14   &25000 & 29.45 & -16.85 & -48.95 & -36.80 \\
II(432) & 5000 & 150 & 1.269 & -46.42 &25000 & 29.47 & -17.27 & -49.08 & -36.91 \\
III(324) & 5000 & 150 & 1.229 &-44.89 &25000 & 29.46 & -19.10 & -45.66 & -35.73 \\
IV(432) & 5000 & 150 & 1.353 & -43.97 &25000 & 29.47 & -18.23 & -45.54 & -34.73 \\
V(504) & 5000 & 150 & 1.316 & -44.39  &25000   &29.47 & -20.11 & -44.17 & -35.23 \\
VI(360) & 25000 & 150 & 1.492 &-43.29 &25000 & 29.46 & -19.78 & -42.81 & -33.56 \\  
VI(360) & 25000 & 225 & 1.474 & -27.88&25000 & 29.46 & -19.94 & -28.69 & -19.60 \\  
VII(432) & 81690 & 443 & 1.700 &-9.79 &9000  & 26.40 & -16.58 & -13.81 & -4.41  \\  
VIII(600) & 60000 & 225 & 1.743 &-23.92&25000 & 29.47 & -19.38 & -24.68 & -14.61 \\  
IX(324)   & 5000  & 150 & 1.244&-46.31&25000 & 29.46 & -19.10 & -46.93 & -36.60 \\  
XI(360)   & 500  & 150 & 0.971&-46.26&25000 & 29.46 & -18.01 & -48.08 & -36.65 \\  
\hline
\end{tabular}
\end{table}

\begin{table}[!hbt]\centering
\caption{\small\sf Same as table \ref{tablaAspce} for the TIP4P model.\vspace{0.3cm}}         
\label{tablaAtip4p}
\begin{tabular}{cccccccccc}
\hline 
Ice & 
$p(bar)$  & 
$T(K)$ & 
$d(g/cm^{3})$  & 
$ \frac{U}{Nk_BT} $  & 
$\frac{\Lambda_E}{k_BT}$({\AA}$^{-2}$) & 
$\frac{A_{0}}{Nk_BT}$ & 
$\frac{\Delta A_{2}}{Nk_BT}$ & 
$\frac{\Delta A_{1}}{Nk_BT}$ & 
$\frac{A_{sol}}{Nk_BT}$\\
\hline
Ih(288) & 1365 & 150 & 0.966 &-42.50  &25000 & 29.46 & -16.92 & -45.55 & -33.45 \\
Ic(216) & 500 & 150 & 0.958 &-42.50   &25000 & 29.45 & -16.96 & -45.52 & -33.47 \\
II(432) & 8230 & 150 & 1.269 & -41.80 &25000 & 29.46 & -16.98 & -44.75 & -32.28 \\
III(324) & 5000 & 150 & 1.237 &-41.43 &25000 & 29.46 & -18.82 & -42.63 & -32.39 \\
IV(432) & 5000 & 150 & 1.353 & -40.86 &25000 & 29.47 & -18.18 & -42.60 & -31.74 \\
V(504) & 5000 & 150 & 1.315 & -41.13  &25000   &29.47 & -19.62 & -41.56 & -32.11 \\
VI(360) & 25000 & 150 & 1.498 &-40.26 &25000 & 29.46 & -19.20 & -40.48 & -30.65 \\  
VI(360) & 25000 & 225 & 1.480 & -25.91&25000 & 29.46 & -19.58 & -27.12 & -17.67 \\  
VII(432) & 78350 & 443 & 1.700 &-8.66 &9000  & 26.40 & -16.76 & -12.85 & -3.63  \\  
VIII(600) & 60000 & 225 & 1.743 &-21.97&25000 & 29.47 & -19.11 & -23.24 & -12.90 \\  
IX(324)   & 5000  & 150 & 1.238&-42.25&25000 & 29.46 & -18.85 & -42.56 & -31.98 \\  
XI(360)   & 500  & 150 & 0.959&-42.53&25000 & 29.46 & -17.03 & -45.50 & -33.09 \\  
XII(540)  & 5000 & 150 & 1.358 &-40.94&25000 & 29.47 & -18.80 & -42.20 & -31.94 \\  
\hline 
\end{tabular}
\end{table}

As a consistency check we determined via Einstein crystal calculations the 
free energy of ice VI at two different thermodynamic conditions for the 
SPC/E and TIP4P. Let us discuss the results for TIP4P.
For TIP4P (ice VI) we obtained from free energy calculations 
$A_1$(225~K,1.480~g/cm$^{3}$)=-17.67~$Nk_BT$ and 
$A_2$(150~K,1.498~g/cm$^{3}$)=-30.65~$Nk_BT$ (both 
states having a pressure of 25000bar).
Starting from the value of $A_1$ and performing thermodynamic integration we 
estimated $A_2$(150~K,1.498~g/cm$^{3}$)=-30.66~$Nk_BT$ in excellent agreement
with the value obtained from Einstein crystal calculations. 

\subsection{Determining the initial coexistence points}
Once the free energy of each phase is known, it is possible to find the points in the pressure-temperature plane 
at which two phases have the same chemical potential; {\it i. e.}, the coexistence points. 
Given that most of the free energies and equations of state were obtained either at temperatures 150 or 
225 K and at pressures 500 or 5000 bar, we  focus the search of the coexistence points at these temperatures
and pressures.

\begin{figure}[!hbt]\centering
\includegraphics[clip,height=7cm,width=0.6\textwidth,angle=-0]{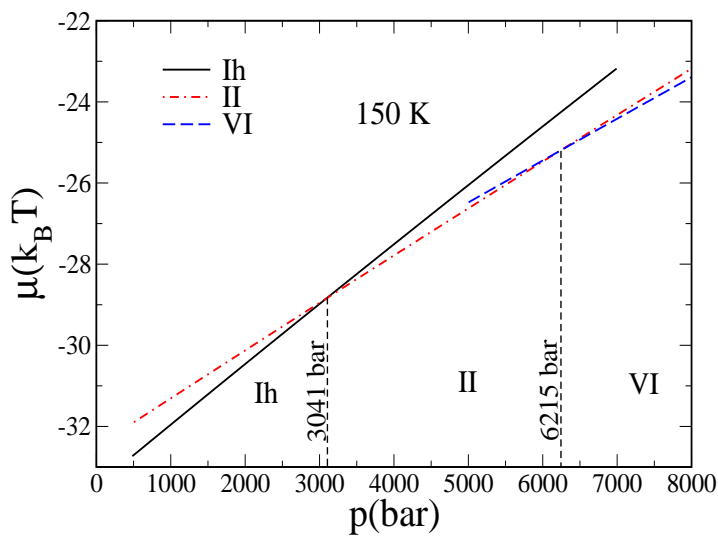}
\caption{\small\sf Chemical potential as a function of pressure at 150~K for ices  Ih, II and VI modeled with TIP4P.}
\label{coex150k}
\end{figure}

Figure \ref{coex150k} shows the chemical potential as a function of the pressure at 150 K
for ices Ih, II, and VI. For a given pressure, the phase of lowest chemical potential is the most stable one.
Ice Ih is the stable phase up to 3041 bar. At that pressure, ices Ih and II coexist. Beyond 3041 bar, 
ice II is the stable phase up to 6215 bar, where the chemical potentials of ices II and VI are equal.     
For higher pressures ice VI is the stable phase. 
\begin{table}[!hbt]\centering
\caption{\small\sf Coexistence  points for TIP4P and SPC/E models from free energy
calculations (and thermodynamic integration).\vspace{0.2cm}} 
\label{puntoscoex}
\begin{tabular}{cccc}
\hline \hline
Model & Phases & T (K) & p (bar) \\
\hline
SPC/E & Ih-II & 150 & -444 \\ 
SPC/E & II-VI & 150 & 25270 \\ 
SPC/E & liquid-II & 250 & 5000 \\ 
SPC/E & liquid-VI & 225 & 20690 \\ 
SPC/E & VI-VIII & 225 & 57860 \\ 
SPC/E & liquid-VIII & 225 & 57500 \\ 
SPC/E & liquid-VII & 443 & 103520 \\ 
SPC/E & liquid-Ih  & 211 &  500  \\
SPC/E & liquid-Ic  & 210 &  500  \\
SPC/E & liquid-XI  & 187 &  500  \\
\hline
TIP4P & liquid-Ih & 228.8 & 500 \\ 
TIP4P & Ih-II & 150 & 3041 \\ 
TIP4P & II-VI & 150 & 6215 \\ 
TIP4P & II-V & 152.6 & 5000 \\ 
TIP4P & II-III & 180.3 & 5000 \\ 
TIP4P & liquid-III & 196.6 & 5000 \\ 
TIP4P & liquid-V & 204.1 & 5000 \\ 
TIP4P & liquid-VI & 225 & 8940 \\ 
TIP4P & VI-VIII & 225 & 57290 \\ 
TIP4P & liquid-VII & 443 & 91940 \\ 
TIP4P & liquid-Ic  & 228.8  & 500     \\
TIP4P & liquid-XI  & 192 &  500 \\
TIP4P & liquid-XII  & 205.0 &  5000 \\
\hline \hline
\end{tabular}
\end{table}
By performing similar plots, coexistence points between different phases could be determined.
In Table \ref{puntoscoex} these coexistence points are presented (for TIP4P and SPC/E). 
The relative stability between ices Ih and Ic  (or between ices V and XII) could not be determined 
since the free energy difference between these solids was smaller
than the typical uncertainty of our free energy calculations ($0.05Nk_BT$).
As  to the stability of ices Ih(Ic) with respect to ice XI (we used the antiferroelectric
version of ice XI of Davidson and Morokuma \cite{morokuma84} rather than the true ferroelectric version), we found 
that the XI-Ih transition occurs at 84 K for SPC/E and 18 K for TIP4P (being the
proton ordered ice XI the stable phase at low temperatures). 

\begin{figure}[!hbt]
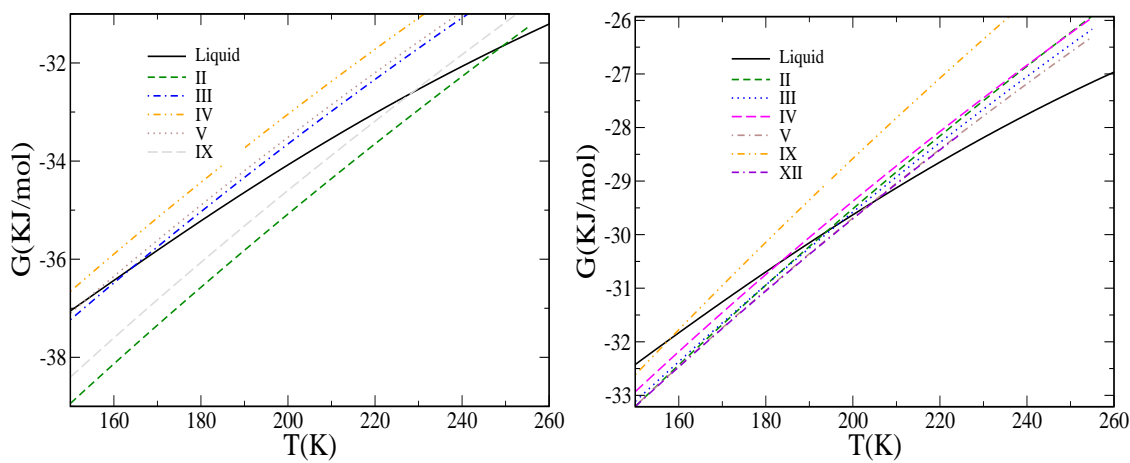
\centering
\includegraphics[clip,height=6cm,width=0.47\textwidth,angle=-0]{5000barGvsTspce.eps}
\includegraphics[clip,height=6cm,width=0.47\textwidth,angle=-0]{GvsT5000barTip4p.eps}
\caption{\small\sf Gibbs free energy versus temperature at 5000 bar for different phases of SPC/E (left) and TIP4P (right).}
\label{5000barGvsT} \end{figure}

The region of the phase diagram corresponding to 5000 bar (figure \ref{5000barGvsT}) is the 
most problematic given that there are as many as seven phases
competing; namely, ices II, III, IV, V, IX and XII and the liquid.
For  SPC/E, ice II is clearly the most stable phase among the solid polymorphs. The liquid is again the stable phase
at high temperatures (beyond 250 K). 
For the case of TIP4P, ices V and XII are the solid phases of lower chemical potential. The free energy difference 
between both polymorphs (0.008 $Nk_BT$) is smaller that the error bar (0.05 $Nk_BT$), so we could not determine which one is the most stable. 
Liquid water coexists either with ice XII or with ice V at 205 K at 5000 bar.

\subsection{The phase diagram of water}
\begin{figure}
\includegraphics[clip,height=500pt,width=200pt,angle=-90]{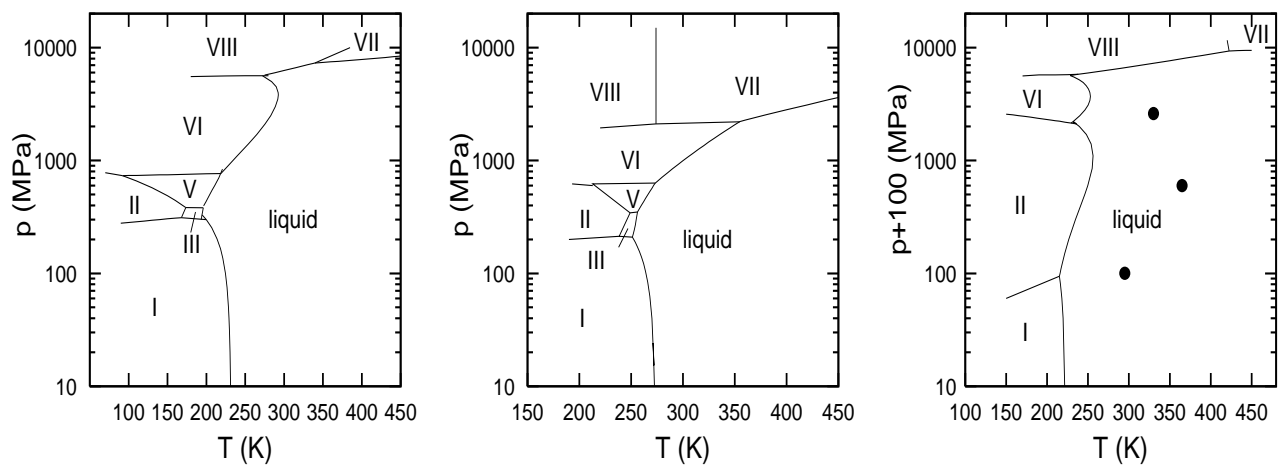}
\caption{ \label{prl_figure} Phase diagram of water as obtained from experiment (center) and from computer
simulations for the TIP4P model (left) and for the SPC/E model (right). The filled circles on
the right panel indicate the stability limit of the solid phases in NpT simulations (without
interfaces). Notice the shift of 100 MPa in the right panel.}
\end{figure}
Once an initial coexistence point has been determined,  by using Gibbs Duhem integration 
(either $dp/dT$ or $dT/dp$  ) it is then possible to draw the complete phase diagram. In certain 
cases the point where two coexistence lines met (triple point) was used as origin of the 
third coexistence line emerging from the triple point. 
The complete phase diagram of SPC/E and TIP4P is presented in figure \ref{prl_figure}.
As can be seen SPC/E fails in reproducing the phase diagram of water (notice for 
instance that ice Ih 
is stable for this model only at negative pressures), whereas TIP4P
provides a qualitatively correct description of the phase diagram (except for the high pressure
region of the phase diagram ). 
The reason of the failure of SPC/E and success of TIP4P has been identified recently.
The low quadrupole moment of SPC/E and the high value of the ratio dipole/quadrupole 
of this model is the cause of the failure \cite{abascal07a,abascal07c}. 
In fact TIP4P provides a quadrupole moment 
and a ratio dipole/quadrupole in much better agreement with experiment.
The effect of a quadrupole moment on the vapor-liquid equilibria of molecular models is well 
known \cite{JCP_1994_101_04166,boublik_kihara,fischer_cuadrupolo}. However it seems that the role of the quadrupole on 
water properties has been overlooked 
in spite of some  warnings about its 
importance \cite{patey76,carnie82,finney85,rick04JCP_2004_120_06085,blumwater,ichiye06}.  
Figure \ref{prl_figure} illustrates how the evaluation of the phase diagram of water by
computer simulation is indeed possible.

\subsection {Hamiltonian Gibbs Duhem simulations for water}

\begin{figure}[!hbt]
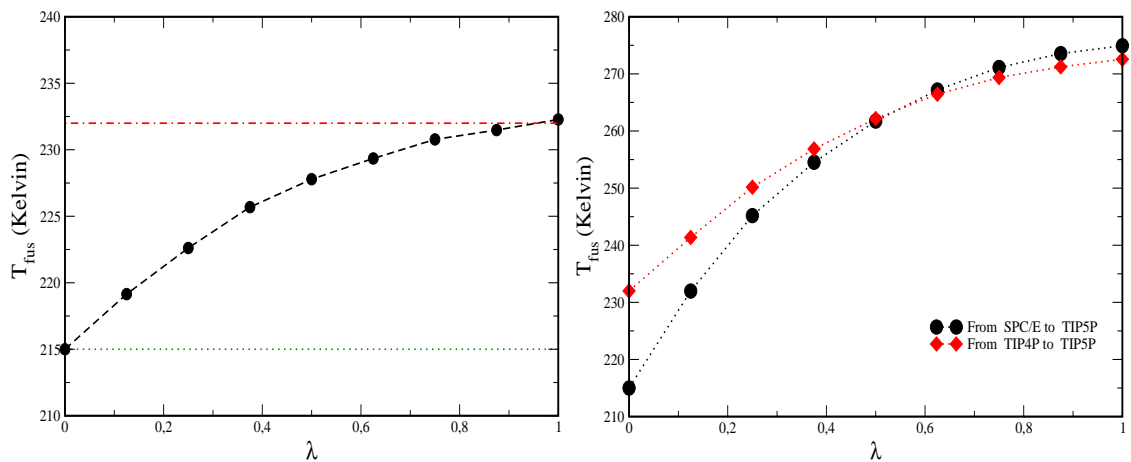

\centering
\includegraphics[clip,height=6cm,width=0.47\textwidth,angle=-0]{despcealtip4p.eps}
\includegraphics[clip,height=6cm,width=0.47\textwidth,angle=-0]{tmtip5p.eps}
\caption{ Hamiltonian Gibbs Duhem integration results. Left.
Melting temperature of ice Ih as a function of the parameter
$\lambda$ connecting the models SPC/E  ($\lambda=0$) and TIP4P ($\lambda=1$).
The points were obtained by using Hamiltonian Gibbs Duhem integration.
The dashed line is a guide the eye. The horizontal lines correspond to the
melting temperatures of SPC/E (dotted line) and TIP4P (dashed-dotted line) as obtained
from free energy calculations. Right: Melting temperature of ice Ih for the TIP5P model ($\lambda=1$)
obtained from Hamiltonian Gibbs Duhem integration starting from SPC/E  or TIP4P models. When connecting two water models by Hamiltonian Gibbs Duhem integration, the position of the oxygen atom and of the 
HOH bisector was the same in both models.}
\label{figura_hamiltonian}
\end{figure}

The liquid-Ih solid coexistence temperatures at $p=1$~bar for TIP4P and SPC/E have been
estimated  from free energy calculations to be $T=232\pm 5$~K and 
$T=215\pm 5$~K, respectively. These numbers are in relatively good agreement 
with estimates from other authors for TIP4P \cite{gao00,tanaka04,wang05}
and for SPC/E \cite{bryk02,bryk04,jungwirth_spce,brodskaya07}.
An interesting question is whether these two values (for TIP4P and SPC/E) are 
mutually consistent. 
Starting from the SPC/E model and performing constant pressure Hamiltonian 
Gibbs Duhem simulations (integrating the generalised Clapeyron equation as 
described previously) one should recover the melting temperature 
of the TIP4P. In fact starting from the SPC/E ice Ih melting point we obtain $T=232.3$~K  for 
TIP4P (see fig.\ref{figura_hamiltonian} ) in very good agreement with the result obtained through 
free energy calculations. 

Once the melting point of ice Ih for TIP4P and SPC/E seems to be firmly established
one could use these values to estimate (by using Hamiltonian Gibbs Duhem 
simulations) the melting point of another water model, as
for instance TIP5P \cite{mahoney00}. Obviously, the properties of the final model
should be independent of the reference model.
When the starting model is SPC/E we obtain $T=275$~K 
for TIP5P whereas the calculated result using the TIP4P model as a reference is
$T=273$~K. 
The agreement between both estimates is satisfactory taking into account that 
the error of the Gibbs-Duhem integration is about $3$~K.
This is is further illustrated in the right panel of figure \ref{figura_hamiltonian} 
which shows the results of the integration. 
By using Hamiltonian Gibbs Duhem integration the melting point of ice Ih for other models
of water was determined. They are presented in Table \ref{tab_tm}. Notice that
most of the water models tend to give low melting points. 

\begin{table}[htbp]
\caption{Melting points obtained from 
free energy calculations (TIP4P and SPC/E), 
Hamiltonian Gibbs 
Duhem integration (rest of the 
models)\protect \cite{abascal05b,vegafilosofico,vega05b,abascal05a},
and from direct fluid-solid coexistence.
The last column is the value of $T_{s}$ (see text)  
obtained from simulations of ice $I_{h}$ with a free interface.  
TIP4P-Ew is a water model proposed by Horn {\em et al.} \cite{horn04} and  NvdE is 
a six sites model proposed by  Nada and van der Eerden \cite{nada03}.\vspace{0.2cm}}
\label{tab_tm}
\centering
\begin{tabular}{lc@{\hspace{1em}}cc}
\hline \hline 
Model &  Free energy& Direct coexistence & Free surface  \\
\hline
TIP4P/Ice   & 272(6)  & 268(2) &  271(1)   \\
TIP4P/2005  & 252(6)  & 249(2) &  249(3) \\
TIP4P-Ew    & 245.5(6)& 242(2) &  243(2)   \\
TIP4P       & 232(4)  & 229(2) & 230(2)   \\
TIP5P       & 274(6)  & 271(2) &  -   \\
SPC/E       & 215(4)  & 213(2) & 217(2)    \\
NvdE        & 290(3)  & 288(3) &  -   \\       
\hline
\end{tabular}
\end{table}

\begin{figure}[ht]
\centering
\includegraphics[clip,height=6cm,width=0.450\textwidth,angle=0]{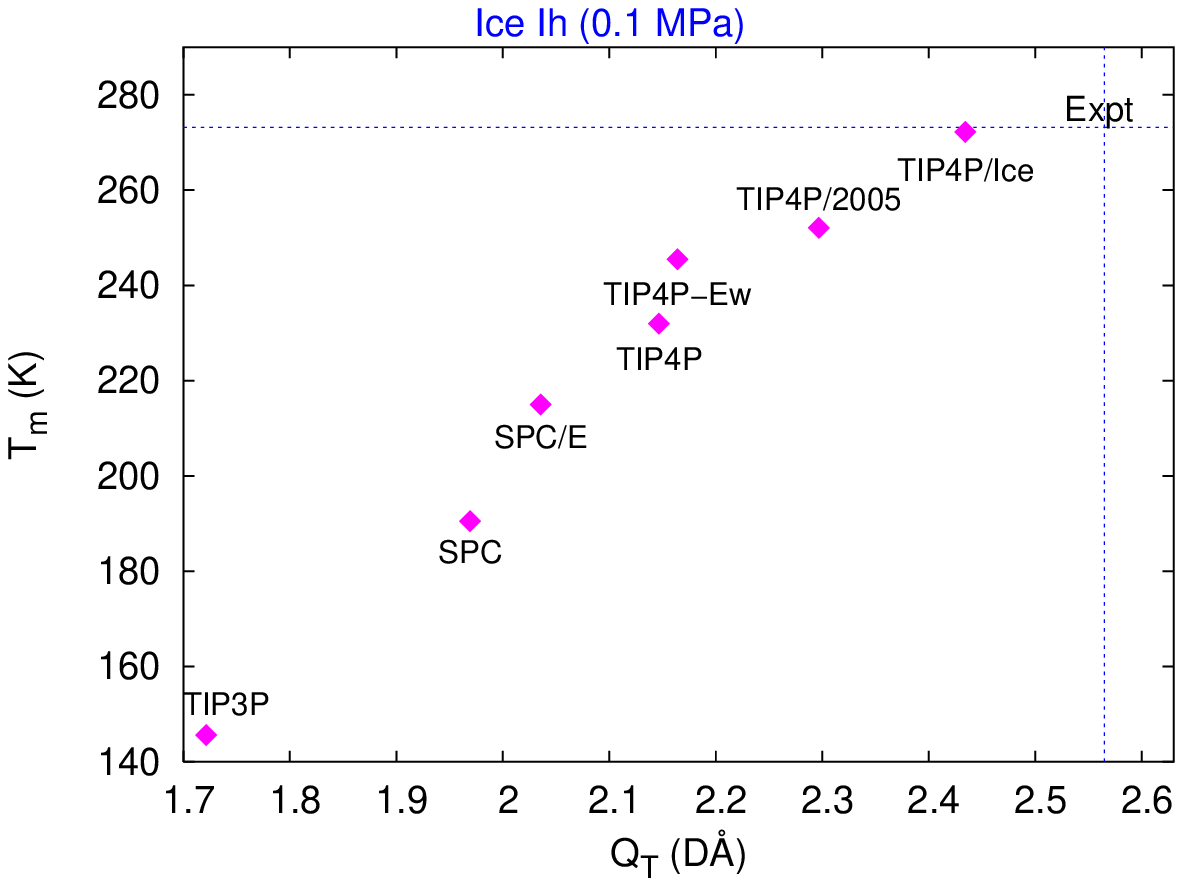}
\includegraphics[clip,height=6cm,width=0.450\textwidth,angle=0]{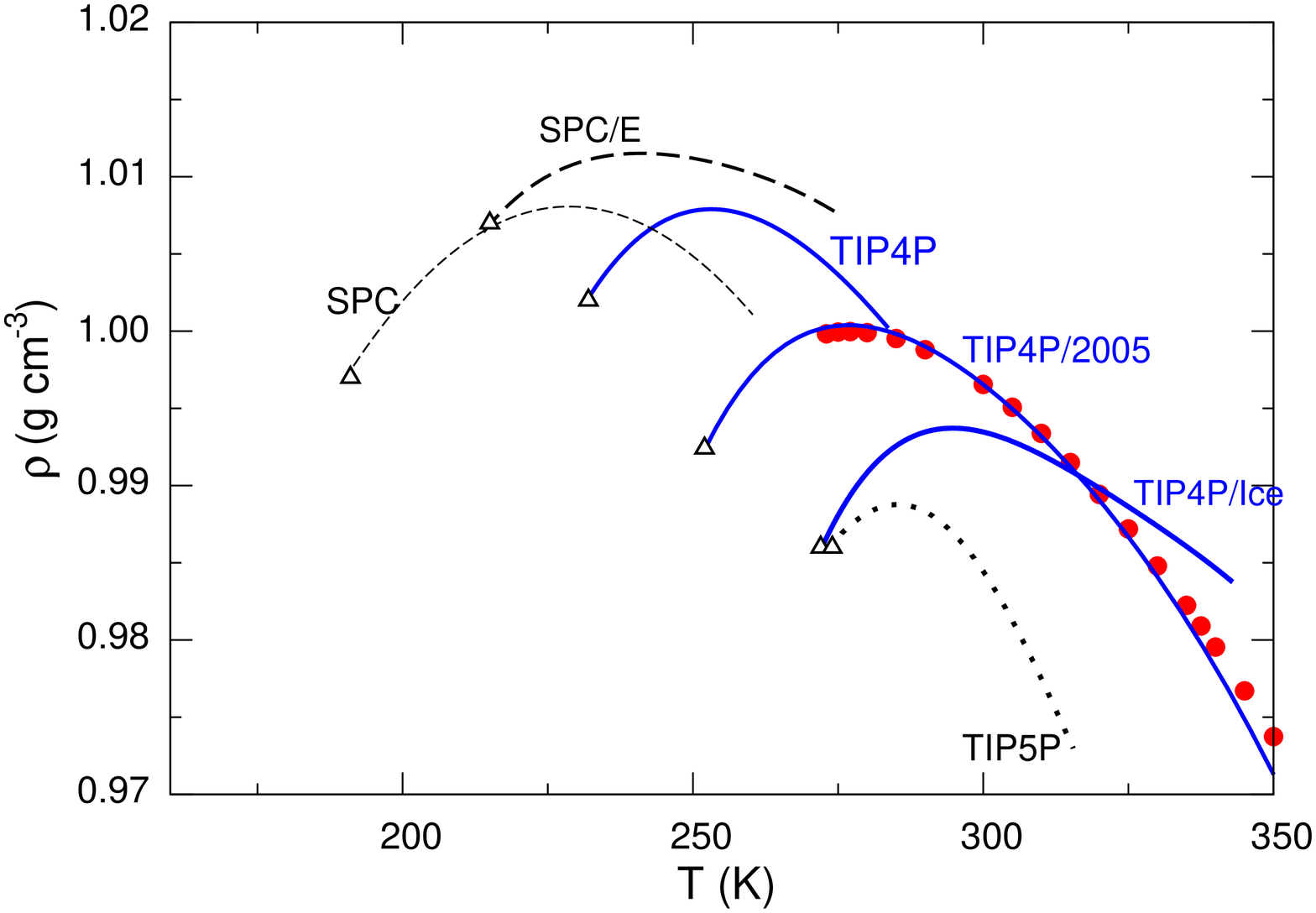}
\caption{Left: melting temperature of ice Ih at 1 bar plotted 
as a function of the quadrupole moment $Q_T$ (taken from \cite{abascal07a}). 
$Q_T$ is defined as the value of $(Q_{xx}-Q_{yy})/2$ of the quadrupolar tensor, 
where the x axis joins the two H atoms and the y axis is perpendicular to the 
molecular plane. 
Right: density of water at room pressure as a function of temperature as 
obtained from experiment (filled circles) and from computer simulations of 
several water models (lines).
The open triangles indicate the melting point of ice Ih for each model.}
\label{tm_versus_q}
\end{figure}

\begin{figure}[!hbt]\centering
\includegraphics[clip,height=6cm,width=0.47\textwidth,angle=-0]{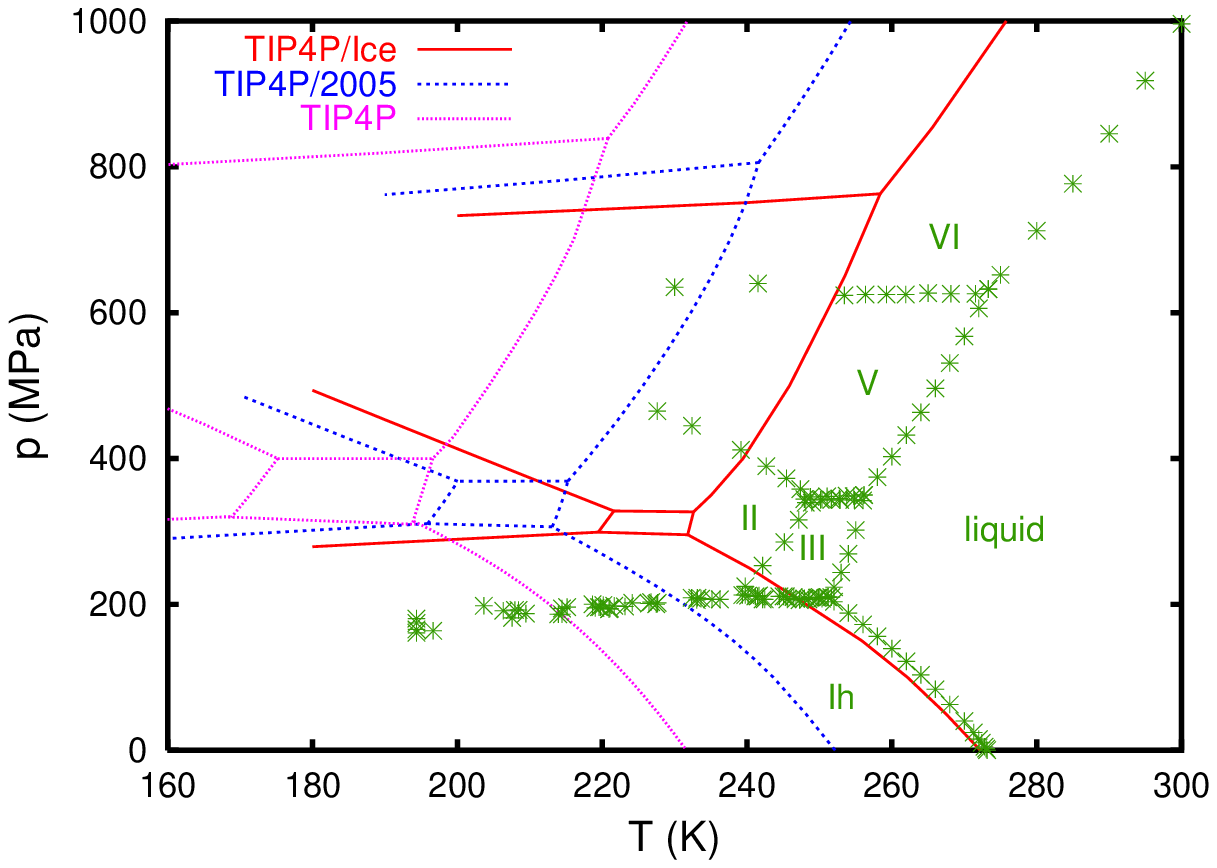}
\includegraphics[clip,height=6cm,width=0.47\textwidth,angle=-0]{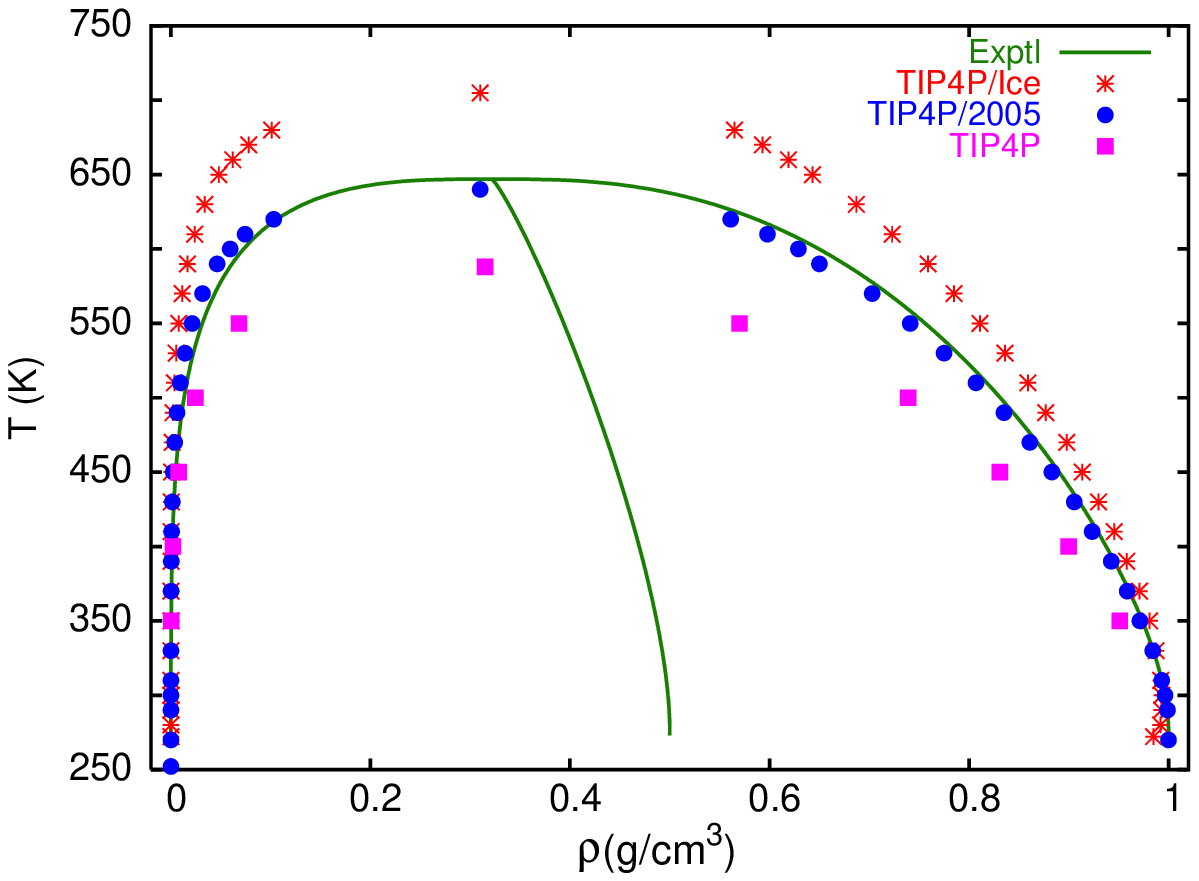}
\caption{ Phase diagram for the TIP4P family.
a) Left panel: fluid-solid equilibria (the lines show the predictions for 
several TIP4P-like models and the symbols represent the experimental data).
b) Right panel: vapor-liquid equilibria (the symbols show the predictions for 
TIP4P-like models and the lines represent the experimental data).}
\label{tip4p_family}
\end{figure}

For models with three charges (SPC, SPC/E, TIP3P, TIP4P, TIP4P/Ew, TIP4P/Ice, TIP4P/2005) 
a correlation between the melting point and the quadrupole moment has been found. 
This is illustrated in figure \ref{tm_versus_q}.
It is seen that models with rather low quadrupole moment (TIP3P, SPC, SPC/E) provide
rather low melting points. The melting temperature of TIP4P is closer to the experimental
value. Motivated by this we have proposed a new modified TIP4P model, with a higher quadrupole moment,
able to reproduce the experimental melting point of water. We have denoted this
new model as TIP4P/Ice \cite{abascal05a}. A second finding 
was that for three charge models, the temperature 
at which the maximum in density occurs at room pressure (TMD) is about 20-25~K 
above the melting temperature \cite{vega05b,paschek}. 
Experimentally for water, the
maximum in density occurs 4 degrees above the melting point.  Therefore it is impossible
with three charge models to reproduce simultaneously the melting point and the TMD. It 
is likely that the inclusion of quantum effects and/or polarizability \cite{jedlovszky_tmd,rick_tmd_tip4p_fq,sadus_water,cummings05,cummings06,cosnew,nezbeda_polarizabilidad} may be needed to 
reproduce these two properties simultaneously. 
For this reason we have proposed the TIP4P/2005, which reproduces the TMD of real water better
than any other water model proposed so far (see figure9b).
An interesting question is to analyze whether these new models still predict correctly 
the phase diagram of water. 
By using Hamiltonian Gibbs Duhem integration it is possible to estimate the phase diagram
of a certain water model, starting from the phase diagram of another reference model.
Thus by using TIP4P as reference, we have estimated the complete phase diagrams for
TIP4P/Ice and TIP4P/2005.
The obtained phase diagrams are presented in figure \ref{tip4p_family}.
As can be seen these models predict quite well the fluid-solid equilibria of water
improving the predictions of TIP4P. The TIP4P/2005 yields also an excellent prediction
of the vapor-liquid equilibria.  

\subsection{Direct coexistence simulations}
To estimate the melting point of ice Ih for several water models by direct coexistence,  NpT 
simulations 
were performed with  870 molecules and the MD program Gromacs \cite{lindahl01,spoel05}. 
In the initial configuration half of them
formed ice, and the other half were in the liquid state. Both phases were in contact so that these
are direct coexistence simulations. 
\begin{figure}[!hbt]\centering
\includegraphics[clip,height=6cm,width=0.47\textwidth,angle=-0]{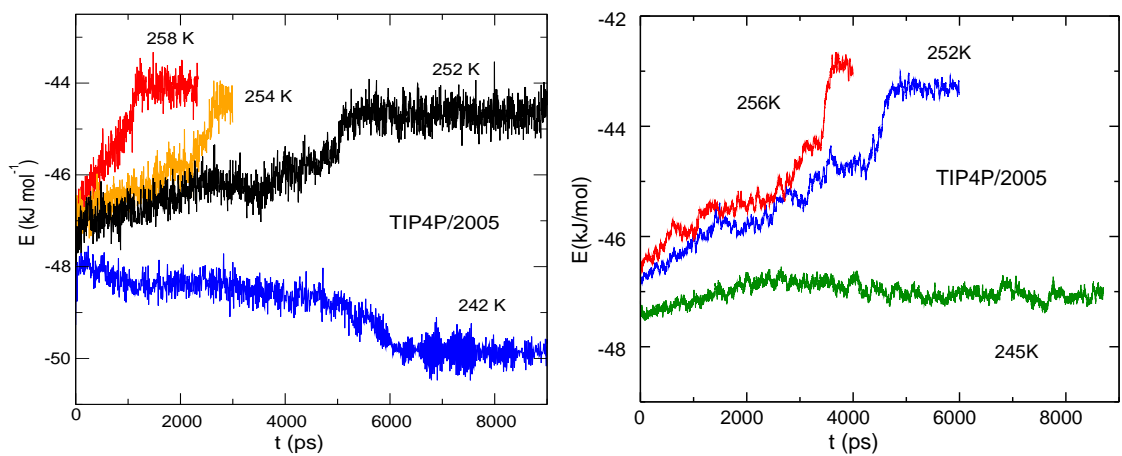}
\includegraphics[clip,height=6cm,width=0.47\textwidth,angle=-0]{fig4_maria_jpcm.eps}
\caption{a) Left panel. Evolution of the total energy (per mole of molecules)
in $NpT$ MD simulations of a box containing
ice and liquid water at 1~bar for the TIP4P/2005 model. 
b) Right panel. 
Total energy as a function of time obtained at several
temperatures by performing MD simulations of TIP4P/2005 for a block of ice $I_h$ with a free
surface. }
\label{evolucion_energia}
\end{figure}

The evolution of the energy for the TIP4P/2005 model with time is presented  
in fig.\ref{evolucion_energia}. 
As can be seen the energy increases with time for $T=252, 254, 256$~K 
reaching a plateau (the plateau indicates the complete melting of the ice). 
The energy decreases for $T=242$~K reaching a plateau (the plateau indicates
the complete freezing of the water). Snapshots of these final configurations can be 
found in Ref. \cite{ramon06}. At a temperature of 249~K the energy does not change with time,
and the interface is stable after 10ns. Therefore this is the estimate of the melting point 
by direct coexistence for TIP4P/2005. 
Similar runs were performed for other water models. 
In Table \ref{tab_tm} the melting points 
of different water models as estimated from direct coexistence are compared
to those obtained from free energy calculations. The agreement between both 
techniques is quite good. Direct interface simulations have been used by several
authors to estimate melting points or ice growth rates for different water 
models \cite{carignano05,carignano07,nada04,nada05,wang05,tension_ice_water_zeng}.

%

\subsection{ Melting point as estimated from simulations of the free surface}
In figure \ref{evolucion_energia} the evolution  
of the total energy of ice Ih (having a free surface) with time is presented \cite{maria06}. 
At high temperatures, the total energy of the system increases
continuously and then reaches a plateau (that corresponds to the complete melting 
of the solid). 
The behaviour at low temperatures is different. At the beginning (first 1-2ns), 
there is an increase of the
energy but after that the energy remains approximately constant, apart from the thermal fluctuations.
The analysis of the 
configurations of the TIP4P/2005 at $T=245$~K, shows that the increase of energy during the first 1ns 
is due to the formation of a thin liquid layer at the surface of
ice, which may indicate the onset of surface melting, mentioned already in the Introduction,
and first proposed by Faraday \cite{physics_today}. 
The formation of a quasi liquid layer on the surface of ice below $T_{m}$ has been found 
both in experiment (see \cite{dash,dash_2,henson,jcp_experimental_liquid_layer,ice_surface_melting_sio2} and references therein) and
in computer computer simulation for several potential models of 
water \cite{kroes92,nada97,nada00,kawamura04,carignano05} and it has been explained by several 
theoretical treatments \cite{dash,henson}.
By repeating the simulation at several temperatures it is possible to determine
the lowest temperature at which the block of ice melts $T_{+}$, and the highest
temperature at which it does not melt $T_{-}$. 
By taking the average of these two temperatures we obtain what we call  
$T_{s}= (T_{+} + T_{-})/2 $. $T_{s}$ provides an estimate of the melting point.
The values of $T_{s}$ obtained for water models are presented in Table 
\ref{tab_tm}. 
As can be seen, $T_{s}$ is identical to $T_{m}$, within the error bar. 
Thus for ice Ih, the presence of a free surface suppresses superheating and ice 
melts at the equilibrium melting temperature (although runs of about 10ns or
longer may be needed). 
\begin{figure}[!hbt]\centering
\includegraphics[clip,width=80mm,angle=0]{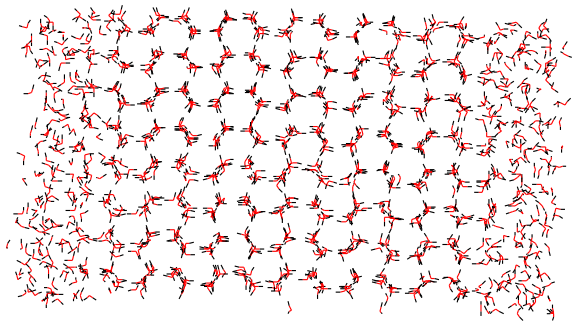}
\caption{\label{snapshot_surface_melting} 
Instantaneous configuration of the TIP4P/Ice system at
$T=268$~K at the end of a 8ns run. Although the temperature is well
below the melting point of the model, a quasi-liquid layer is clearly
present in the ice-vacuum interface. }
\end{figure}
In figure \ref{snapshot_surface_melting} the final configuration (after a 
8~ns run) obtained for the TIP4P/Ice at a temperature well below the melting 
point of the model ($T=264$~K). 
As can be seen, a quasi liquid layer is already present in the system.

\subsection{Properties at 0~K}

\begin{table}[!h]
\caption{\label{zerot_zerop_properties} Residual internal energy (in Kcal/mol) 
of several ice polymorphs at $T=0~K$ and $p=0$ for popular water models.
The results for the most stable phase of each model are presented in bold.\vspace{0.2cm}}
\centering
\begin{tabular}{ccccc}
\hline
Ice & TIP4P/Ice&TIP4P/2005& SPC/E & TIP5P \\
\hline \hline
\cline{2-5}
Ih & {\bf
-16.465}& {\bf
-15.059}& -14.691 & -14.128 \\
II  & -16.268 & -14.847 & {\bf
-14.854}& {\bf
-14.162}\\
III & -16.140 & -14.741 & -14.348 & -13.320 \\
V   & -16.049 & -14.644 & -14.169 & -13.101 \\
VI  & -15.917 & -14.513 & -13.946 & -12.859 \\
\hline
\end{tabular}
\end{table}

\begin{figure}[h!]
\centering
\includegraphics*[scale=0.4,angle=0]{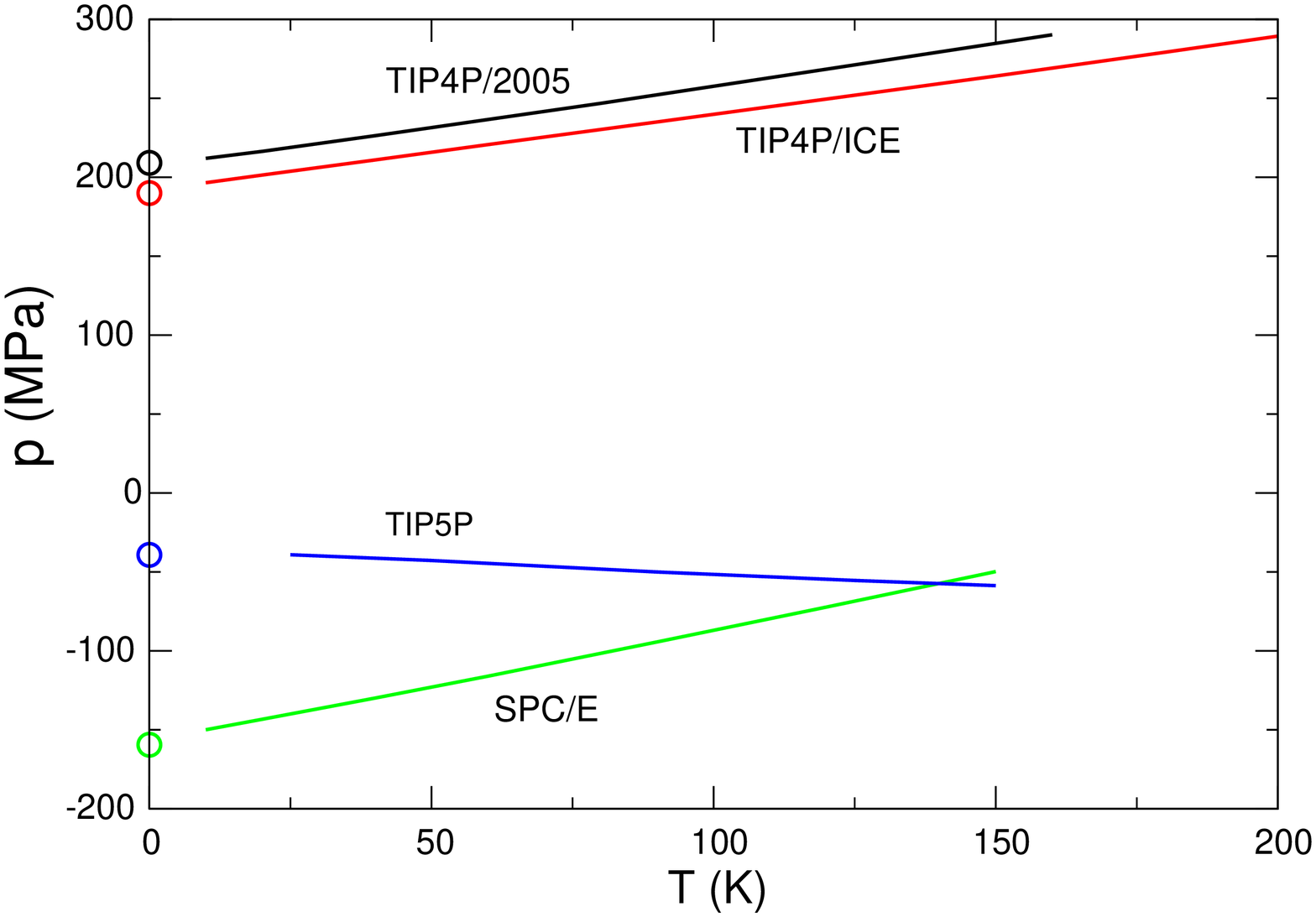}
\caption{Coexistence lines between ices Ih and II as obtained from
Gibbs-Duhem simulations for TIP4P/2005, TIP4P/Ice, SPC/E and TIP5P models
(solid lines).
The symbols represent the coexistence pressures as obtained from the properties
of the systems at zero temperature. For each water model, ice Ih is the stable phase below the
coexistence line (low pressures) and ice II is the stable phase above the coexistence line
(high pressures).}
\label{gd_I_II}
\centering
\end{figure}

In Table \ref{zerot_zerop_properties} the residual internal energies at zero
T and p  are given for the TIP4P/Ice, TIP4P/2005, SPC/E and TIP5P models.
For TIP4P/Ice and TIP4P/2005 ice Ih is the structure with the lowest energy (probably 
ice XI which is a proton ordered version of ice Ih would have an slightly lower energy
but it was not considered for this study).  
However for SPC/E and TIP5P the lowest internal energy corresponds to ice II.
Thus, for TIP4P/2005 and TIP4P/Ice ice Ih is the stable phase at zero 
pressure and temperature whereas for SPC/E and TIP5P the stable phase
is ice II.  From the properties at 0~K it is simple to 
locate the Ih-II transition pressure at 0~K (see \cite{aragones1}). 
In figure~\ref{gd_I_II} the predicted pressures using the 
calculations at 0~K are presented (circles).
The lines represent the results obtained from free energy calculations (used to
obtain an initial coexistence point at 150~K) and Gibbs Duhem integration.
It may be seen that both sets of calculations agree quite well so both sets of 
results are mutually consistent (i.e., the estimated coexistence pressure at 0~K 
is the same). 
It is clear that ice II is more stable than ice Ih at zero temperature and 
pressure for the SPC/E and TIP5P models.
This example illustrates how 0~K properties can be used to test for self
consistency in phase diagram predictions. They are also quite useful to 
test the performance of water models \cite{whalley84,baranyai05,baranyai06,s_price07,aragones1}. 

\section{ Phase diagram for a primitive model of electrolyte }

\begin{figure}[!hbt]\centering
\includegraphics[scale=0.45]{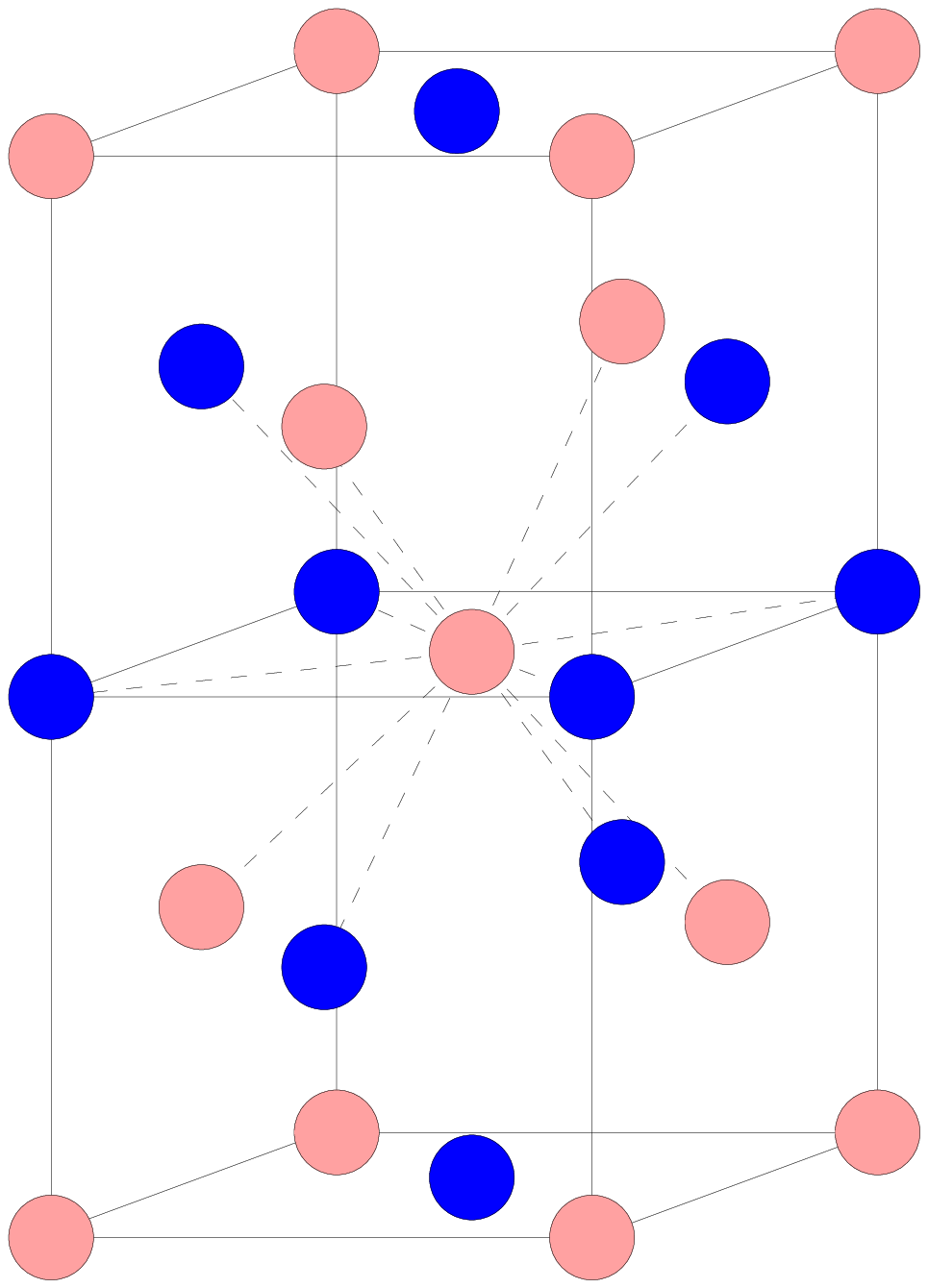} 
\hspace{1cm}
\includegraphics[clip,height=7cm,width=0.47\textwidth,angle=-0]{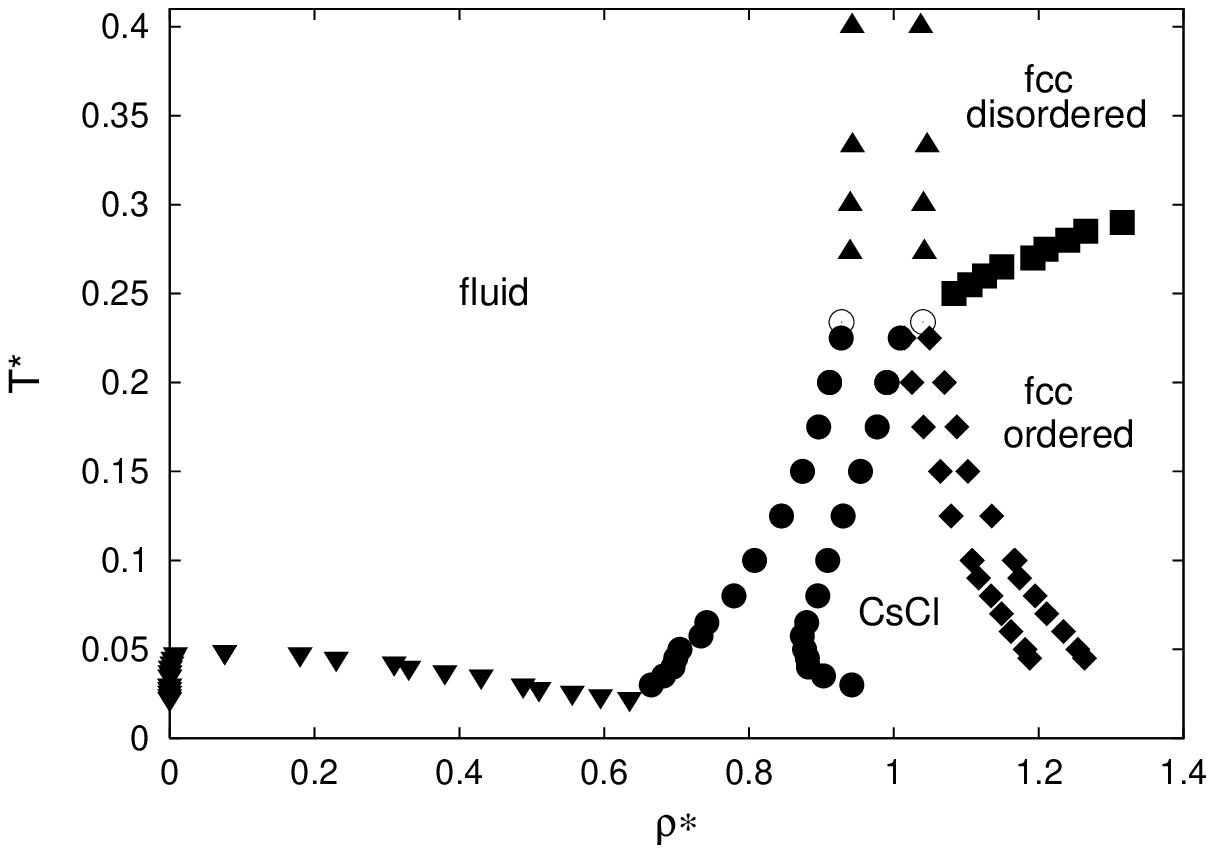}
\caption{Left: Tetragonal ordered structure of the RPM model. The atoms form an 
fcc structure, but the ions are ordered. 
Right: Phase diagram of the RPM model showing the equilibria between vapor
and liquid (inverted triangles), fluid and CsCl like structure (filled circles), 
CsCl like structure and tetragonal (fcc ordered) phase (rhombs), fluid and 
tetragonal structure (open circles), fluid and fcc disordered structure
(triangles), and ordered-disordered fcc phases (squares).} 
\label{rpm_figure}
\end{figure}

Let us now present another example of a phase diagram calculation for a completely 
different model, the restricted primitive model (RPM). 
The restricted primitive model (RPM) is one the simplest model of electrolytes.
It this model the cations are represented by hard spheres of diameter $\sigma$ having 
a charge $+q$ and anions represented by hard spheres of diameter $\sigma$ having a charge
$-q$. The model is quite simple and
for this reason it can also be studied theoretically \cite{ciach1,ciach2}.
The system has vapor-liquid equilibrium \cite{rpm_liq_vapor,lv_rpm,JCP_2002_116_03007} 
(in spite of the absence of dispersive 
forces). Several solid structures can be considered \cite{lovett}, the simplest being the 
CsCl like structure (a bcc type of structure with the anions occupying the 
vertices of a cube and the cations occupying the center of the unit cell). Another
possible structure is the fcc disordered structure. In this structure the ions occupy 
an fcc lattice, but with positional disorder (cations and anions occupy the lattice 
points in a disordered way). At low temperatures it is possible to conceive a
solid structure with an fcc arrangements of the ions, but with positional order.
The symmetry of the phase is tetragonal. This phase will be labelled as the tetragonal
phase \cite{bresme00}. It is presented in figure \ref{rpm_figure}.
Free energy calculations (Einstein crystal) were performed to determine the 
free energy of the CsCl and tetragonal structures. 
Due to the presence of partial disorder the Einstein crystal method can not be 
applied directly (to a snapshot) to get the free energy of 
the fcc disordered structure (although one may suspect that an strategy similar 
in spirit to that proposed for the plastic crystal phases 
can be also successful here if the external field is able to lead the system from disordered
configurations to an ordered solid without crossing first order transitions). 
The RPM system becomes a hard sphere at infinitely high T 
for which the free energy is known (a trivial mixing contribution should be added). 
For this reason the free energy of the fluid and of the fcc disordered solid can be 
obtained by thermodynamic integration.  
Exchange moves (where a cation and an anion exchange their positions) were used to sample
correctly the disorder, both in the fluid and in the fcc disordered structure. 

After computing the free energies, some initial coexistence points were determined, 
and then by using Gibbs Duhem integration the complete phase diagram was computed.
The resulting phase diagram \cite{MP_1996_87_0159,pre_madrid,bresme00,vega03} 
is presented in figure \ref{rpm_figure}. 
At  high  temperatures, freezing leads to the  substitutionally disordered
close packed structure.
By decreasing the temperature this solid structure undergoes an order-disorder
transition  transforming into the tetragonal solid.
At low temperatures freezing leads to the caesium
chloride structure (CsCl) which undergoes a phase transition to the
tetragonal structure at high pressures.
The tetragonal solid is the stable solid phase at low temperatures
and high densities.
In a narrow range of temperatures  coexistence between the fluid and
the tetragonal solid is observed.
Three triple points are found for the model considered:
the usual vapor-liquid-CsCl, the fluid-CsCl-tetragonal and the fluid-fcc 
disordered-tetragonal triple point (notice on figure~\ref{rpm_figure} the narrow
range of the  fluid- tetragonal solid coexistence line).

Although initially conceived to describe electrolytes and ionic salts, the RPM has been found to 
be quite useful to describe certain colloidal mixtures, consisting in equimolar mixtures
of colloidal particles of equal size, but with charges of different sign. Thus a colloidal
version of the RPM exists \cite{caballero_coloides_rpm}. For these colloidal mixtures, three different solid phases
have been found experimentally, the fcc disordered solid, the CsCl solid, and a fcc ordered 
structure \cite{rpm_nature,prl_british}. The fcc ordered structure was 
found to be of CuAu type, rather than the 
tetragonal structure of figure \ref{rpm_figure}. This difference with the 
RPM phase diagram seems to be due to the fact that in colloidal mixtures the interaction between 
charged particles is of Yukawa type rather than being purely Coulombic. This is so
because the interaction between charged particles is screened by the presence of an ionic
atmosphere  due to the counter ions of the colloids. In fact, when this screening is incorporated
in the potential with a Yukawa type model,
the CuAu structure was found to be stable in a thermodynamic region between the CsCl and
the tetragonal structure \cite{hynninen,caballero}. 

In summary the RPM has proved to be quite useful in the description of mixtures of charged 
colloidal particles. It would be of interest to determine the phase diagram for a model
where the two spheres of the model present different size. This is usually called the 
primitive model. In fact the primitive model (where cations and anions have different size) 
may indeed be a more general model than RPM, since in real
ionic fluids the size of the ions is usually different. A colloidal
realization of the PM model has been obtained experimentally \cite{rpm_nature}. Finally, there is an increasing interest in 
determining the properties of ionic liquids. It would be of interest to determine the 
factors affecting the melting point of ionic liquids \cite{ionic_liquids} which 
are regarded as new solvents. \cite{greensolvents}. Work in this direction has appeared 
recently \cite{melting_ionic_liquid}.
\section{Phase diagram of a simplified model of globular proteins}

A final  example of the application of the techniques described here
is provided by the calculation of the phase diagram of a simple
model of globular proteins. 
During the last few years there has been an increasing number
of studies of the phase behaviour of globular proteins
using very simplified models. 
The first studies have been performed using short ranged
isotropic potentials and, even with these very simple models, it was already
possible to reproduce some of the features of the phase
diagram of proteins, e.g., the existence of a metastable 
critical point \cite{tenwolde}. 
However, proteins
are known to form very low density crystals, with densities below 
those of the close-packed crystals typically formed by isotropic potentials,
which is indicative of highly directional interactions \cite{Matthews68}.
Further evidence of the importance of anisotropy in the interactions
among proteins has been recently obtained in a theoretical study of the 
fluid-fluid equilibria. This study has shown that a quantitative
agreement with experiments is obtained by the introduction of anisotropy \cite{sciortino_patchy},
as opposition to isotropic models that only provided a qualitative description.
Moreover, theoretical studies of the phase behaviour of anisotropic models 
are also acquiring much interest due to the fact that 
several experimental groups have recently been able to produce colloids that
are anisotropic either in shape or in their 
interactions \cite{pine_science2003,blaaderen_science2006,Levy06b,Roh06}.
So far there has been already a few theoretical studies of the 
phase behaviour of simple anisotropic 
models \cite{vega98,sear_jcp1999,song_pre2002,dixit_jcp2002,frenkel_jcp2003,sandler2,sandler_jpcb2005,talanquer_jcp2005},
although most of them were concerned with the fluid-fluid equilibria rather
than with the solid-fluid and solid-solid equilibria.

We have used a very simplified model which consists of
a repulsive core (the Lennard-Jones repulsive core),
plus an attractive tail modulated by Gaussian functions
located at some given positions or patches, which will be specified
by some vectors. The total energy between two interacting particles 
will be given by \cite{jon,alex,eva_patchy}:
\begin{equation}
V({\bf r}_{ij},{\bf \Omega}_i,{\bf \Omega}_j)  = 
\cases{
V_{LJ}(r_{ij})  & $r_{ij} <  \sigma_{LJ}$ \cr
V_{LJ}(r_{ij})V_{ang}(\widehat{\mathbf{r}}_{ij},{\bf \Omega}_i,{\bf \Omega}_j) & $r_{ij} \ge  \sigma_{LJ}$ \cr }
\label{eq2}
\end{equation}
\begin{equation}
V_{ang}(\widehat{r}_{ij},{\bf \Omega}_i,{\bf \Omega}_j)= \exp \left(-\frac{\theta_{k_{min},ij}^2}{2\sigma_{patchy}^2 } \right)
\exp \left(-\frac{\theta_{l_{min},ji}^2}{2\sigma_{patchy}^2 } \right),
\label{eqn4}
\end{equation}
where $V_{LJ}$ is the Lennard-Jones potential, $\sigma_{patchy}$ is the standard deviation of the Gaussian,
$\theta_{k,ij}$ ($\theta_{l,ji}$) is
the angle formed between patches $k$ ($l$) on atom $i$ ($j$) and
the interparticle vector
${\bf r}_{ij}$ ( ${\bf r}_{ji}$), and $k_{min}$ ($l_{min}$) is the patch
that minimises the magnitude of this angle.
The interaction is a maximum when both patches are pointing at
each other along the interparticle vector ${\bf r}_{ij}$ and it will decrease
as the particles deviate further from this equilibrium orientation.
We have chosen to study a model with 6 patches distributed in an octahedral 
symmetry. A relatively narrow width of
the patches ($\sigma_{patchy}$=0.3 radians), for which it is expected that the 
low density simple cubic (sc) crystal becomes stable.

\begin{figure}[!h]
\begin{center}
\includegraphics[height=5cm,angle=-0]{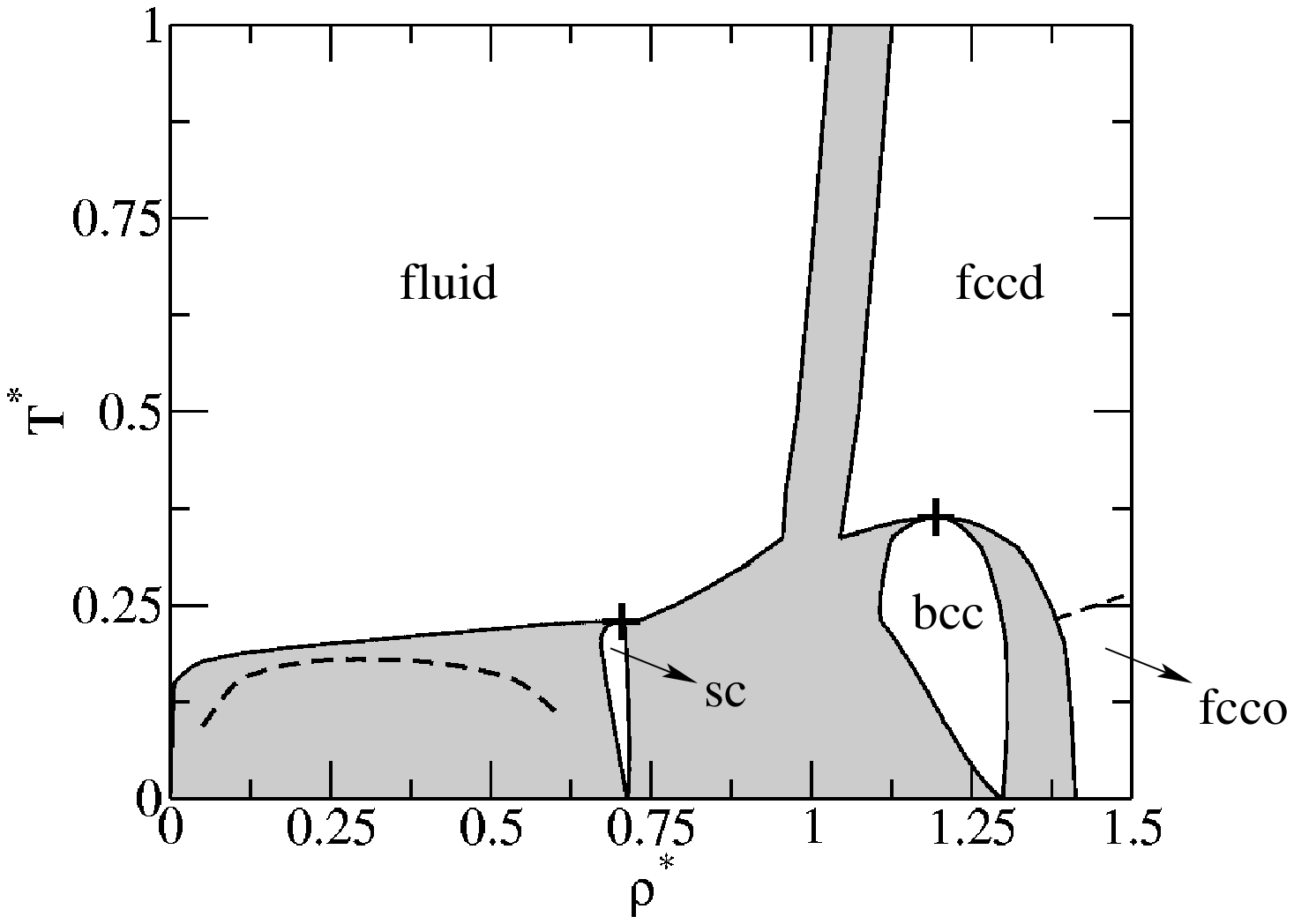} \qquad
\includegraphics[bb= 170 242 461 540,width=5cm,angle=-0]{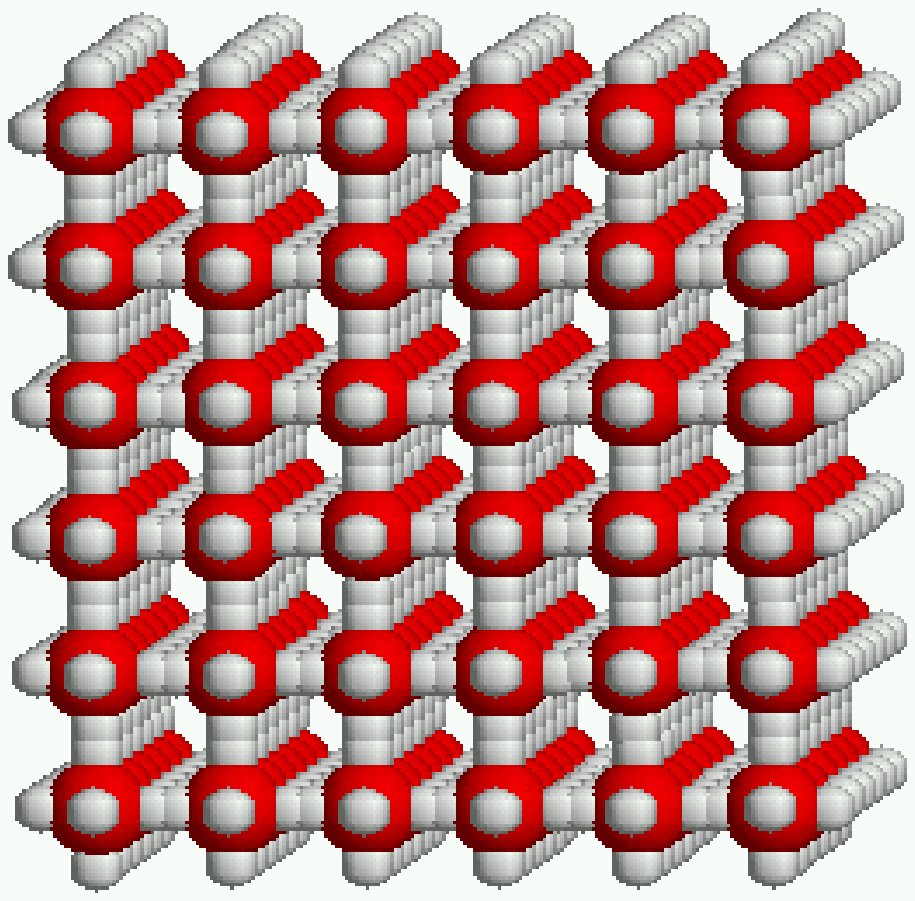}
\caption{\label{fig_patchy} Left: Phase diagram of the octahedral six-patches model.
Labels show the region of stability of each phase. The dashed line
in the fluid-sc coexistence region signals the expected fluid-fluid
binodal (see Ref. \cite{eva_patchy} for more details).
Right: Structure of the low density sc solid for the octahedral
anisotropic model.}
\end{center}
\end{figure}

Besides the sc crystal in which each of a particle's patch is pointing at each
one of its six nearest neighbours (see figure \ref{fig_patchy}), there are other
three solid phases that might be formed with this model and at this patch width,
$\sigma_{patchy}=0.3$ radians.
The first structure is a body centred cubic (bcc) solid, in which
each patch is aligned with the second neighbours. 
This structure can be also seen as two interpenetrated sc lattices
that almost do not interact between each other (similar to
the behavior of high density ice polymorphs). Therefore, a higher
density crystal is obtained with a low penalty in the energy.
At high pressures it is expected that a closed packed 
face centred cubic (fcco) structure will also appear. In this
case the patches will be also pointing to the second neighbours.
This structure exhibits a much higher energy than both the sc and the bcc solids.
Finally, at high temperatures, it is expected that a plastic
phase will also appear (fccd), i.e., a solid where the center of mass of
each particle is located at the lattice positions of a fcc structure,
but that is orientationally disordered.

Following the procedure described before, the coexistence point between two
phases was computed by imposing the conditions of equal temperature, pressure and
chemical potential. The free energy of the fluid was
computed by thermodynamic integration. In the case of the fluid, 
the equation of state was integrated up to very low densities, where the fluid
can be considered to behave as an ideal gas. The free energy of the solid
was computed by Einstein crystal calculations.
Once that a coexistence point is known the whole coexistence line have been
traced using the Gibbs-Duhem method. 

The resulting phase diagram is plotted in figure \ref{fig_patchy}.
At high temperatures the fluid freezes into the orientationally disordered
plastic crystal phase (fccd), at intermediate temperatures into the bcc solid
and at low temperatures into the sc crystal. 
The sc structure is destabilised at high pressure by the bcc solid
and, at even higher pressures, the ordered fcc solid becomes stable (fcco).
This fcco ordered solid transforms into a plastic crystal fccd as the temperature 
increases.  There are three triple points in the phase diagram:
the fluid-sc-bcc, fluid-fccd-bcc and bcc-fccd-fcco. Finally, it is also
worth noting that in the neighbourhood of the fluid-sc-bcc and
fluid-fccd-bcc triple points, reentrant melting occurs. Coexistence points
from free energy calculations were found to be in agreement with those found
from direct coexistence calculations. 

In summary, even for a relatively simple model potential, 
we have  obtained
a complex phase diagram with many solid phases and
unusual behaviour such as reentrant melting.
But the most interesting
finding is that even with a very simple model as the one described here, it is
possible to reproduce two important features
of the phase diagram of globular proteins, namely, the existence of
a metastable critical point and the stabilisation of a low density crystal.
Similar behavior has been found for a primitive model of water \cite{nezbeda_pm,vega98,sciortino_patchy_2}. 
Studies of nucleation of these models can 
be very relevant to understand the crystallization of globular
proteins.

\section{Conclusions}
In this work we have reviewed the methodology of free energy calculations and 
given examples of recent work on the determination of fluid-solid and
solid-solid equilibria by computer simulations. Free energy calculations are used to
compute initial coexistence points between phases, and then Gibbs Duhem integration
is used to compute coexistence lines. Other procedures to estimate fluid-solid equilibria
as direct coexistence and simulations of the free surface of the solid have been  discussed.
The Einstein crystal methodology and the Einstein molecule approach have been presented
in a rather comprehensive way. Both methodologies yield identical values of the 
free energy. It is shown that the free energy presents a strong N dependence, and 
that finite size corrections are needed to estimate properties of solids in the thermodynamic
limit. The issue of the symmetry of the orientational
field in Einstein crystal calculations for molecular fluids has been discussed.
We do hope that the extensive discussion of all these aspects helps
other researchers in the area to perform free energy calculations and phase diagram 
determination. In fact there are at least six areas where one can predict intense 
activity in the future. The first one is the determination of the phase diagram of 
molecular fluids. In this work, the procedure to obtain free energies for water is
presented. Besides, free energies and coexistence points for SPC/E and TIP4P models
of water are presented by the first time. These results lead to the determination 
of the full phase diagram for water, performed recently by our group \cite{sanz1}. 
The example of water illustrates clearly that phase diagram calculations for molecular
fluids is indeed feasible nowadays and that it can help to improve current models.
In fact these free energy calculations led to the proposal of an improved version
of TIP4P denoted as TIP4P/2005. This model 
is able to describe correctly the phase diagram of water, the maximum 
in density of water, the density of
the ice polymorphs including the methane hydrate \cite{sloan_review,hugh1,hugh2}, 
the vapor-liquid equilibria \cite{vega06}, 
the surface tension \cite{vega07a}, the diffusion 
coefficient, and the structure of water \cite{abascal05b,soper00} over a wide range of temperatures and pressures. 
The determination of the phase diagram of TIP4P was a crucial step in the development 
of TIP4P/2005. We do not see any difficulty  in performing similar studies to improve potential
models for other molecular fluids. The work on water, shows that  melting points obtained
from free energy calculations, direct coexistence simulations and free surface simulations,
are almost identical (taking into account the statistical uncertainties and the slightly
different implementations of the potential). 
The second area where phase diagram calculations can be useful is in the study of 
ionic systems. Here we reviewed the phase diagram of the RPM model, but it is clear that
it would be of great interest the determination of the phase diagram of  PM models and 
of other models of salts (including probably ionic liquids which are becoming 
increasingly important from a technological point of view). Some ionic systems can 
also be used to describe charged colloids. 

The third area is the area of crystallization 
of proteins. It is now clear that models with short range anisotropic forces can be 
regarded as primitive models of proteins. In such models the liquid-liquid separation 
is metastable with respect to freezing, and the competition between phase separation and 
crystallization is of great interest to understand the presence or absence of 
crystallization in proteins. The fourth area is the study of freezing under confinement 
due to the interest in understanding fluid-solid equilibria 
on the nanometer scale \cite{koga_nanotube,perla_balbuena,gubbins_review}. The fifth area is the study 
of the solubility of salts(or solids in general) in water (or solvents in general) where 
the knowledge of the chemical potential of the solute in the solid phase is required.
Very little effort has been devoted to this problem \cite{ferrariokf,sanz_nacl}.
Finally studies on nucleation \cite{trout03} should be benefit from the knowledge of the equilibrium
melting temperatures.
In summary, the study of fluid-solid and solid-solid
equilibria of molecular and complex systems by computer simulation is now feasible,
and the procedures to do it seem well established. 
The study of fluid-solid and solid-solid equilibria by 
computer simulation can play
a central role in developing potential models for condensed phases and for providing 
molecular understanding of a number of phenomena  involving solid and liquid phases. 
The enormous sensitivy of phase diagrams to interaction potentials allows to 
test the performance of the different potentials available for a certain 
substance, and offers a unique opportunity for their improvement. 

\section{Appendices}
\subsection{Appendix A. Partition function of the Einstein crystal with fixed center of mass}
\label{apendice_aeins}
The translational contribution to the partition function of an Einstein crystal
with fixed center of mass is:
\begin{eqnarray}
Q_{Ein,t}^{CM}=\frac{1}{h^{3(N-1)}} \int \exp\left[-\beta \sum ^{N}_{i=1} \frac
{{\bf p}^{2}_{i}}{2m_{i}}\right]\delta(\sum^{N}_{i=1}
{\bf p}_{i})d{\bf p}_{1}...d{\bf p}_{N}\nonumber\\
\int \exp\left[-\beta \Lambda_E \sum ^{N}_{i=1} ({\bf r}_{i}-{\bf r}_{io})^{2}\right]\delta(\sum^{N}_{i=1} \mu_{i}
({\bf r}_{i}-{\bf r}_{io}))d{\bf r}_1...d{\bf r}_N.
\end{eqnarray}
The integral over the space of momenta
is not relevant to compute the free energy and, therefore,
we will leave aside this contribution,
that we will include simply as a factor $P^{CM}$:
\begin{equation}
\label{pcm}
P^{CM}=   \frac{1}{h^{3(N-1)}} \int \exp\left[-\beta \sum ^{N}_{i=1} 
\frac{{\bf p}^{2}_{i}}{2m_{i}}\right]\delta(\sum^{N}_{i=1}
{\bf p}_{i})d{\bf p}_{1}...d{\bf p}_{N}
\end{equation}
We will focus on the integral over the configurational space:
\begin{equation}
\label{mucm}
Z^{CM}_{Ein,t}=\int \exp\left[-\beta \Lambda_E \sum ^{N}_{i=1} 
({\bf r}_{i}-{\bf r}_{io})^{2}\right]\delta(\sum^{N}_{i=1}
\mu_{i}({\bf r}_{i}-{\bf r}_{io}))d{\bf r}_1...d{\bf r}_N.
\end{equation}
This integral can be expressed in a simpler way by defining
a change of variable, ${\bf r}_{i}-{\bf r}_{io}={\bf r}^{'}_{i}$.
The Jacobian of this change of variable is 1, and the configurational
integral can be written as: 
\begin{equation}
Z^{CM}_{Ein,t}=\int \exp\left[-\beta \Lambda_E \sum ^{N}_{i=1} 
({\bf r}^{'}_{i})^{2}\right]\delta(\sum^{N}_{i=1}
\mu_{i}{\bf r}^{'}_{i})d{\bf r}^{'}_1...d{\bf r}^{'}_N.
\end{equation}
The Dirac delta function can be written \cite{frenkel_Delta_Dirac} as:
\begin{equation}
\delta(\sum^{N}_{i=1} \mu_{i}{\bf r}^{'}_{i})=\frac{1}{(2\pi)^{3}}
\int \exp\left[i{\bf k}\left(\sum^{N}_{i=1} 
\mu_{i}{\bf r}^{'}_{i}\right)\right]d{\bf k},
\end {equation}
the configurational integral can be written as:
\begin{equation}
Z^{CM}_{Ein,t}=\frac{1}{(2\pi)^{3}}\int \exp\left[-\beta 
\Lambda_E \sum ^{N}_{i=1} \left(({\bf r}^{'}_{i})^{2}-
\frac{i{\bf k}}{\beta \Lambda_E}\mu_{i}{\bf r}^{'}_{i}\right)\right] 
d{\bf k}d{\bf r}^{'}_{1}...d{\bf r}^{'}_{N}.
\end{equation}
Each term in the summation can be rewritten in a more convenient form:
\begin{eqnarray}
({\bf r}^{'}_{i})^{2}-\frac{i{\bf k}}{\beta \Lambda_E}\mu_{i}{\bf r}^{'}_{i}&=&
({\bf r}^{'}_{i})^{2}-\frac{2i{\bf k}}{2\beta \Lambda_E}\mu_{i}{\bf r}^{'}_{i}+
\frac{i^{2}k^{2}\mu^{2}_{i}}{4\beta^{2}\Lambda_E^{2}}-\frac{i^{2}k^{2}
\mu^{2}_{i}}{4\beta^{2}\Lambda_E^{2}}\\
&=&\left({\bf r}^{'}_{i}-\frac{i{\bf k}\mu_{i}}{2\beta\Lambda_E}\right)^{2}
+\frac{k^{2}\mu^{2}_{i}}{4\beta^{2}\Lambda_E^{2}}
\end{eqnarray}
Notice that in equations (84-86) there is an implicit scalar product between 
the vector ${\bf k}$ and the accompanying vector. 
The integral can then be expressed as:
\begin{equation}
\fl \qquad
Z^{CM}_{ein,t}=\frac{1}{(2\pi)^{3}}\int \exp\left[-\beta \Lambda_E 
\sum ^{N}_{i=1}\left(\left({\bf r}^{'}_{i}-
\frac{i{\bf k}\mu_{i}}{2\beta\Lambda_E}\right)^{2}+
\frac{k^{2}\mu^{2}_{i}}{4\beta^{2}\Lambda_E^{2}}\right)
\right] d{\bf k} d{\bf r}^{'}_{1}...d{\bf r}^{'}_{N}
\end{equation}
It is now convenient to do another change of variable:
\begin{equation}
{\bf r}^{''}_{i}={\bf r}^{'}_{i}-\frac{i{\bf k}\mu_{i}}{2\beta\Lambda_E}
\end{equation}
To compute the Jacobian associated to this transformation,
we will consider the simple case of a system in one-dimension
and with two particles. The Jacobian J is given by
the following determinant:
\begin{eqnarray}
J = \left|
\begin{array}{ccc}
\frac{\partial {\bf r}^{'}_{1}}{\partial {\bf r}^{''}_{1}} &
\frac{\partial {\bf r}^{'}_{1}}{\partial {\bf r}^{''}_{2}} &
\frac{\partial {\bf r}^{'}_{1}}{\partial {\bf k}}\\
\frac{\partial {\bf r}^{'}_{2}}{\partial {\bf r}^{''}_{1}} &
\frac{\partial {\bf r}^{'}_{2}}{\partial {\bf r}^{''}_{2}}&
\frac{\partial {\bf r}^{'}_{2}}{\partial {\bf k}}\\
\frac{\partial {\bf k}}{\partial {\bf r}^{''}_{1}} &
\frac{\partial {\bf k}}{\partial {\bf r}^{''}_{2}}&
\frac{\partial {\bf k}}{\partial {\bf k}}
\end{array}
\right|
\end{eqnarray}

\begin{eqnarray}
J = \left|
\begin{array}{ccc}
1 & 0 & 0\\
0 & 1 & 0\\
\frac{-2\beta\Lambda_E}{i\mu_1} & \frac{-2\beta\Lambda_E}{i\mu_2} & 1
\end{array}
\right| =1
\end{eqnarray}
After applying the change of variable, the integral can be
expressed as:
\begin{equation}
\fl \qquad
Z^{CM}_{Ein,t}=\frac{1}{(2\pi)^{3}}\int \exp\left[-\beta \Lambda_E 
\sum ^{N}_{i=1}\left(({\bf r}^{''}_{i})^{2}+\frac{k^{2}\mu^{2}_{i}}
{4\beta^{2}\Lambda_E^{2}}\right)
\right] d{\bf k} d{\bf r}^{''}_{1}...d{\bf r}^{''}_{N}
\end{equation}
This integral can be split in two Gaussian integrals:
\begin{equation}
\fl \qquad
Z^{CM}_{Ein,t}=\frac{1}{(2 \pi)^{3}}\int \exp\left[-\beta \Lambda_E 
\sum ^{N}_{i=1}({\bf r}^{''}_{i})^{2}\right]d{\bf r}^{''}_{1}...d{\bf r}^{''}_{N}
\int \exp\left[-\frac{k^{2}\sum ^{N}_{i=1}\mu^{2}_{i}}{4\beta \Lambda_E} \right] d{\bf k}
\end{equation}
whose solution is:
\begin{equation}
Z^{CM}_{Ein,t}=\frac{1}{(2\pi)^{3}}\left(\frac{\pi}{\beta 
\Lambda_E}\right)^{N3/2}\left(\frac{4\beta \pi \Lambda_E}
{\sum ^{N}_{i=1}\mu^{2}_{i}}\right)^{3/2}
\end{equation}
Doing a bit of algebra it can be shown that:
\begin{equation}
Z^{CM}_{Ein,t}=\left(\frac{\beta \Lambda_E}{\pi}\right)^{3/2}\left(\frac{\pi}{\beta \Lambda_E}\right)^{N3/2}
\left(\sum ^{N}_{i=1}\mu^{2}_{i}\right)^{-3/2}
\end{equation}
or, more simply:
\begin{equation}
\label{derivation}
Z^{CM}_{Ein,t}=\left(\frac{\pi}{\beta \Lambda_E}\right)^{3(N-1)/2}
\left(\sum ^{N}_{i=1}\mu^{2}_{i}\right)^{-3/2}
\end{equation}
Summarising, we have obtained that the partition function of
an Einstein crystal with fixed center of mass is given by:
\begin{equation}
\label{ecuacionfinal1apendiceA}
Q^{CM}_{Ein,t}= P^{CM} \left(\frac{\pi}{\beta \Lambda_E}\right)^{3(N-1)/2}
\left(\sum^{N}_{i=1}\mu^{2}_{i}\right)^{-3/2}.
\end{equation}
When all molecules are identical the reduced mass $\mu_{i}$ is simply $1/N$. Therefore the
previous equation can be simplified to:
\begin{equation}
\label{ecuacionfinal2apendiceA}
Q^{CM}_{Ein,t}= P^{CM} \left(\frac{\pi}{\beta \Lambda_E}\right)^{3(N-1)/2}
\left(N\right)^{3/2}.
\end{equation}
which is the final expression for the free energy of an ideal Einstein crystal with fixed 
center of mass. 
An explicit expression for $P^{CM}$ is not needed to get the free energy of the solid since
it cancels out with a similar term in equation (\ref{deltaa3}). 
However, it is not difficult to obtain $P^{CM}$  by realizing that equation 
(\ref{pcm}) is formally identical to equation (\ref{mucm}) (with  $\mu_i=1$ and
$\Lambda_E=1/(2m_i)$ and ommiting the prefactor $h^{3(N-1)}$). 
A derivation similar to that used to get equation (\ref{derivation}) from 
equation (\ref{mucm}) leads to:
\begin{equation}
\label{pcm_final}
P^{CM}= \frac{1}{\Lambda^{3(N-1)}} N^{-3/2}
\end{equation}
to be compared with 
\begin{equation}
P = \frac{1}{\Lambda^{3N}} 
\end{equation}
so that the ratio $P^{CM}/P$ adopts the value $\Lambda^3 N^{-3/2}$. 
If equation (\ref{pcm_final}) is replaced into equation 
(\ref{ecuacionfinal2apendiceA}) one obtains:
\begin{equation}
Q^{CM}_{Ein,t}= \frac{1}{\Lambda^{3(N-1)}} \left( \frac{\pi}{\beta \Lambda_E} \right)^{3(N-1)/2}
\end{equation}
which is dimensionless as it should be. 

\subsection{Appendix B: Computing  $U_{Ein}^{CM}$ within a  Monte Carlo program }
\label{apendiceb}
The condition of fixing the center of mass of the reference points is 
quite useful since it eliminates any divergence of the integrand of 
equation (\ref{lqvi}).  A displacement $\vec{d}$ of a given molecule must be accompanied
by a displacement $(-\vec{d}/(N))$ of all the molecules of the system (assuming all particles
are identical),
so that the center of mass remains in its original position. 
In practice, this
is not very convenient. It is more convenient  to perform a simulation
without the restriction over the center of mass but keeping
track of the position of the center of mass \cite{frenkelbook}. 
Let us denote as 
${\bf r}_{io}$ the initial position of the reference point of molecule i 
in the perfect lattice, and  $\Delta {\bf R}_{CM}$ is the difference
between the present position of the center of mass
and its initial position. Let us denote by ${\bf r}^{U}_{i}$  the actual position in the simulation
(without the restriction in
the center of mass) of the reference atom of molecule i. 
Let us define $\Delta{\bf r}_{i} = {\bf r}^{U}_{i}-{\bf r}_{io}-\Delta{\bf R}_{CM}$.
Let us compute the energy change when in a trial move, the random displacement of molecule 
is ${\bf \Delta}_i$.  The energy of the old (prefix $old$) and new (prefix $new$) configurations
is given by:
\begin{eqnarray}
\fl
U^{CM,old}_{Eins}/\Lambda_E & = & (\Delta {\bf r}_{i}^{old})^2+ 
\sum_{j\neq i} (\Delta {\bf r}_{j}^{old})^2 \nonumber \\
\fl
U^{CM,new}_{Eins}/\Lambda_E &=&
\left({\bf r}_{i}^{U,old}+{\bf \Delta}_i-{\bf r}_{io}-\Delta {\bf R}_{CM}^{old}
-\frac{{\bf \Delta}_i}{N}\right)^2 +
\sum_{j\neq i}
\left({\bf r}_{j}^{U,old}-{\bf r}_{jo}-\Delta {\bf R}_{CM}^{old}-
\frac{{\bf \Delta}_i}{N}\right)^2 \nonumber \\
&=& \left(\Delta {\bf r}_{i}^{old}+{\bf \Delta}_i-\frac{{\bf \Delta}_i}{N}\right)^2+
\sum_{j\neq
i}\left(\Delta {\bf r}_{j}^{old}-\frac{{\bf \Delta}_i}{N}\right)^2
\end{eqnarray}
We have assumed that all particles have the same mass, so that a displacement
${\bf \Delta}_i$ of one of them leads to a displacement ${\bf \Delta}_i/N$ of the
center of mass. We have not included the orientational contribution since it cancels out when 
computing the energy change (i.e., the orientational energy is not affected by the translation 
of molecule i).  Therefore, the potential energy change will be given by:
\begin{equation}
\fl
\Delta U^{CM}_{Eins}/\Lambda_E=2\Delta {\bf r}_{i}^{old}.{\bf \Delta}_i - 
2\Delta {\bf r}_{i}^{old}.\frac{{\bf \Delta}_i}{N}+\left(\frac{N{\bf \Delta}_i-
{\bf \Delta}_i}{N}\right)^2 + \sum_{j\neq i}\left[\left(\frac{{\bf \Delta}_i}{N}\right)^2 
- 2{\bf \Delta} {\bf r}_{j}^{old}\frac{{\bf \Delta}_i}{N}\right]
\end{equation}
This equation can be simplified since, as the center of mass is constrained, 
it holds that $\sum_{i=1}^{N}\Delta {\bf r}_{i}=0$
(in the right hand side, the second and last terms cancel out):
\begin{equation}
\fl
\Delta U^{CM}_{Eins}/\Lambda_E=2\Delta {\bf r}_{i}^{old}.{\bf \Delta}_i
+\left(\frac{N{\bf \Delta}_i-{\bf \Delta}_i}{N}\right)^2 +
\sum_{j\neq i}\left(\frac{{\bf \Delta}_i}{N}\right)^2.
\end{equation}
It is easy to show that this expression can be also written as:
\begin{equation}
\Delta U^{CM}_{Eins}/\Lambda_E=2\Delta {\bf r}_{i}^{old} . {\bf \Delta}_i+{\bf \Delta}_i^{2}\left(\frac{N-1}{N}\right)
\end{equation}
In this way, it is possible to perform a MC simulation
without keeping the center of mass fixed, but including this constraint
through the Monte Carlo acceptance rule.

\subsection {Appendix C. The Frenkel-Ladd expression}
\label{apendicec}
In 1984 Frenkel et Ladd derived an expression for the 
free energy of a solid. It is not the same as that given by Polson {\em et al.} \cite{polson00} and presented 
also in this paper. The reason of this difference is the following:
\begin{itemize}
\item{ 1. The expression used for $\Delta A_3$  by Frenkel and Ladd was 
$(1/N) \ln N$ (in $Nk_BT$ units) lower than the correct one. }
\item{ 2. The expression used for $A^{CM}_{Ein-id}$ by Frenkel and Ladd was 
$(3/N) \ln N$ (in $Nk_BT$ units) higher than the correct one.}
\end{itemize}
Taking into account both contributions it turns out that the 
Frenkel Ladd expression gives an energy (in $Nk_{B}T$ units) $(2/N) \ln N$ higher
than the correct one. 
Let us describe briefly the source of these two discrepancies. 
For $\Delta A_3$ Frenkel and Ladd used :
\begin{equation}
\label{deltaa3_2}
\Delta A_{3}^{FL} =
A_{sol}-A^{CM}_{sol}=k_BT (\ln (P^{CM}/P) -   \ln (V) )
\end{equation}
instead of :
\begin{equation}
\label{deltaa3_3}
\Delta A_{3} =
A_{sol}-A^{CM}_{sol}=k_BT (\ln (P^{CM}/P) -   \ln (V/N) )
\end{equation}
which is the expression to be used when all permutations, $N!$, are included in the reference
ideal Einstein crystal. 
The second discrepancy is due to the fact that the constraint of fixing the 
center of mass was implemented by Frenkel and Ladd as :
\begin{eqnarray}
\sum^{N}_{i=1}  ({\bf r}_{i}-{\bf r}_{io})&=&0
\end{eqnarray}
instead of :
\begin{eqnarray}
\sum^{N}_{i=1} \mu_{i} ({\bf r}_{i}-{\bf r}_{io})&=&0
\end{eqnarray}
with $\mu_{i}=1/N$. 
One can simply say that to fix the center of mass 
Frenkel and Ladd used $\mu_i=1$ instead of $\mu_i=1/N$. 
It is simple to analyze the mathematical consequences of that.
It is just enough to look Appendix A, and to see what happens in the final expression 
(equation (\ref{ecuacionfinal1apendiceA})) when $\mu_i=1$  is used instead of
$\mu_i=1/N$.
Then one obtains:
\begin{equation}
\label{ecuacionfinal2apendiceC}
Q^{CM}_{Ein,t}= P^{CM} \left(\frac{\pi}{\beta \Lambda_E}\right)^{3(N-1)/2}
\left(N\right)^{-3/2}.
\end{equation}
By comparing equation (\ref{ecuacionfinal2apendiceC}) (Frenkel-Ladd) with  
equation (\ref{ecuacionfinal2apendiceA}) it is simple
to see how the Frenkel Ladd expression for $Q^{CM}_{Ein,t}$ gives a contribution (in $Nk_BT$ units) $(3/N) \ln N$ higher 
than the correct one. 

In summary, when all factors are considered, the Frenkel-Ladd expression gives an energy 
(in $Nk_BT$ units) $(2/N) \ln N$ higher than the correct one.

\ack
One of us (CV) would like to thank Peter Monson (Amherst,USA) for introducing 
him, during his post-doctoral stay in 1991-1992 into the fascinating world of 
fluid-solid equilibria and to Daan Frenkel for teaching him the first steps in 
computer simulation. 
It is a pleasure to acknowledge to L G MacDowell, C McBride, and R G Fernandez 
for many helpful discussions and for their contribution to different parts of 
this work.
The same is true for the younger members of our group MM Conde and JL Aragones. 
Discussions with E de Miguel (Huelva), A Patrykiejew (Lublin), I Nezbeda 
(Prague) F Bresme (London), A A Louis and J P K Doye (Oxford) are also 
gratefully acknowledged. 
This work was funded by grants FIS2007-66079-C02-01
from the DGI (Spain), S-0505/ESP/0229 from the CAM, MTKD-CT-2004-509249
from the European Union and 910570 from the UCM. 
We would like to thank Francesco
Sciortino for the invitation to write this review. 
E.G.N. wishes to thank the Ministerio de Educaci\'on y Ciencia
and the Universidad Complutense de Madrid for a Juan de la Cierva fellowship.

\section{References}
\bibliographystyle{./apsrev}

\end{document}